\documentclass[12pt,lineno]{article}

\usepackage{amsmath,amsthm,amssymb,amsfonts}
\usepackage{hyperref,lineno,graphicx,subcaption}
\usepackage{natbib}
\setcitestyle{authoryear}

\usepackage{xcolor}
\usepackage{pgfplots}
\usepackage{titletoc}
\usepackage{comment}
\usepackage{booktabs}
\usepackage{floatrow}
\usepackage{threeparttable}
\usepackage{float}
\usepackage{bm}

\usepackage{setspace}

\usepackage[margin=0.98in]{geometry}

\newtheorem{lemma}{Lemma}
\newtheorem{proposition}{Proposition}
\newtheorem{remark}{Remark}

\newtheorem{theorem}{Theorem}

\newtheorem{assumption}{Assumption}

\numberwithin{equation}{section}
\numberwithin{proposition}{section}
\numberwithin{corollary}{section}
\numberwithin{lemma}{section}
\numberwithin{remark}{section}
\numberwithin{assumption}{section}

\hypersetup{
	colorlinks,
	linkcolor={red!50!black},
	citecolor={blue!50!black},
	urlcolor={blue!80!black}
}

\title{Learning about Treatment Effects with Prior Studies:\\ A Bayesian Model Averaging Approach\thanks{We thank Marc Agusti and Gevorg Khandamiryan for excellent research assistance. Usual disclaimer applies.} } 

\author{Frederico Finan \thanks{Department of Economics, 508-1 Evans Hall, Berkeley, California 94720-3880. Email: ffinan@berkeley.edu; and BREAD, IZA, NBER}\\UC Berkeley \and Demian Pouzo\thanks{Department of Economics, 508-1 Evans Hall, Berkeley, California 94720-3880. Email: dpouzo@econ.berkeley.edu }\\UC Berkeley }

\date{\today}

\begin{document}

\maketitle
\thispagestyle{empty}

\begin{abstract}
We establish concentration rates for estimation of treatment effects in experiments that incorporate prior sources of information --- such as past pilots, related studies, or expert assessments --- whose external validity is uncertain. Each source is modeled as a Gaussian prior with its own mean and precision, and sources are combined using Bayesian model averaging (BMA), allowing data from the new experiment to update posterior weights. To capture empirically relevant settings in which prior studies may be as informative as the current experiment, we introduce a nonstandard asymptotic framework in which prior precisions grow with the experiment's sample size. In this regime, posterior weights are governed by an external-validity index that depends jointly on a source’s bias and information content: biased sources are exponentially downweighted, while unbiased sources dominate. When at least one source is unbiased, our procedure concentrates on the unbiased set and achieves faster convergence than relying on new data alone. When all sources are biased, including a deliberately conservative (diffuse) prior guarantees robustness and recovers the standard convergence rate.

\vspace{.2cm}

\end{abstract}

\newpage

\setcounter{page}{1}




\doublespacing

\section{Introduction}

Governments, firms, and researchers often conduct experiments to estimate the causal effects of a given policy or intervention. In practice, they rarely run a single, isolated experiment; instead, policies are piloted, refined, and expanded in sequential waves. In these settings, new experiments begin with substantial prior information---results from earlier pilots, studies conducted in related populations, or expert assessments about likely effect sizes. Incorporating this information can dramatically reduce the cost of experimentation and accelerate inference. However, prior evidence typically varies in relevance and external validity, and na\"{i}vely pooling it with new data can lead to biased estimates and misguided policy decisions. A central question is therefore how to systematically incorporate prior sources for learning the expected effect of treatment, while allowing for the possibility that some may be biased, misspecified, or only partially informative for the current environment.


This paper provides a principled solution by merging standard estimation of treatment effects with treating each prior experiment (or expert assessment) as a distinct ``model'' in the Bayesian model averaging (BMA) tradition.\footnote{For excellent reviewss of BMA, see \cite{KassRaftery1995,Hoeting1999,Wasserman2000,Steel2020} among others.} The experimenter uses new data to update the posterior probability that each source is externally valid for the current environment, so the resulting estimator --- a weighted average of source-specific posterior means for each treatment-covariate pair, where the weights are BMA posteriors --- automatically assigns more weight to sources supported by the data and downweights those that are inconsistent or biased. This approach offers desirable properties: it yields faster learning when externally valid sources exist, and it remains robust when they do not. These properties arise under an asymptotic framework that fundamentally differs from the standard asymptotics used in Bayesian model averaging.

While our analysis is formally expressed using BMA and Bayes model posteriors, their role here differs fundamentally from that in the standard BMA literature. Classical theoretical results rely on an asymptotic regime in which the current experiment grows large while the information content of each prior source remains fixed \citep[e.g.,][]{Schwarz1978,Walker2004,Wasserman2000,Steel2020}. Although this regime is useful in many model-selection settings, it can be a coarse or even misleading approximation in environments such as multi-site RCTs, sequential pilots, and phased rollouts, where prior studies may be comparable in size to—or larger than—the ongoing experiment. In these cases, standard BMA asymptotics offer limited guidance on how posterior weights evolve across sources and on the resulting speed of learning of the estimator. 

To address this gap, the paper introduces a new asymptotic framework in which the precision of each prior source is allowed to grow with the sample size of the ongoing experiment. This device captures settings where all available evidence---past and present---contains substantial information, and it allows us to characterize how uncertainty from both the new data and the priors is resolved. This delivers a novel form of discrimination that reflects learning about external validity of sources rather than differences in likelihood fit. This discrimination yields oracle-type and robustness results that emerge as implications of modeling empirically relevant environments with large and heterogeneous sources of prior information, rather than as objectives built into the asymptotic design. 

The oracle-type result implies that when at least one prior source is unbiased---its prior mean coincides with the true treatment effect---the posterior asymptotically identifies these sources and assigns weight only to them. As a result, our estimator converges strictly faster than the standard estimator based solely on the new data. The magnitude of this improvement depends on the relative size of the unbiased prior sources: when such sources are of comparable scale to the current experiment, the estimator achieves a convergence rate that is twice as fast, representing a substantial reduction in estimation error.

Equally important, the framework delivers a form of robustness. If all prior sources are biased but at least one diffuse (low-precision) source is included, the procedure automatically downweights the biased sources and the estimator converges to the truth at the standard rate. In this way, the procedure never performs worse than using only the new experiment, and performs strictly better whenever externally valid sources exist. This robustness property is, to our knowledge, absent from the classical BMA literature, which --- under global misspecification --- lacks a mechanism that guarantees performance no worse than using the new experiment alone. 
Here, robustness follows from a simple safeguard: the inclusion of a deliberately diffuse source, which can always be incorporated by the researcher.

The mechanism behind these results differs fundamentally from existing BMA theory. In classical applications---such as Bayesian variable selection---models differ in their likelihood or parameter dimension, and asymptotic selection is driven by differences in goodness-of-fit penalized by model complexity \citep[e.g.,][]{KassRaftery1995,RafteryMadiganHoeting1997,RafteryMadiganHoeting2002}. In our setting, by contrast, all sources correspond to the same scalar parameter (the expected outcome for each treatment-covariate pair) and share the same likelihood; they differ only in their prior means and precisions. Under standard asymptotics, these differences vanish asymptotically and offer no discriminatory power. Under our asymptotic regime, however, posterior discrimination operates through a continuous external-validity index that depends jointly on a source's bias and effective precision. Biased sources are exponentially downweighted, while unbiased sources dominate with weights proportional to their asymptotic information content. These results demonstrate how, under our nonstandard asymptotic framework, Bayesian model averaging provides a coherent and transparent method for incorporating prior experimental evidence when external validity is uncertain.


Why focus on learning/convergence rates? From a theoretical perspective, convergence rates are the basic building blocks for downstream frequentist results, including asymptotic normality, coverage guarantees, and the behavior of plug-in decision rules. Establishing sharp convergence rates is therefore an important step to determine the statistical precision available at each sample size that underpins all subsequent inferential and decision-theoretic guarantees. In sequential experimentation, however, rates are also interesting in their own right. They determine how quickly the true treatment effects are learned, and therefore how soon an experimenter can credibly stop, scale up, or revise a policy. At a given sample size, faster rates translate directly into tighter policy recommendations, and they quantify the value of borrowing from prior evidence relative to collecting additional observations. 

The paper also extends these results in two directions that are especially relevant in practice. First, we allow outcomes to be binary and develop a Bernoulli version of the model. In that case, the same logic continues to govern posterior source weights, but external validity is no longer summarized by a quadratic loss. Instead, it is measured by a Kullback-Leibler projection index that captures the likelihood cost of reconciling a source's prior belief with the target environment. This delivers the same qualitative conclusions as in the Gaussian benchmark: biased sources are exponentially downweighted, while unbiased or sufficiently diffuse sources determine the asymptotic behavior of the estimator.

Second, we allow different sources to bring different sets of control variables in a Gaussian linear regression framework. This changes the geometry of learning because posterior updating is no longer arm-by-arm: treatment effects and nuisance coefficients are learned jointly, and the relevant external-validity object becomes matrix-valued after partialling out controls. Even so, the same oracle and robustness logic survives. 
Together, these extensions show that our framework is not tied to the simplest Gaussian setup but applies more broadly to environments researchers actually face.

\paragraph{Related literature.} Perhaps closest in spirit to our asymptotic framework are the ideas behind Zellner's $g$-prior \citep{Zellner1986}. The $g$-prior is used in a different problem than ours---variable selection in a Gaussian linear regression model---but it has the notable feature of a parameter (the ``$g$'' factor) that regulates the prior precision relative to sample size. The literature has explored many choices and hyperpriors for $g$, often motivated by model-selection consistency and predictive performance; see, for example, \citet{FernandezLeySteel2001} and \citet{LiangEtAl2008} for overviews and prominent proposals. For fixed values of $g$, the influence of the prior does not vanish, even asymptotically; in that sense, this feature resembles ours. However, the motivations and implications of these choices are different from ours: in our setting, scaling prior precision with the experiment is a device to formalize learning about external validity across heterogeneous prior sources. 

One might ask whether our framework is equivalent to BMA with a Zellner $g$-prior, possibly under a different scaling of $g$. This is not the case. In classical variable-selection settings, Zellner priors are centered at zero for all models, so prior means coincide across specifications. As a result, all models are equally biased whenever the true parameter differs from zero, and Bayes factors discriminate models only through differences in model dimension. In contrast, our framework allows prior means to differ across sources and treats these means as potentially misspecified objects. Combined with an asymptotic regime in which prior precision grows at the same order as the experiment, this makes bias itself an object of learning: biased sources are exponentially downweighted, while unbiased sources dominate. This selection mechanism has no analogue under Zellner-type priors.

Regarding model selection results in BMA, there is a large body of results developed in the context of covariate selection in regression and moment‐condition models.\footnote{See \cite{LI2016132,Johnson01062012} and references therein, as well as \cite{Wasserman2000} for a review.} However, as noted above, the mechanism driving those selection results is fundamentally different from ours. In standard BMA, model‐selection consistency arises from differences in likelihood fit and model dimension under fixed‐prior asymptotics. In contrast, selection in our setting emerges from a nonstandard asymptotic regime in which prior precision grows with sample size, allowing the posterior to learn about external validity and to exponentially downweight biased sources while concentrating on unbiased ones. Moreover, to the best of our knowledge, existing literature does not deliver an analogue of our finding that such selection directly translates into faster concentration rates for the resulting estimator.

\paragraph{Roadmap.} Section \ref{Sec:Setup} describes the setup. Section \ref{sec:results} presents the main results. Section \ref{sec:scaled-error-sims} presents simulation evidence. Section \ref{sec:extensions} develops the binary-outcome and control-variable extensions. Section \ref{sec:conclusion} concludes. All proofs are related to  Appendix \ref{sec:proofs}.

\section{Setup}\label{Sec:Setup}

In this section, we describe the experiment and how prior information is incorporated into the estimation of our parameter of interest. The design is intentionally general, applying both to sequential experiments and to standard randomized controlled trials.

\paragraph{The Experiment.}

Consider an experiment in which individuals are assigned to a set of treatments, $\mathbb{D} := \{0,\dots,M\}$ based on observable characteristics, $x \in \mathbb{X}$, where both sets are finite. For each treatment–covariate pair $(d,x)$, let $Y(d,x) \in \mathbb{R}$ denote the potential outcome, and define the parameter of interest as the mean potential outcome
\begin{equation*}
\theta(d,x) := \mathbb{E}[Y(d,x)].    
\end{equation*}
At each instance $t \in \mathbb{N}$, the observed outcome covariate profile, $x$ is $Y_t(x) := Y_t(D_t(x),x)$ where $D_{t}(x) \in \mathbb{D}$ is the assigned treatment, which is assigned according to a policy rule
\begin{equation*}
(y^{t-1}, d^{t-1}) \mapsto \delta_t(y^{t-1}, d^{t-1})(\cdot \mid x) \in \Delta(\mathbb{D}), 
\end{equation*}
which specifies a probability distribution over treatments as a function of the past history of outcomes and assignments.\footnote{The index $t$ should be interpreted as indexing experimental stages rather than calendar time. In a standard RCT, $t$ can be viewed as labeling independent experimental units or cohorts drawn under a fixed randomization scheme, so that $\delta_t$ does not vary with the realized history. In sequential or adaptive experiments, by contrast, $t$ indexes decision rounds, and the assignment rule $\delta_t$ is allowed to depend on past outcomes and treatment assignments. Our analysis accommodates both interpretations.} Thus, $\delta_t(d \mid x)$ gives the probability that an individual with covariates, $x$, receives treatment, $d$, at instance, $t$. When no confusion arises, we omit the conditioning on past history. 

The policy rule encompasses standard randomized controlled trials (RCTs), in which the assignment rule $\delta_t(y^{t-1}, d^{t-1})(\cdot \mid x)$ is independent of past outcomes and treatments, as well as more sophisticated sequential experimentation designs in which treatment assignment adapts to accumulated information. We deliberately refrain from taking a stand on the desirability of any particular policy rule. Instead, our objective is to maintain sufficient generality so that the learning rates derived below apply uniformly across a wide class of commonly used assignment mechanisms, including both static and adaptive designs.

\paragraph{Prior sources of information.}

For each $(d,x) \in \mathbb{D} \times \mathbb{X}$, the experimenter has access to a collection of prior sources $\mathcal{S} := \{0,\dots,L\}$. Each source $s \in \mathcal{S}$ is represented by a Gaussian prior
\begin{equation*}
\phi(\cdot; \zeta^{s}_{0}(d,x),1 / \nu^{s}(d,x)),    
\end{equation*}
where $\phi(\cdot;a,b)$ denotes the normal density with mean $a$ and variance $b$. The quantity $\zeta^{s}_{0}(d,x)$ represents the prior value of the expected outcome under treatment $d$ for individuals with covariates $x$, while $\nu^{s}(d,x)$ reflects the precision of source $s$, with larger values indicating greater confidence.

The prior sources may derive either from previous experiments or from expert judgments. When $s$ corresponds to a past experiment, the experimenter simply collects the estimated average outcome for each treatment–covariate pair, which becomes $\zeta^{s}_{0}(d,x)$, along with the number of units with covariates $x$ assigned to treatment $d$, which becomes $\nu^{s}(d,x)$. In contrast, when $s$ represents an expert opinion or recommendation, the experimenter elicits the expert’s assessment of the expected outcome -- serving as $\zeta^{s}_{0}(d,x)$ -- and the expert’s confidence in that assessment, which naturally maps into the precision parameter $\nu^{s}(d,x)$. In all cases, the pair $(\zeta^{s}_{0}, \nu^{s})_{s \in \mathcal{S}}$ is treated as non-random.

\paragraph{Posterior updating.}

Assume that the experimenter models the outcome distribution as belonging to the Gaussian family
\begin{equation*}
\{\phi(\cdot;\theta(d,x),1) : \theta(d,x)\in \mathbb{R} \}.    
\end{equation*}
After observing treatments and outcomes up to instance ($t$), the experimenter computes, for each source $s$, the posterior distribution for $\theta(d,x)$ using Bayes’ rule. Under conjugacy, this posterior is Gaussian with mean\begin{align}	\notag 
	\zeta^{s}_{t}(d,x) =  &   \frac{ N_{t}(d,x)  }{ 	N_{t}(d,x)   + \nu^{s}(d,x)  } N_{t}(d,x)^{-1} \sum_{i=1}^{t} Y_{i}(d,x) 1\{ D_{i}(x)  = d \}   \\ \label{eqn:zeta.o}
    & + \frac{ 	\nu^{s}(d,x)  }{ 	N_{t}(d,x)   + \nu^{s}(d,x)  } \zeta^{s}_{0}(d,x),~where~N_{t}(d,x) : = \sum_{i=1}^{t} 1\{ D_{i}(x)  = d \}, 
\end{align}
and the posterior precision is $N_t(d,x) + \nu^{s}(d,x)$.

\paragraph{Model posterior probabilities.}
Faced with $L+1$ sources for each $(d,x)$, the experimenter assigns posterior model weights
\begin{align*}
	\alpha^{s}_{t}(d,x) : =  \frac{ \int \prod_{i=1}^{t}  \phi(Y_{i}(d,x) ; \theta , 1 ) ^{1\{ D_{i}(x) =d  \}}  \phi (\theta; \zeta^{s}_{0}(d,x) , 1/\nu^{s}(d,x))(\theta) d\theta   }{ \sum_{s \in \mathcal{S}} \int \prod_{i=1}^{t}  \phi (Y_{i}(d,x) ; \theta , 1)  ^{1\{ D_{i}(x) =d  \}}  \phi (\theta; \zeta^{s}_{0}(d,x) , 1/\nu^{s}(d,x))(\theta) d\theta    }.
\end{align*}
These quantities are Bayesian model posterior probabilities in the sense of Bayesian model averaging (BMA). In our context, they can be interpreted as the probabilities that each source is externally valid for the pair $(d,x)$. A formal and detailed discussion is provided in Section~\ref{sec:weights.EV}.

\paragraph{Estimator of average effects.}

For each treatment–covariate pair $(d,x)$, the estimator for $\theta(d,x)$ is given by the BMA-weighted average
\begin{equation}\label{Eq:estimator}
\widehat{\theta}_{t}(d,x)
:= \sum_{s \in \mathcal{S}} \alpha^{s}_{t}(d,x)  \zeta^{s}_{t}(d,x).    
\end{equation}
The Gaussian–Gaussian framework yields a simple and intuitive estimator: a weighted average of the source-specific posterior means, where the weights adaptively reflect each source’s fit to the observed data. Importantly, we do not assume that the true distribution of outcomes is Gaussian. The Gaussian likelihood is a working model used only for tractable inference. Because the parameter of interest is the mean outcome, and because sample averages are unbiased under very general conditions (e.g., finite second moments), the estimator remains consistent even if the Gaussian working model is misspecified.

\section{Theoretical Results}
\label{sec:results}

In this section, we present our asymptotic framework and discuss how the posterior weights, $\alpha^{s}_{t}$, can be interpreted as a measure of the external validity of source $s$ for the current experiment. We then derive the learning rates of our estimator and show that its rate of convergence is proportionally faster than the standard one whenever at least one unbiased source exists.

To establish these results, we impose the following assumptions. The first assumption describes the data-generating process for potential outcomes.

\begin{assumption}\label{ass:IID}
For each $t \in \mathbb{N}$ and each $x \in \mathbb{X}$, the collection $\{Y_{t}(d,x)\}_{d \in \mathbb{D}}$ is drawn IID from a distribution $P(\cdot \mid d,x) \in \Delta(\mathbb{R})$ with finite second moment: $\mathbb{E}\big[\,|Y(d,x)|^{2}\,\big] < \infty$.\footnote{For each $(d,x)$, the expectation $\mathbb{E}$ over (functions of) $Y(d,x)$ is taken with respect to $P(\cdot \mid d,x)$. } 
\end{assumption}

The second assumption concerns the assignment mechanism. Beyond the structure discussed in the setup, we impose the following minimal restriction.

\begin{assumption}\label{ass:PR}
For each $(d,x) \in \mathbb{D} \times \mathbb{X}$, $\sum_{i=1}^{t} \delta_{i}(d \mid x)$ diverges almost surely as $t \to \infty$.
\end{assumption}

Assumption~\ref{ass:PR} guarantees that the number of times treatment $d$ is assigned to covariate profile $x$ diverges almost surely as $t$ grows. Intuitively, the assumption allows the probability of assignment to decay but not too quickly.\footnote{See Lemma~\ref{lem:N.diverge} in Appendix~\ref{app:AlmostSure} for a formal statement and a discussion of its role.} 
It generalizes the standard overlap assumptions in randomize control trials and it is satisfied by standard heuristic policies widely used in sequential experimentation. For generalized $\epsilon$-greedy algorithms, we have $\delta_{i}(d \mid x) \ge \epsilon_{i}$, so Assumption~\ref{ass:PR} requires only that $(\epsilon_{i})_{i}$ decays more slowly than $1/i$. For Thompson Sampling and UCB algorithms, it is well known that
\[
\sum_{i=1}^{t} \delta_{i}(d \mid x) \asymp t \quad \text{for the optimal arm}, 
\qquad
\sum_{i=1}^{t} \delta_{i}(d \mid x) \asymp \log t \quad \text{for suboptimal arms},
\]
under mild regularity conditions; see, for example, \cite{auer2002finite} for UCB and \cite{pmlr-v23-agrawal12} and \cite{Kaufman2012} for Thompson Sampling. Hence, all three heuristics satisfy Assumption~\ref{ass:PR}.

In general, Assumption~\ref{ass:PR} allows for decreasing assignment probabilities but rules out excessively fast decay. The experimenter must continue to explore each treatment--covariate pair infinitely often, though possibly at a slowly vanishing rate.

\subsection{Asymptotic Framework}

To simplify the analysis, we rely on asymptotic techniques. However, in order to better approximate the empirical settings we consider, we deviate from the standard asymptotic framework.

In many applications, the size of the treatment groups in the target experiment and in the prior source $s$ are of comparable magnitude. In such cases, the usual asymptotics --- where $\nu^{s}(d,x)$ is fixed while $N_{t}(d,x)$ diverges --- may not provide an accurate approximation to finite-sample behavior. To address this issue, we adopt a nonstandard asymptotic framework in which the prior precision $\nu^{s}(d,x)$ is allowed to depend on $t$. We write $\nu^{s}_{t}(d,x)$ to make this dependence explicit.

We assume that $(\nu^{s}_{t}(d,x))_{t}$ diverges and satisfies
\begin{align}\label{eqn:c.limit}
	\lim_{t \rightarrow \infty} \frac{\nu^{s}_{t}(d,x)}{N_{t}(d,x)} = : c^{s}(d,x) \in \mathbb{R}_{+},~a.s.
\end{align}

Allowing the prior precision to depend on $t$ should be understood as a mathematical device used to approximate empirically relevant settings in which both the prior sample size and the sample size of the current experiment are large. This asymptotic framework nests the standard one as a special case: setting $c^{s}(d,x)=0$ corresponds precisely to the usual assumption that the prior precision remains fixed while $N_{t}(d,x)$ diverges.

\subsection{Weights and External Validity}
\label{sec:weights.EV}

As presented in the setup, the estimator uses BMA to aggregate across the $L+1$ sources. The resulting weights $\alpha^{s}_{t}(d,x)$ --- the posterior probability that model $s$ best fits the observed data --- can be interpreted as the experimenter's subjective probability that source $s$ is externally valid for the pair $(d,x)$ in the current experiment. To formalize this, we introduce a quantitative measure of external validity and relate it to the asymptotic behavior of the weights $(\alpha^{s}_{t}(d,x))_{s=0}^{L}$.

Let $E : \mathbb{R}_{+} \times \mathbb{R}_{+} \to \overline{\mathbb{R}}$ be defined by
\begin{equation*}
E(mc,p) := -p \times mc + \log p.    
\end{equation*}
The first argument represents the ``misspecification cost" (mc) associated with a source, and the second represents the precision associated with its marginal likelihood for $(d,x)$. The value $E(mc,p)$ provides a continuous measure of external validity: unbiased sources (with $p\geq 1$) satisfy $E(0,p)>0$, and this value increases with precision, with the extreme case corresponding to a degenerate source with arbitrarily large $p$. Biased sources instead satisfy $E(mc,p)<0$, and their external validity worsens as precision increases, since their probability mass becomes more tightly concentrated around an incorrect value. Thus, unlike frequentist treatments where external validity is typically binary, this Bayesian measure varies continuously with both the bias and the precision of the source.

For the Gaussian model, for any source $s$ and any treatment-covariate pair $(d,x)$, the misspecification cost is captured by $mc^{s}(d,x) : = (bias^{s}(d,x))^{2} : =  (\theta(d,x) - \zeta^{s}_{0}(d,x))^{2}$ --- the square of the difference between the prior value and the true expected outcome.\footnote{As it turns out, what matters for external validity is not the bias in the sense of a difference in parameters, but bias in the sense of systematic likelihood loss. This loss is quantified by the Kullback–Leibler divergence. In the Gaussian case with known variance, this divergence is exactly proportional to the squared difference in means, so squared bias emerges as a convenient and exact summary of misspecification. We refer the reader to Section \ref{sec:extensions} for a more thorough explanation and additional instances of the misspecification cost for other settings.}

The next result provides an asymptotic equivalence between the Bayesian posterior weights and the mapping $E$. 

\begin{proposition}\label{pro:Bweights.EV}
For any $(d,x) \in \mathbb{D} \times \mathbb{X}$,\footnote{Henceforth, let $o_{as}(\cdot)$ denotes a random variable that converges to zero almost surely at the indicated rate and  $O_{as}(\cdot)$ denotes a random variable that remains almost surely bounded at the indicated rate. The underlying probability distribution is the one generated by the true distribution over outcome, $P$, and the policy rule $\delta$.}
\[
\alpha^{s}_{t}(d,x)
=
\frac{
\exp\!\left(
\frac{1}{2} E\!\left((bias^{s}(d,x))^{2}, \frac{\nu^{s}_{t}(d,x)}{1+c^{s}_{t}(d,x)}\right)(1+o_{as}(1)) + o_{as}(1)
\right)
}{
\displaystyle
\sum_{s'=0}^{L}
\exp\!\left(
\frac{1}{2} E\!\left((bias^{s'}(d,x))^{2}, \frac{\nu^{s'}_{t}(d,x)}{1+c^{s'}_{t}(d,x)}\right)(1+o_{as}(1)) + o_{as}(1)
\right)
}
\]
where $c^{s}_{t}(d,x) : = \nu^{s}_{t}(d,x)/N_{t}(d,x)$.
\end{proposition}

\begin{proof}
See Appendix \ref{sec:proofs}.
\end{proof}

The intuition behind Proposition~\ref{pro:Bweights.EV} is as follows. The weight 
$\alpha^{s}_{t}(d,x)$ represents the posterior probability assigned to source $s$, 
and its behavior is governed by the marginal likelihood of that source relative to 
the others. Under standard asymptotics—where $N_{t}(d,x)$ diverges while 
$\nu^{s}_{t}(d,x)$ remains fixed—it is well known that $\alpha^{s}_{t}(d,x)$ becomes 
asymptotically proportional to the marginal likelihood evaluated at the true 
parameter. In this regime, one source of uncertainty, the sampling 
uncertainty, is resolved at rate $N_{t}(d,x)$, but the other source, the prior uncertainty, 
never disappears because the prior precision is fixed. The latter prevents perfect 
separation across sources, and asymptotically the weights remain proportional to $e^{\frac{1}{2} E((bias^{s}(d,x))^{2},\nu^{s}_{t}(d,x)) }$.  

Under our asymptotics, however, $\nu^{s}_{t}(d,x)$ is permitted to diverge, which 
changes this behavior. Let us first analyze the case where $\nu^{s}_{t}(d,x)$ diverges but at a slower rate than 
$N_{t}(d,x)$, so that $c^{s}(d,x)=0$. In this case, perfect discrimination among sources becomes possible asymptotically, and the determining 
factor in the asymptotic behavior of $\alpha^{s}_{t}(d,x)$ becomes the bias --- distance 
between $\theta(d,x)$ and the prior mean $\zeta^{s}_{0}(d,x)$ ---, with smaller bias 
yielding exponentially larger posterior weight. The rate of decay is given by $\nu^{s}_{t}(d,x)$, the rate at which precision grows. Thus, although the weights remain asymptotically proportional to $e^{\frac{1}{2} E(bias^{s}(d,x),\nu^{s}_{t}(d,x)) }$, the interpretation differs because the prior precision is no longer fixed.

A third case arises when $\nu^{s}_{t}(d,x)$ diverges proportionally with 
$N_{t}(d,x)$, so that $\nu^{s}_{t}(d,x)/N_{t}(d,x) \to c^{s}(d,x) > 0$. In this setting, 
the precision stemming from the prior remains relevant asymptotically, but now the 
effective precision of the marginal likelihood becomes 
$N_{t}(d,x)\,\frac{c^{s}(d,x)}{1+c^{s}(d,x)}$. This quantity reflects the combined 
contribution of both the experimental data and the prior source, with the factor 
$\frac{c^{s}(d,x)}{1+c^{s}(d,x)}$ determining how much of the total precision is 
attributable to the prior. When $c^{s}(d,x)$ is large, the overall precision approaches 
that of the current experiment; when $c^{s}(d,x)$ is small, residual prior 
uncertainty persists. The term 
$E(bias^{s}(d,x), \nu^{s}_{t}(d,x)/(1+c^{s}(d,x)))$ captures this relationship exactly: 
the bias penalizes the source, and the effective precision amplifies or mitigates 
this penalty depending on the informativeness of source $s$. For this reason, 
$\nu^{s}_{t}(d,x)/(1+c^{s}(d,x))$ is an appropriate asymptotic measure of the 
precision of source $s$ when its precision grows proportionally to that of the 
target experiment.

Based on this discussion, we define the external validity of source $s$, for pair $(d,x)$ at instance $t$ as
\begin{align}\label{eqn:EV}
\mathbb{EV}^{s}_{t}(d,x) : = E((bias^{s}(d,x))^{2},\nu^{s}_{t}(d,x)/(1+c^{s}_{t}(d,x))),
\end{align}
where the inputs are the bias and the precision. 

Proposition \ref{pro:Bweights.EV} provides, asymptotically and almost surely, an isomorphism between the posterior odds ratio of the sources and their external validity, i.e., 
\begin{align}\label{eqn:BayesOdd.EV}
	\log \frac{\alpha^{s}_{t}(d,x)}{\alpha^{s'}_{t}(d,x)} = \frac{1}{2} \left( \mathbb{EV}^{s}_{t}(d,x)  - \mathbb{EV}^{s'}_{t}(d,x)  \right)(1+o_{as}(1)) + o_{as}(1).
\end{align}

This relationship allow us to obtain a source selection in terms of their external validity. To see this, let, for any pair $(d,x)$, $\mathcal{U}(d,x) : = \{ s \in \mathcal{S} \colon bias^{s}(d,x)  =0  \}$ be the set of \emph{unbiased sources} (which could be empty). For any biased source, $b \notin \mathcal{U}(d,x)$, $\mathbb{EV}^{b}_{t}(d,x)$ diverges to minus infinity with its size, $\nu^{b}_{t}(d,x)$, while an unbiased source is diametrically opposite, diverging to \emph{plus} infinity with its size. Therefore, if unbiased sources exist, equation \ref{eqn:BayesOdd.EV} implies that bias sources receive \emph{exponentially} vanishing weight. 

This result has an oracle-type feature: asymptotically, our estimator assigns weight only to unbiased sources. However, this result is silent about what happens if there are no unbiased sources --- all sources are mis-specified. In this case, we are able to obtain a robustness-type result provided that diffuse sources are included. A source is considered to be \emph{diffuse} if $\sup_{t} \nu_{t}(d,x) \leq K < \infty$ for some small constant $K$.\footnote{Asymptotically, the constant $K$ need not be small, simply finite. However, for finite sample and to capture the idea of a ``diffuse" source, $K$ should be chosen to be small.} For such sources the $\mathbb{EV}^{b}_{t}(d,x)$ remains bounded (recall that for biased sources this quantity diverges to minus infinity). Therefore, when all sources are biased, equation \ref{eqn:BayesOdd.EV} implies the diffuse source will accumulate all the weight \emph{exponentially} fast and thus our estimator will be essentially equal to the standard one. In this sense, we view this result as a robustness property: Our estimator performs no worse than the standard one even if all sources are mis-specified.

Even though the result relies on the presence of a diffuse source, the experimenter can always include one. The condition should therefore be interpreted as a practical recommendation rather than a formal restriction.

The next proposition formalizes this discussion. 
\begin{proposition}\label{pro:alpha.asymp}
	There exists a finite constant $C$ such that the following statements are true  for any $(d,x)$:
\begin{enumerate}
	\item 	Suppose $\mathcal{U}(d,x)$ is non-empty. Then, for any source $b \notin \mathcal{U}(d,x)$,
	\begin{align*}
		\alpha_{t}^{b}(d,x) = O_{as}\left( \frac{\nu^{b}_{t}(d,x)}{ \max_{u \in \mathcal{U}(d,x)} \nu^{u}_{t}(d,x) }  e^{  - C \nu^{b}_{t}(d,x) (bias^{b}(d,x))^{2}  }  \right). 
	\end{align*}
	\item 	Suppose a diffuse source exists and $\mathcal{U}(d,x)$ is empty. Then, for any non-diffuse source $b \in \mathcal{S}$,
	\begin{align*}
		\alpha_{t}^{b}(d,x) = O_{as}\left( e^{  - C \nu^{b}_{t}(d,x) (bias^{b}(d,x))^{2}  }  \right). 
	\end{align*}
\end{enumerate}
\end{proposition}

\begin{proof}
See Appendix \ref{sec:proofs}.
\end{proof}

We conclude by pointing out that this result does not rank sources \emph{within} the class of unbiased sources. For two unbiased sources, $u$ and $u'$, equation \ref{eqn:BayesOdd.EV} implies that 
\begin{align}\label{eqn:BayesOdd.EV1}
 \lim_{t \rightarrow \infty}	\log \frac{\alpha^{u}_{t}(d,x)}{\alpha^{u'}_{t}(d,x)} = \frac{1}{2} \log \frac{c^{u}(d,x)}{c^{u'}(d,x)}~a.s.
\end{align}
So the procedure assigns larger weight to more informative sources, but the weights do not collapse onto a single source.

\subsection{Learning Rates for Average Effects} 

As mentioned above, the object of interest is the average effect of each treatment.  
At each instance $t$ and for each $(d,x) \in \mathbb{D} \times \mathbb{X}$,  
the experimenter estimates this effect using
\begin{equation*}
\widehat{\theta}_{t}(d,x)
:= \sum_{s \in \mathcal{S}} \alpha^{s}_{t}(d,x)\, \zeta^{s}_{t}(d,x).
\end{equation*}

The next result is the main result in the paper and establishes the rate at which this estimator concentrates around the true expected outcome $\theta(d,x)$. Henceforth, let $\ell : [1,\infty) \to \mathbb{R}_{+}$ be any increasing function such that $\int_{1}^{\infty} 1/(x \ell(x)^{2}) dx < \infty $.\footnote{This function serves as a scaling factor for our almost sure concentration rates, and it stems from classical results;  
its role is explained in Lemmas~\ref{lem:N.diverge} and \ref{lem:Y.ASrate} in the Appendix \ref{app:AlmostSure}.}

\begin{theorem}\label{thm:learning.EV}
For any $(d,x) \in \mathbb{D} \times \mathbb{X}$, the following hold:

\begin{enumerate}
    \item If $\mathcal{U}(d,x)$ is non-empty, then
    \begin{align*}
    \big|\, \widehat{\theta}_{t}(d,x) - \theta(d,x) \,\big|
    =  
    o_{as}\!\left(
        \frac{\ell(N_{t}(d,x))}{\sqrt{N_{t}(d,x)}}\,
        \mathcal{A}_{t}(\mathcal{U}(d,x))
    \right)
    +
    O_{as}\!\left( N_{t}(d,x)^{-1} \right),
    \end{align*}
    where
    \[
    \mathcal{A}_{t}(\mathcal{U}(d,x)) 
    :=
    \sum_{s \in \mathcal{U}(d,x)}
    (1+c^{s}(d,x))^{-1}
    \frac{
        \alpha^{s}_{t}(d,x)
    }{
        \sum_{s' \in \mathcal{U}(d,x)} \alpha^{s'}_{t}(d,x)
    }.
    \]

    \item If $\mathcal{U}(d,x)$ is empty but there exists a diffuse source, then
    \begin{align*}
    \big|\, \widehat{\theta}_{t}(d,x) - \theta(d,x) \,\big|
    =  
    o_{as}\!\left(
        \frac{\ell(N_{t}(d,x))}{\sqrt{N_{t}(d,x)}}
    \right)
    +
    O_{as}\!\left( N_{t}(d,x)^{-1} \right).
    \end{align*}
\end{enumerate}
\end{theorem}

\begin{proof}
    See Section \ref{sec:proofs}.
\end{proof}

\paragraph{Remarks.} 

The term $\frac{1}{\sqrt{N_{t}(d,x)}}$ is the standard almost-sure rate for estimating  
$\sum_{i=1}^{t} \mathbf{1}\{D_{i}(x)=d\} Y_{i}(d,x)/N_{t}(d,x)$, and the scaling factor $\ell(N_{t}(d,x))$ is a standard "loss" in almost sure results, e.g., $\ell(x) = \log x$; see Lemma~\ref{lem:Y.ASrate} in Appendix \ref{app:AlmostSure}.%

When $\mathcal{U}(d,x)$ is non-empty, the convergence rate of $\widehat{\theta}_{t}(d,x)$ improves proportionally relative to the standard rate by the factor $\mathcal{A}_{t}(\mathcal{U}(d,x))$.  
This factor is an average of $(1+c^{s}(d,x))^{-1}$ across unbiased sources, weighted by their posterior probabilities.  
It always lies in $[0,1]$ and is strictly less than one whenever at least one unbiased source has size proportional to the target experiment.  
Only unbiased sources contribute to this improvement because, as shown in Propsotion~\ref{pro:alpha.asymp}, the posterior eventually assigns positive weight exclusively to unbiased sources whenever they exist.  
For example, if $c^{s}(d,x)$ takes the common value $c(d,x)$ across all unbiased sources, then the rate is accelerated by the multiplicative factor $(1+c(d,x))^{-1}$.  
Even moderate relative precision of the prior source can therefore generate a meaningful proportional gain. In this sense, our estimator enjoys an oracle-type property by (asymptotically) putting all the weight on unbiased sources.  

If $\mathcal{U}(d,x)$ is empty the concentration rate is asymptotically equal to the standard one, provided a diffuse source is included ---  a diffuse source can always be included by adding a source with arbitrarily small precision. In this case, as shown in Proposition \ref{pro:alpha.asymp}, the diffuse source dominates all biased sources and $\alpha^{diffuse}_{t}(d,x) \to 1$.  The estimator converges at the standard rate, thereby yielding a natural robustness property: even when all sources are biased, the aggregation procedure performs no worse (up to constants) than the usual estimator based solely on the target experiment. 

Thus, with our procedure the experimenter does not need to identify in advance which sources are unbiased or correctly specified.  
When unbiased sources of comparable size exist, the procedure automatically assigns them weight and converges at a strictly faster rate—behaving as if it were an ``oracle'' with knowledge of which sources are unbiased.  
When no unbiased sources exist, the procedure remains robust and achieves the standard convergence rate.

\paragraph{PAC interpretation.} A PAC interpretation of our results further clarifies the gains delivered by incorporating unbiased sources. The convergence rate in Theorem \ref{thm:learning.EV} implies that achieving a target precision $\varepsilon$ requires on the order of $(1/\varepsilon^2)\log(1/\varepsilon)$ observations under the standard regime, which matches the canonical PAC rate for estimating a mean. By contrast, when an unbiased source of relative size $c$ exists, the effective rate is scaled by the factor $A= 1/(1+c) <1$, so the required sample size decreases to approximately $(A/\varepsilon^2)\log(1/\varepsilon)$. Thus the presence of unbiased sources reduces the number of observations needed to attain a given accuracy by a proportional factor $A$, reflecting the larger effective sample size generated by external information. In this sense, the procedure behaves as an adaptive PAC learner: in the presence of unbiased sources it attains a strictly smaller required sample size for a given precision, while in misspecified settings (all sources biased) it reverts to the standard PAC rate.

\section{Model Simulations: Learning and Model Weights}\label{sec:scaled-error-sims}

This section reports Monte Carlo evidence on the finite-sample performance of our estimator with multiple sources that vary in bias and effective sample size --- for simplicity, we focus on the case of no covariates. We focus on the behavior of the \emph{scaled absolute error} in treatment arm $d$,
\[
\sqrt{N_T(d)}\,|\hat{\theta}(d)-\theta(d)|,
\]
where $N_T(0) = N_T(1) = : N_T$ is the number of observations per arm (in our balanced design, $N_T=T/2$). Throughout, potential outcomes satisfy $Y(0)\sim \mathcal{N}(1,1)$ and $Y(1)\sim \mathcal{N}(1.3,1)$, and each experiment is replicated $1{,}000$ times for each design point. For ease of exposition and without loss of generality, we restrict attention to the control arm, $d=0$.

\subsection*{Experimental design and models}

In each experiment, we observe $N_T=T/2$ draws per arm and compute two estimators of $\theta(0)$: (i) the standard sample mean $\bar{Y}(0)$ and (ii) our estimator that averages across sources using posterior model weights as defined in Equation \ref{Eq:estimator}. Each source $s$ is characterized by an initial mean $\zeta_s(0)$ and a precision (effective sample size) parameter $\nu_s(0)$. We parameterize the strength of the non-diffuse sources as scaling with the experiment sample size via $e\in\{0.5,1,2\}$, so that (holding fixed baseline multipliers) $e$ can be interpreted as the source's effective sample size \emph{relative} to the per-arm sample size in the experiment.\footnote{In the simulations, the diffuse source is kept weak with a fixed precision $\nu=1$ to represent a low-information baseline.}

We consider three configurations:
\begin{enumerate}
    \item \textbf{Model 1 (diffuse + unbiased):} a diffuse source centered at $\theta(0)$ and an informative unbiased source centered at $\theta(0)$.
    \item \textbf{Model 2 (diffuse + biased):} a diffuse source centered at $\theta(0)$ and an informative biased source centered at $\theta(0)+1$ (i.e. one standard deviation).
    \item \textbf{Model 3 (diffuse + unbiased + biased):} a diffuse source, an informative unbiased source, and an informative biased source.
\end{enumerate}

\begin{figure}[!h]
\centering
    \begin{subfigure}{0.7\textwidth}
        \centering
        \includegraphics[width=\textwidth]{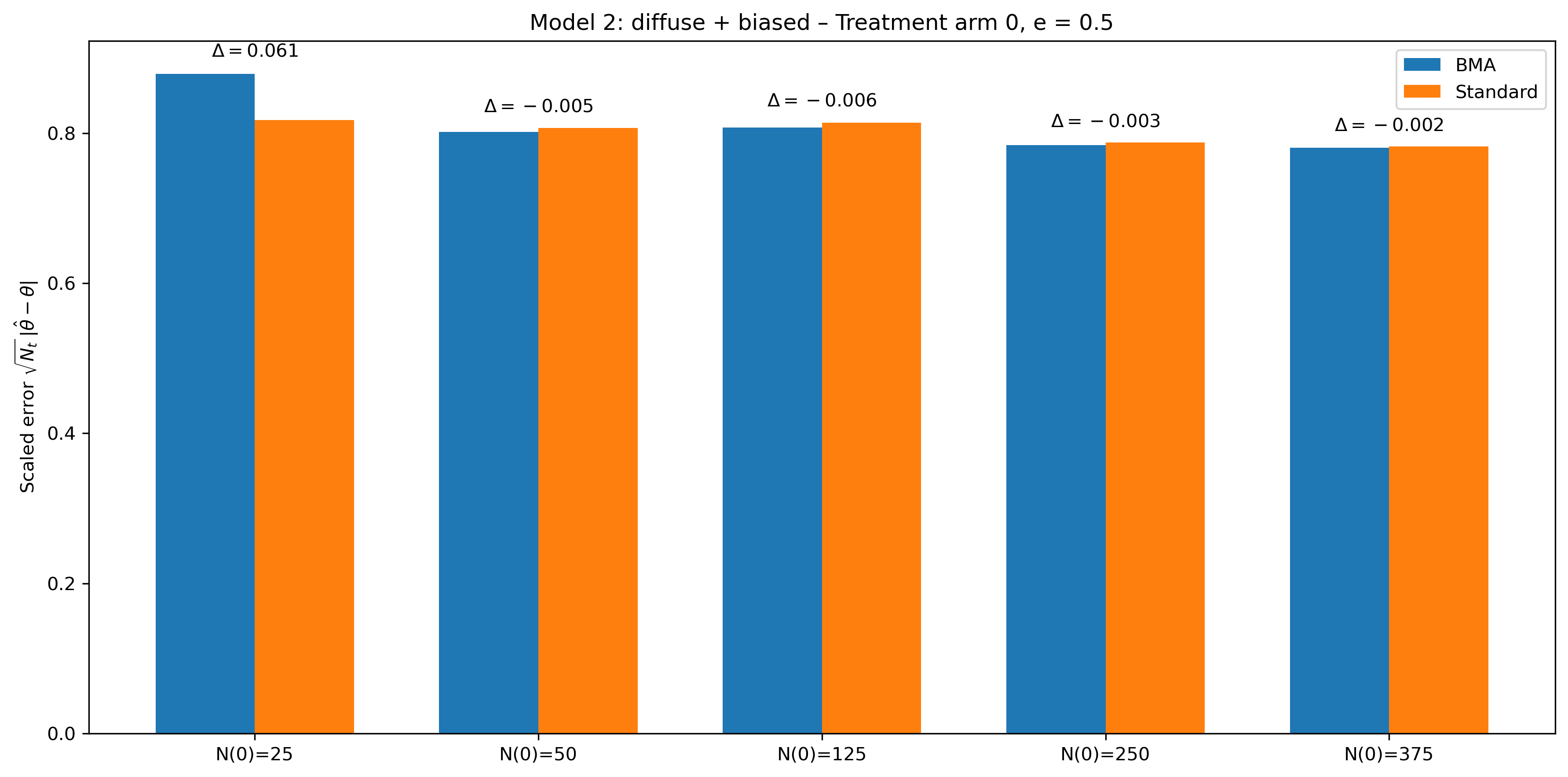}
        \caption{$E=0.5$}
        \label{fig:sub1}
    \end{subfigure}
    \hfill 
    \begin{subfigure}{0.7\textwidth}
        \centering
        \includegraphics[width=\textwidth]{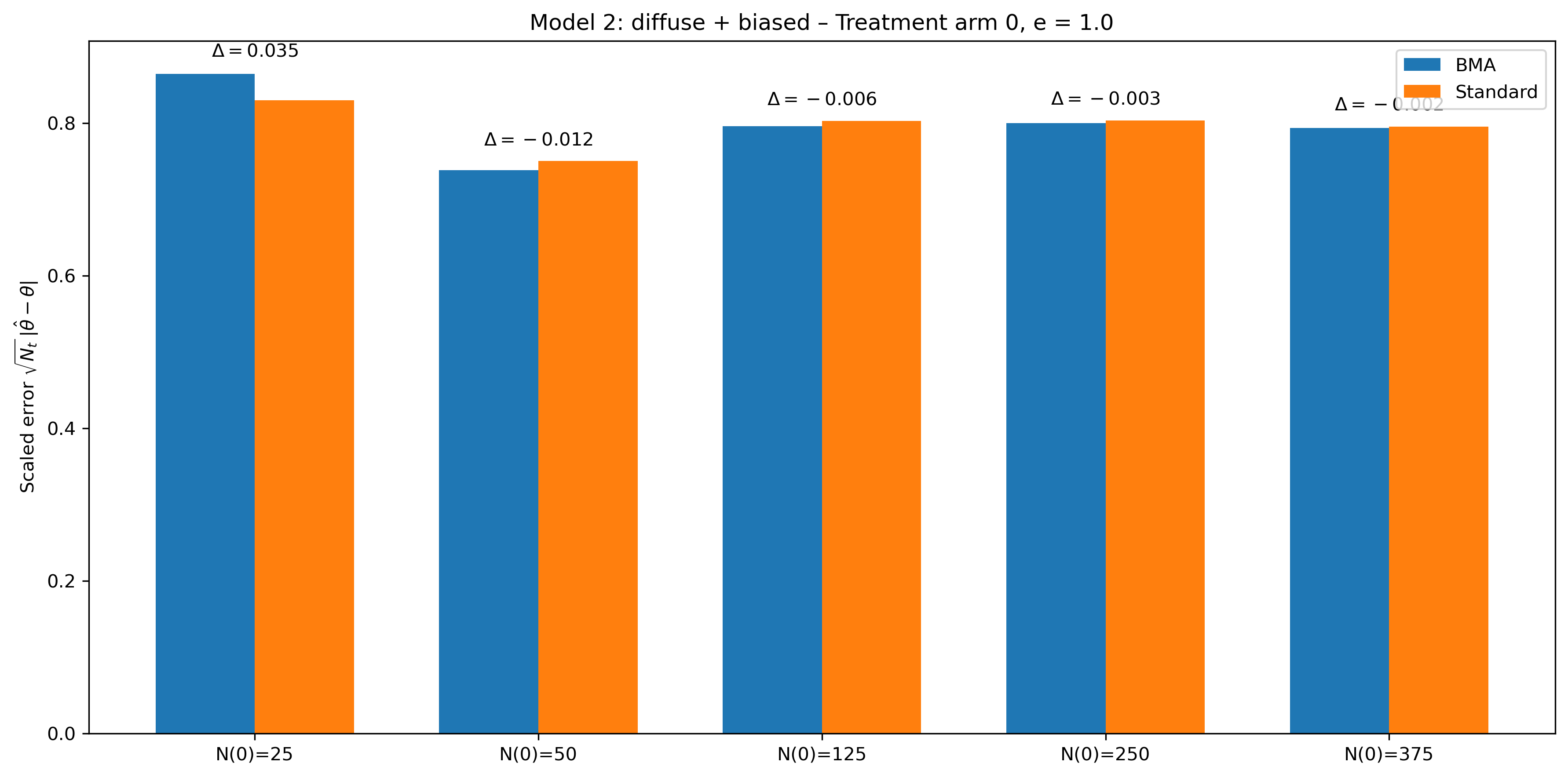}
        \caption{$E=1.0$}
        \label{fig:sub2}
    \end{subfigure}
    \hfill 
    \begin{subfigure}{0.7\textwidth}
        \centering
        \includegraphics[width=\textwidth]{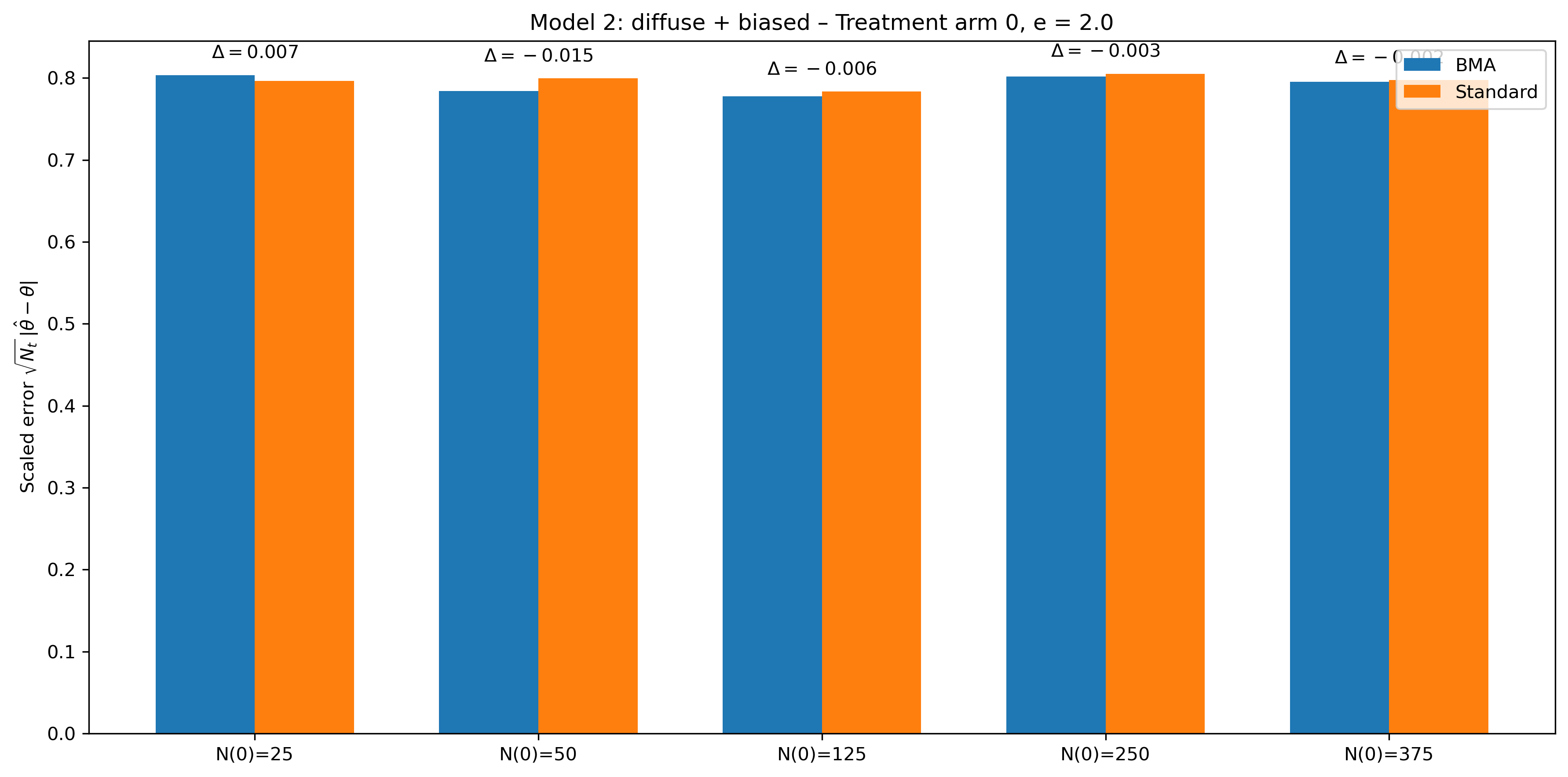}
        \caption{$E=2.0$}
        \label{fig:sub2}
    \end{subfigure}
    \caption{Scaled-Error Rates}
    \label{fig:model2}
  \floatfoot{\footnotesize \textit{Notes:} Each panel reports the mean scaled absolute error in the control arm, $\sqrt{N(0)}\,|\hat\theta(0)-\theta(0)|$, from 1{,}000 Monte Carlo replications. Bars compare the BMA posterior mean estimator (blue) to the standard sample mean (orange) across $N(0)\in\{25,50,125,250,375\}$ (corresponding to $T\in\{50,100,250,500,750\}$). The three panels vary the relative strength of the informative source, $e\in\{0.5,1,2\}$, which scales the source precision proportionally to $N(0)$. Model~2 includes a diffuse baseline source centered at the true mean and an informative \emph{biased} source whose initial mean is shifted by $+1$ relative to the truth. The data-generating process is $Y(0)\sim \mathcal{N}(1,1)$ and $Y(1)\sim \mathcal{N}(1.3,1)$. The $\Delta$ labels denote the difference in mean scaled error (BMA minus standard) at each sample size.}

\end{figure}

\begin{figure}[!h]
\centering
    \begin{subfigure}{0.7\textwidth}
        \centering
        \includegraphics[width=\textwidth]{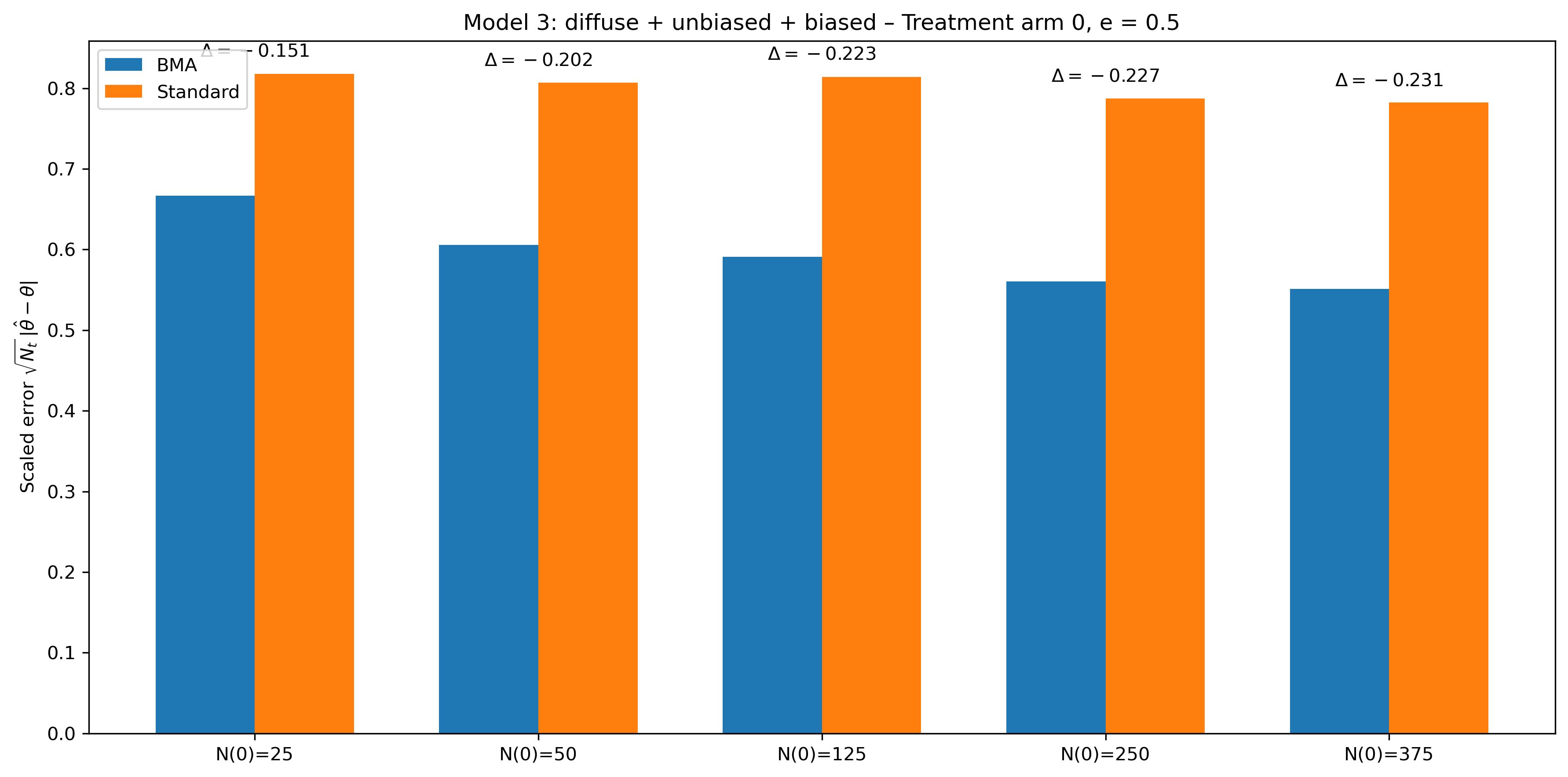}
        \caption{$E=0.5$}
        \label{fig:sub1}
    \end{subfigure}
    \hfill 
    \begin{subfigure}{0.7\textwidth}
        \centering
        \includegraphics[width=\textwidth]{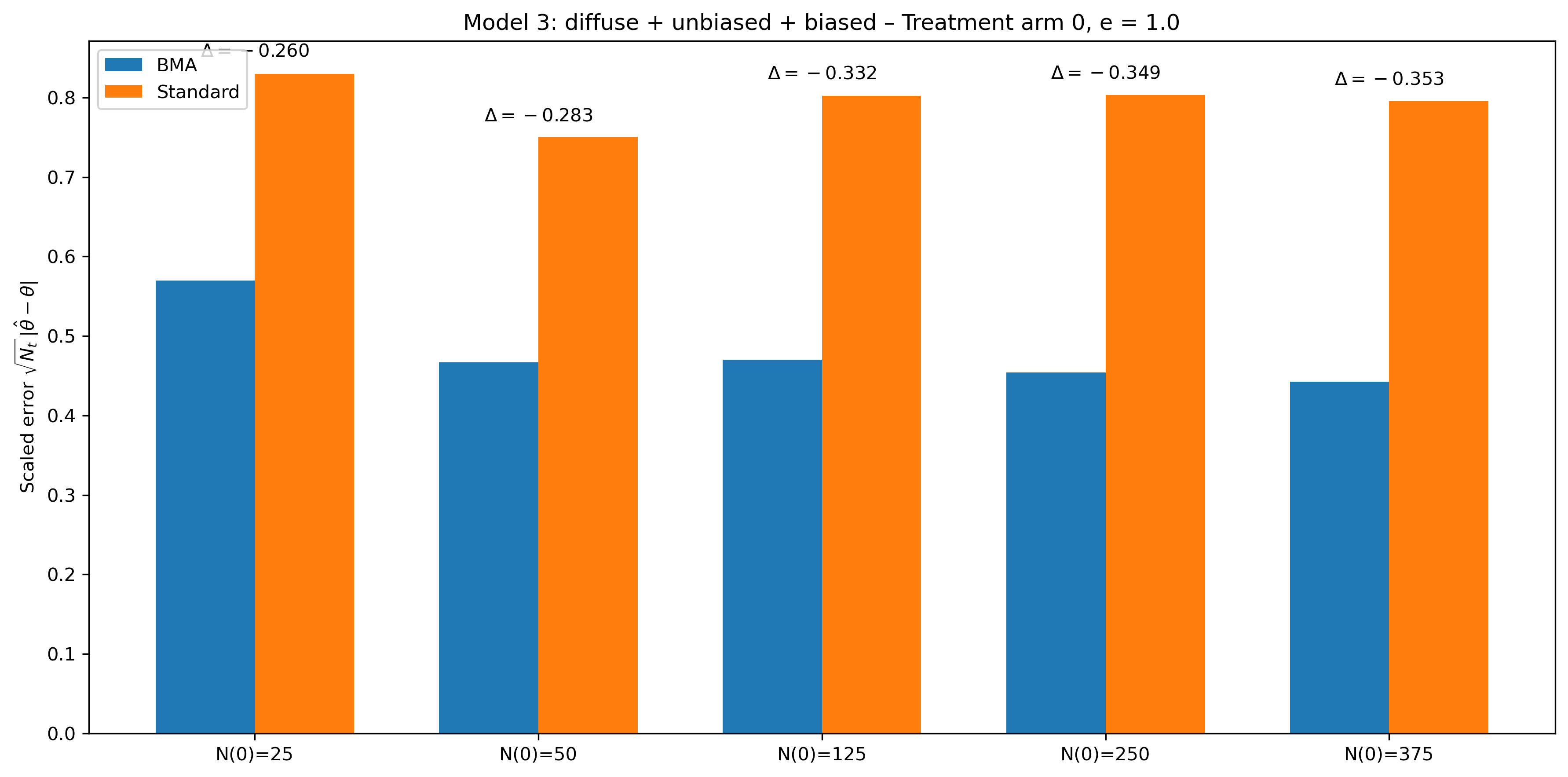}
        \caption{$E=1.0$}
        \label{fig:sub2}
    \end{subfigure}
    \hfill 
    \begin{subfigure}{0.7\textwidth}
        \centering
        \includegraphics[width=\textwidth]{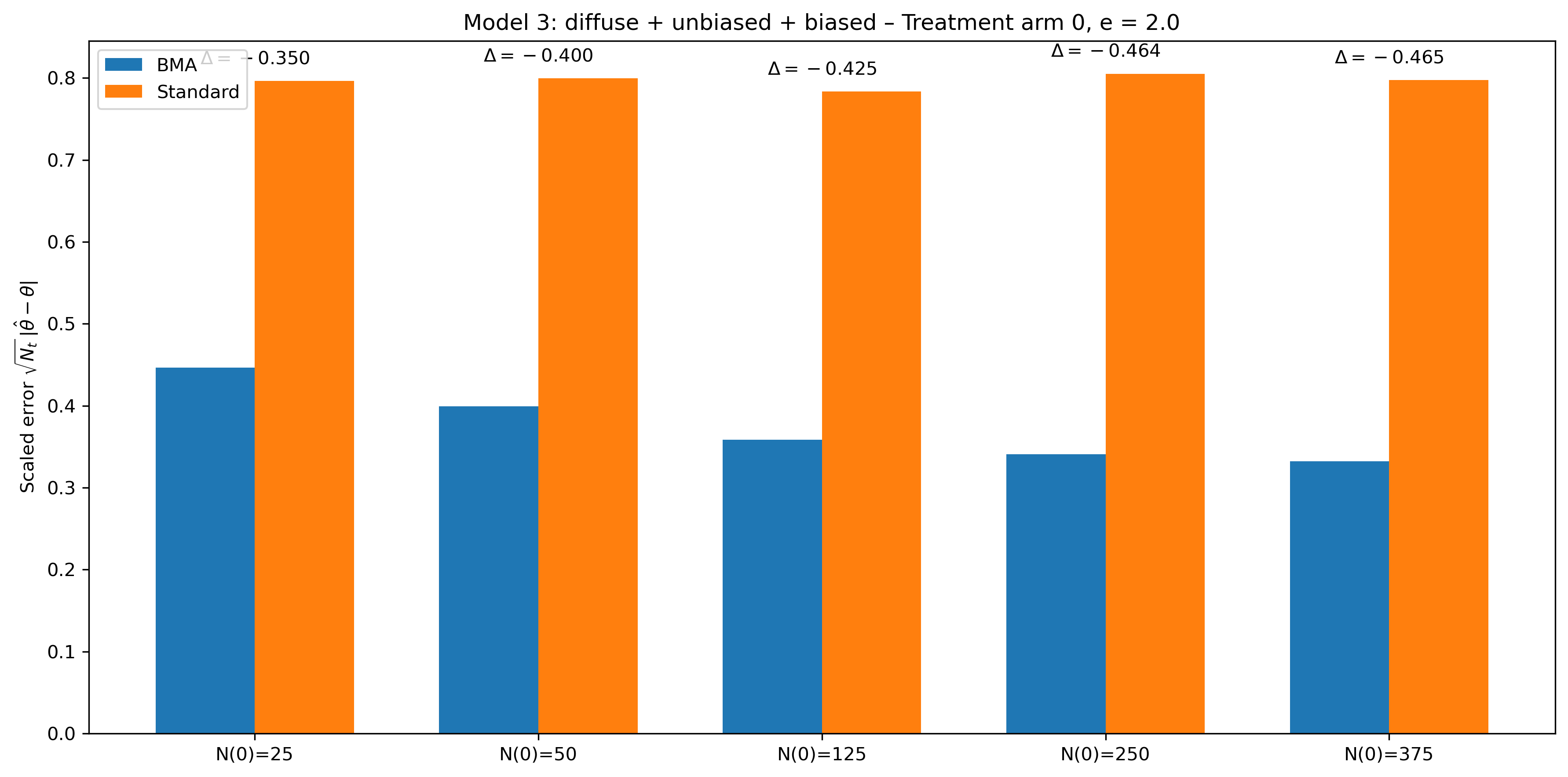}
        \caption{$E=2.0$}
        \label{fig:sub2}
    \end{subfigure}
    \caption{Scaled-Error Rates}
    \label{fig:model3}
   \floatfoot{\footnotesize \textit{Notes:} Each panel reports the mean scaled absolute error in the control arm, $\sqrt{N(0)}\,|\hat\theta(0)-\theta(0)|$, from 1{,}000 Monte Carlo replications. Bars compare the BMA posterior mean estimator (blue) to the standard sample mean (orange) across $N(0)\in\{25,50,125,250,375\}$ (corresponding to $T\in\{50,100,250,500,750\}$). The three panels vary the relative strength of the informative sources, $e\in\{0.5,1,2\}$, which scales source precision proportionally to $N(0)$. Model~3 includes a diffuse baseline source, an informative unbiased source centered at the true mean, and an informative biased source whose initial mean is shifted by $+1$ relative to the truth. The data-generating process is $Y(0)\sim \mathcal{N}(1,1)$ and $Y(1)\sim \mathcal{N}(1.3,1)$. The $\Delta$ labels denote the difference in mean scaled error (BMA minus standard) at each sample size.}

\end{figure}

\begin{figure}[h]
\centering
    \begin{subfigure}{0.7\textwidth}
        \centering
        \includegraphics[width=\textwidth]{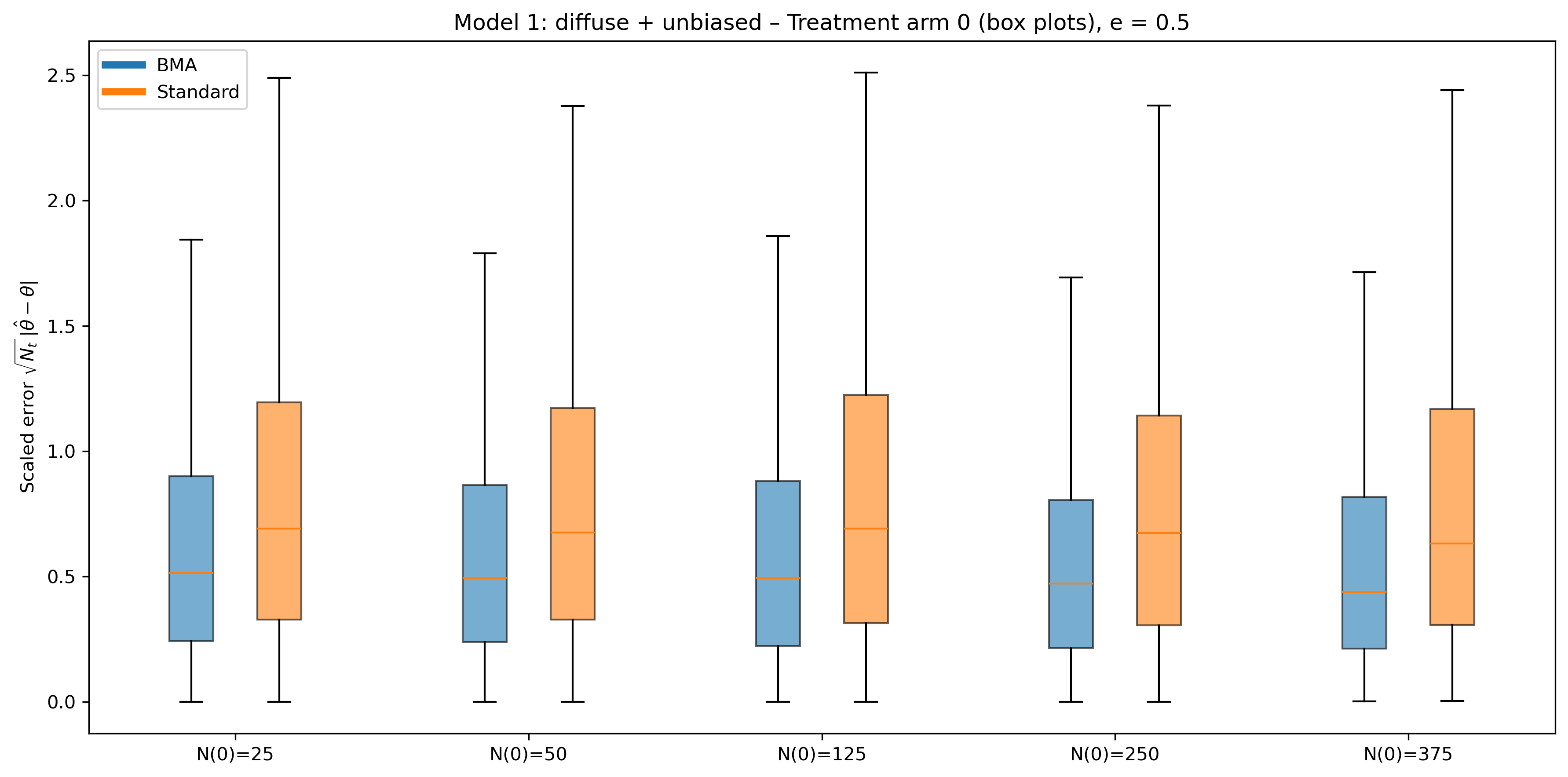}
        \caption{$E=0.5$}
    \end{subfigure}
    \hfill 
    \begin{subfigure}{0.7\textwidth}
        \centering
        \includegraphics[width=\textwidth]{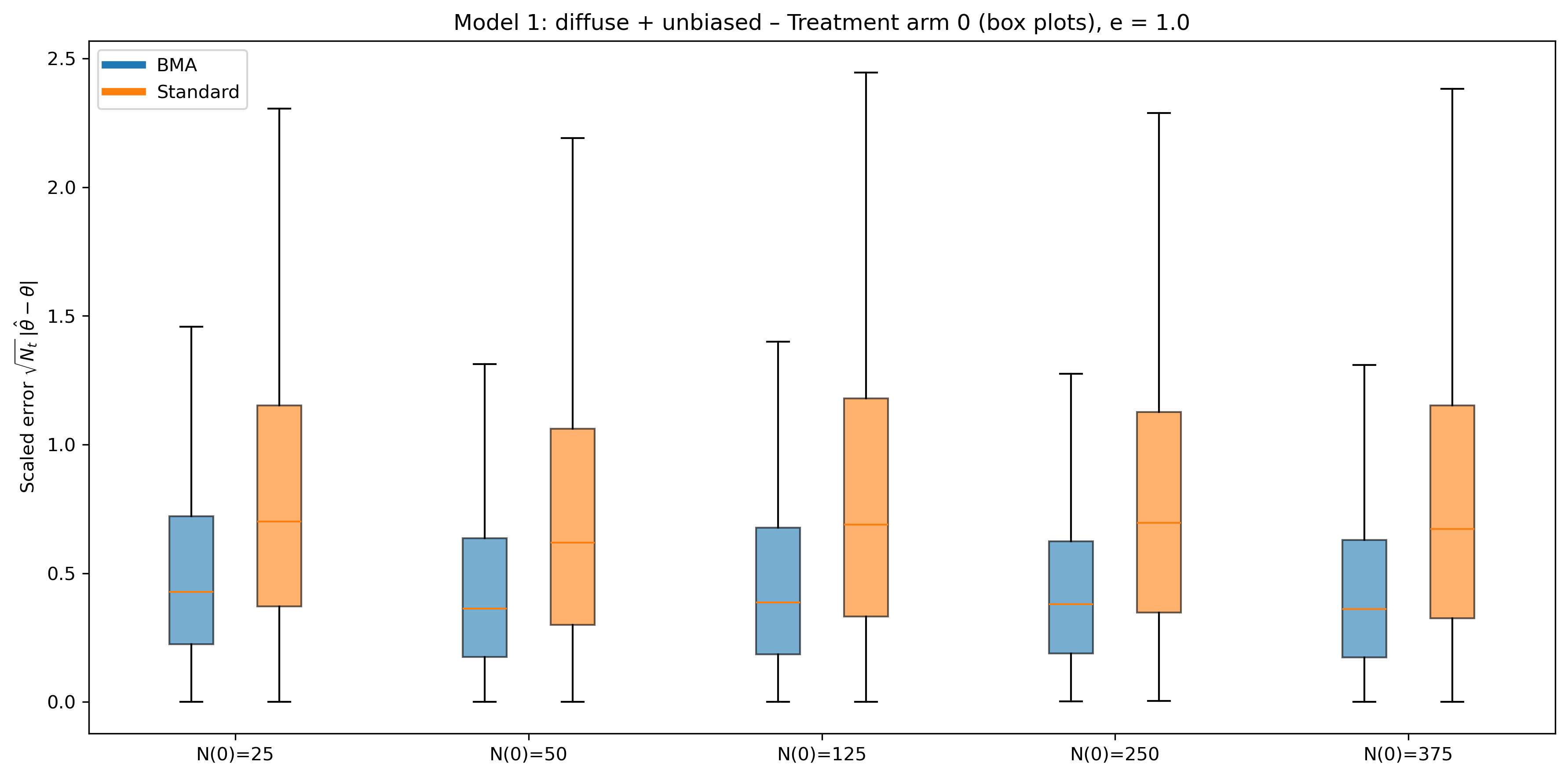}
        \caption{$E=1.0$}
    \end{subfigure}
    \hfill 
    \begin{subfigure}{0.7\textwidth}
        \centering
        \includegraphics[width=\textwidth]{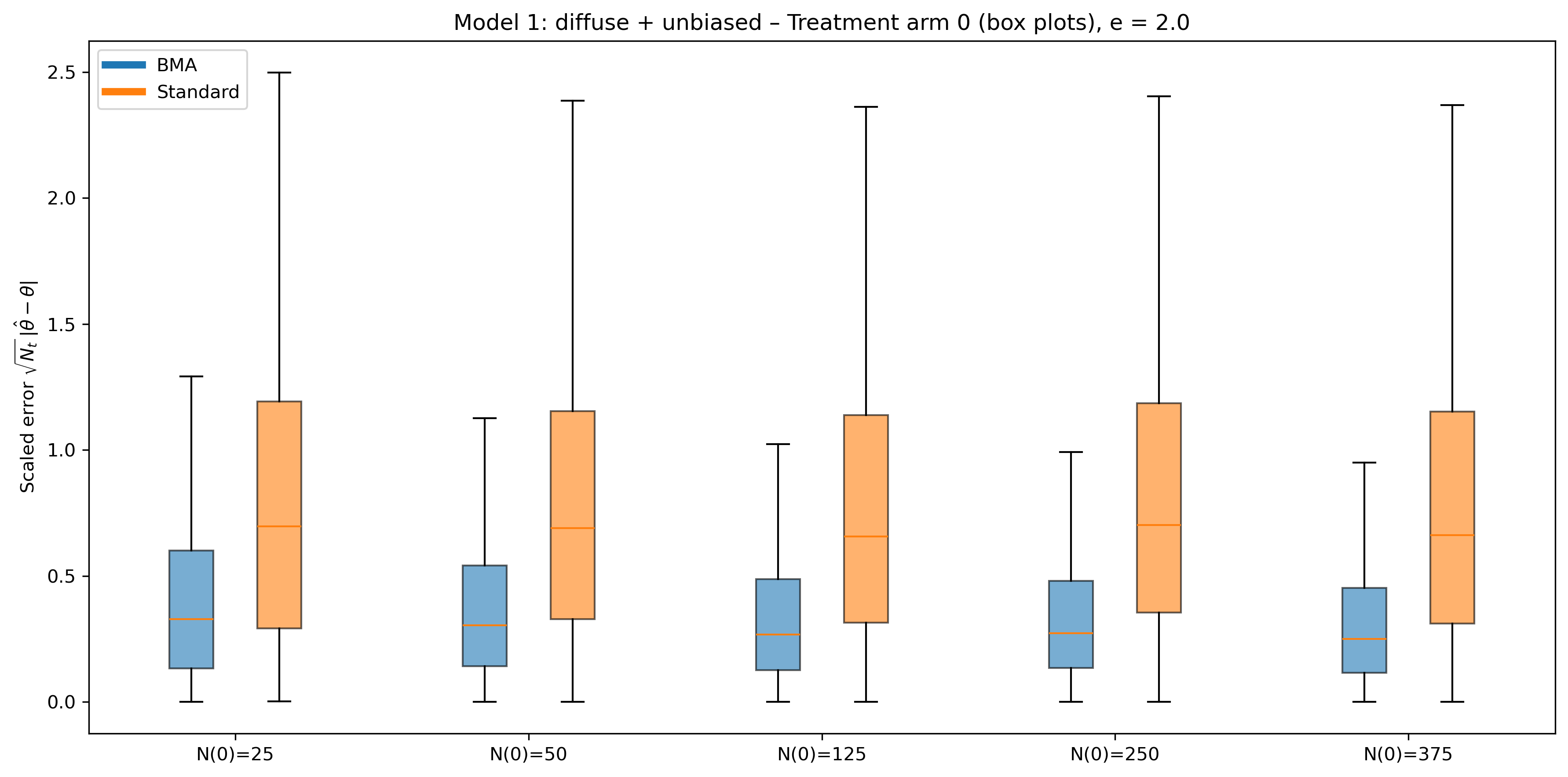}
        \caption{$E=2.0$}
    \end{subfigure}
    \caption{Scaled-Error Rates}
    \label{fig:boxplots_model1}
   \floatfoot{\footnotesize \textit{Notes:} Each panel reports box plots of the scaled absolute error in the control arm, $\sqrt{N(0)}\,|\hat\theta(0)-\theta(0)|$, across 1{,}000 Monte Carlo replications. For each $N(0)\in\{25,50,125,250,375\}$ (corresponding to $T\in\{50,100,250,500,750\}$), the figure shows the distribution of scaled errors for the BMA posterior mean estimator and the standard sample mean. The three panels vary the external-evidence parameter $e\in\{0.5,1,2\}$, which scales the precision of the informative source proportionally to $N(0)$. Model~1 includes a diffuse baseline source (fixed precision $\nu=1$) and an informative \emph{unbiased} source centered at the true mean. The data-generating process is $Y(0)\sim \mathcal{N}(1,1)$ and $Y(1)\sim \mathcal{N}(1.3,1)$. Outliers are suppressed in the box plots.}

\end{figure}

\begin{figure}[h]
\centering
    \begin{subfigure}{0.7\textwidth}
        \centering
        \includegraphics[width=\textwidth]{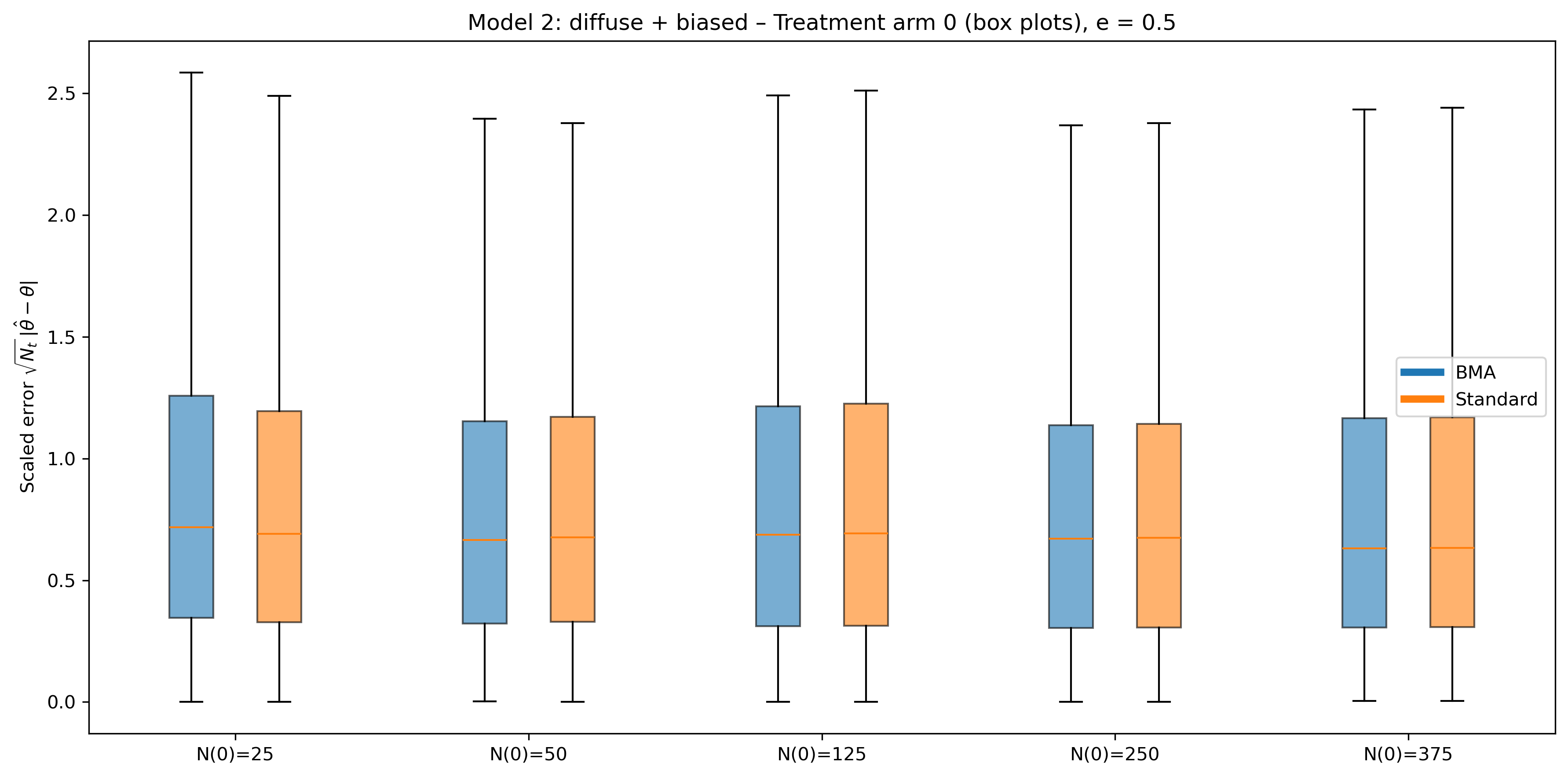}
        \caption{$E=0.5$}
    \end{subfigure}
    \hfill 
    \begin{subfigure}{0.7\textwidth}
        \centering
        \includegraphics[width=\textwidth]{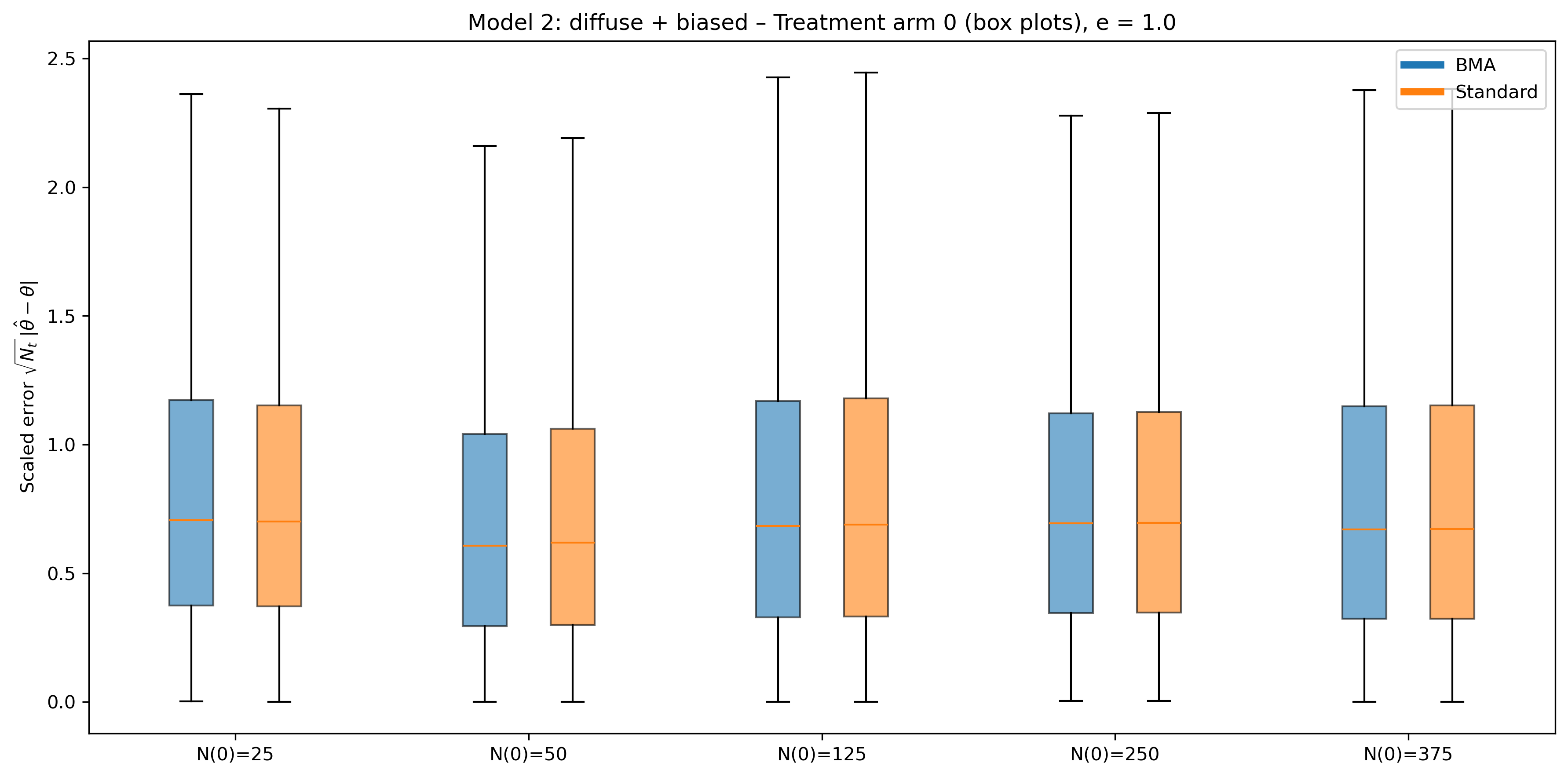}
        \caption{$E=1.0$}
    \end{subfigure}
    \hfill 
    \begin{subfigure}{0.7\textwidth}
        \centering
        \includegraphics[width=\textwidth]{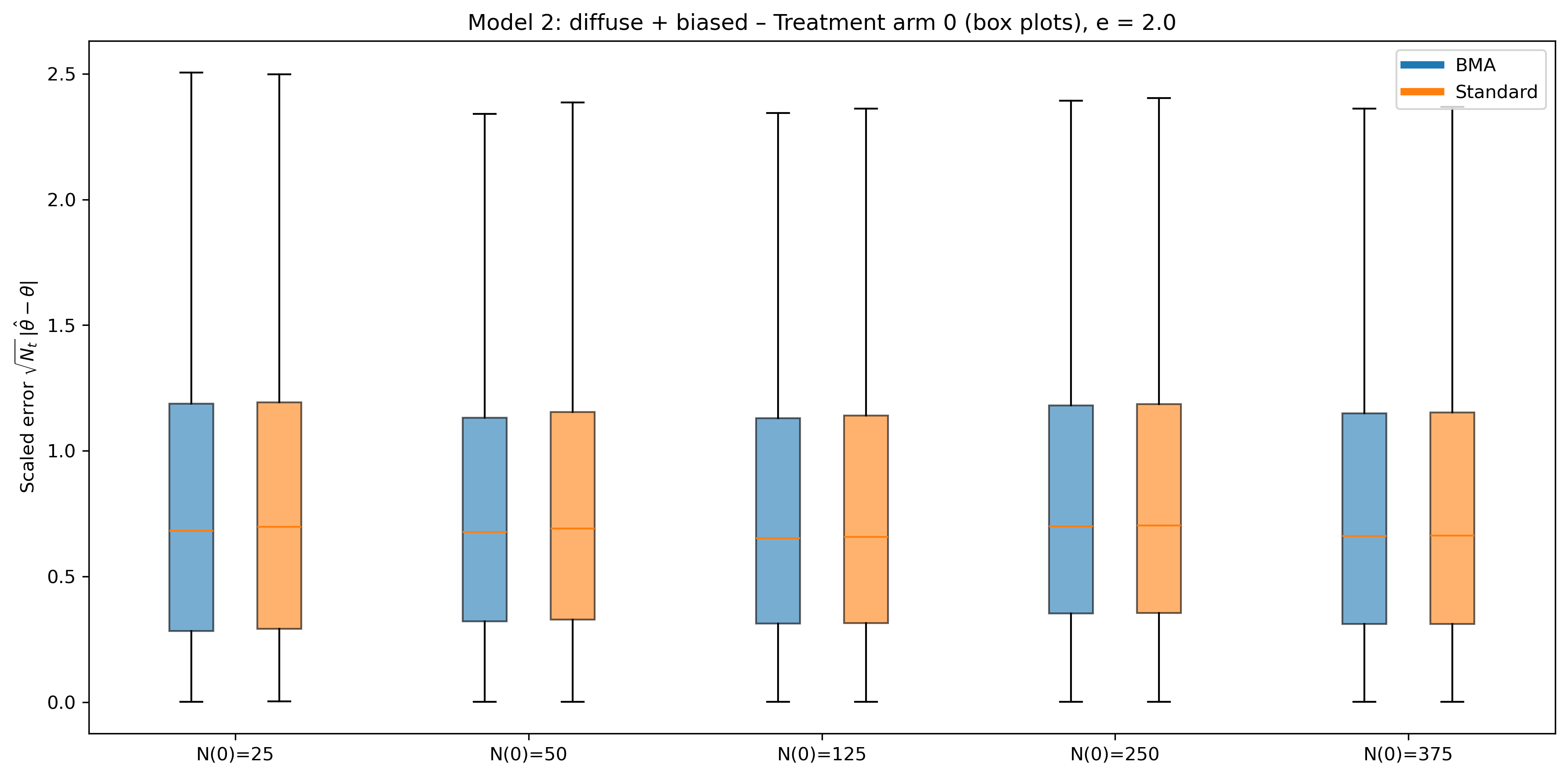}
        \caption{$E=2.0$}
    \end{subfigure}
    \caption{Scaled-Error Rates}
    \label{fig:boxplots_model2}
    \floatfoot{\footnotesize \textit{Notes:} Each panel reports box plots of the scaled absolute error in the control arm, $\sqrt{N(0)}\,|\hat\theta(0)-\theta(0)|$, across 1{,}000 Monte Carlo replications. For each $N(0)\in\{25,50,125,250,375\}$ (corresponding to $T\in\{50,100,250,500,750\}$), the figure shows the distribution of scaled errors for the BMA posterior mean estimator and the standard sample mean. The three panels vary the external-evidence parameter $e\in\{0.5,1,2\}$, which scales the precision of the informative source proportionally to $N(0)$. Model~2 includes a diffuse baseline source (fixed precision $\nu=1$) and an informative \emph{biased} source whose initial mean is shifted by $+1$ relative to the truth. The data-generating process is $Y(0)\sim \mathcal{N}(1,1)$ and $Y(1)\sim \mathcal{N}(1.3,1)$. Outliers are suppressed in the box plots.}

\end{figure}

\begin{figure}[h]
\centering
    \begin{subfigure}{0.7\textwidth}
        \centering
        \includegraphics[width=\textwidth]{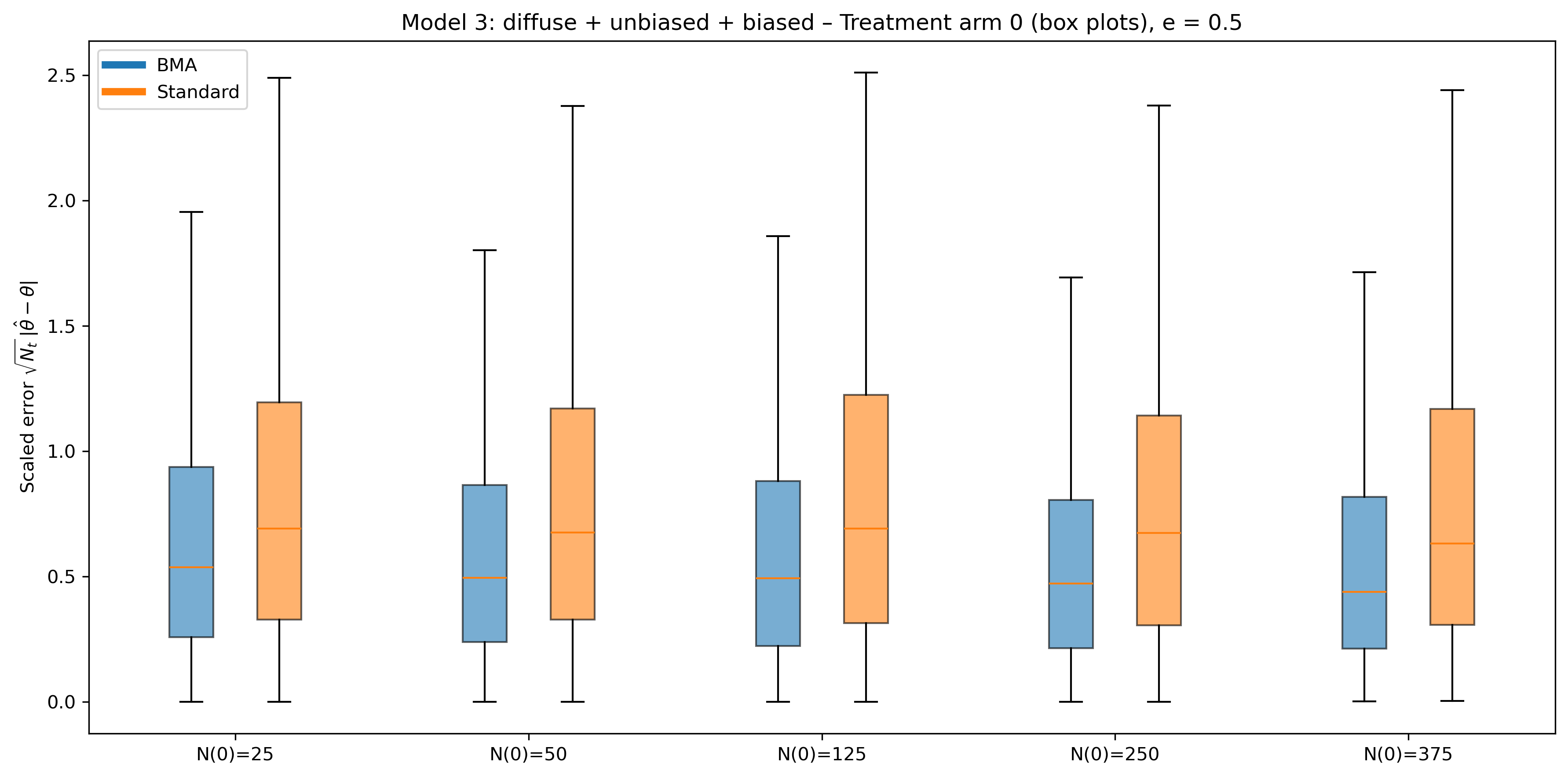}
        \caption{$E=0.5$}
    \end{subfigure}
    \hfill 
    \begin{subfigure}{0.7\textwidth}
        \centering
        \includegraphics[width=\textwidth]{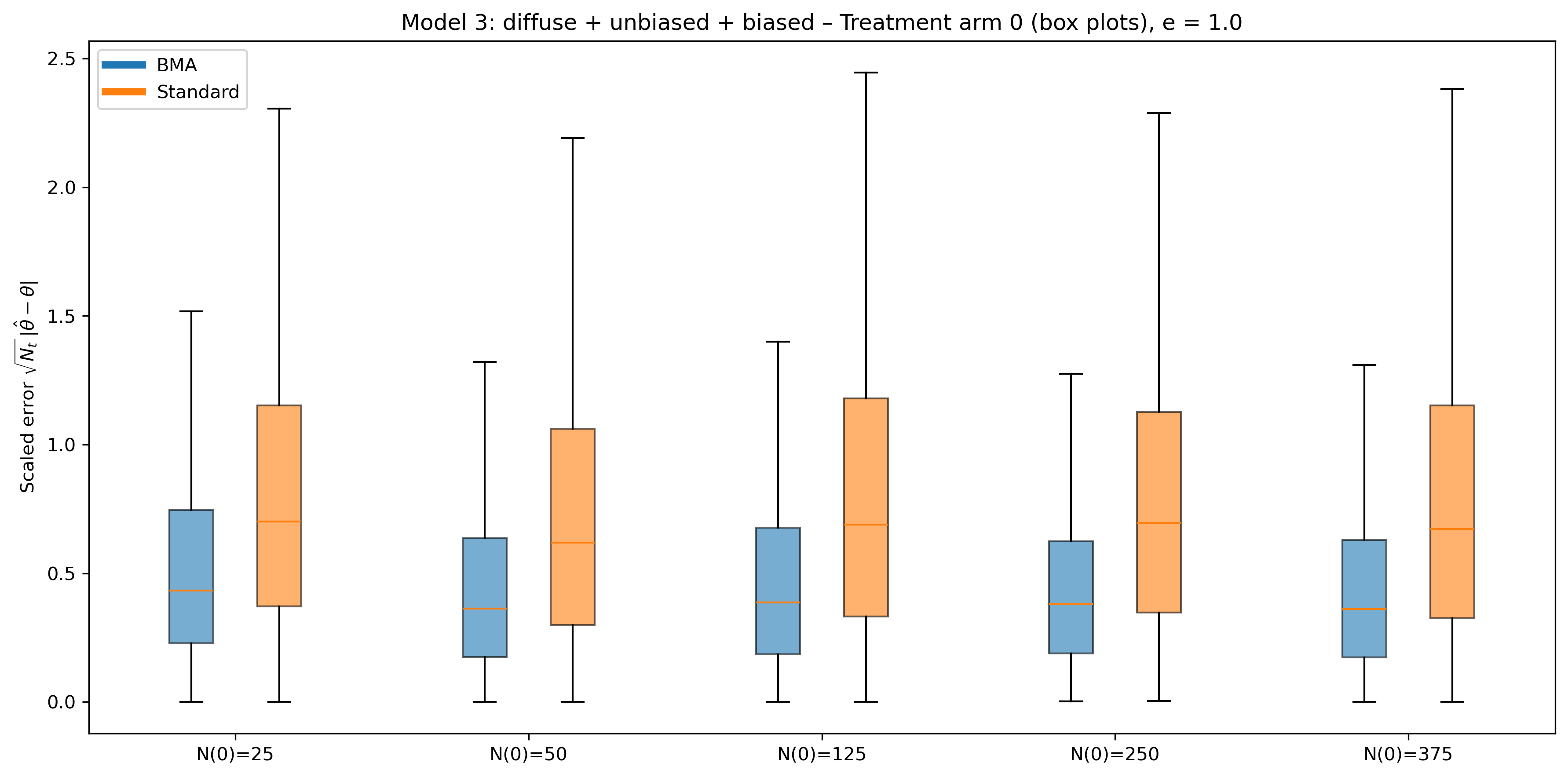}
        \caption{$E=1.0$}
    \end{subfigure}
    \hfill 
    \begin{subfigure}{0.7\textwidth}
        \centering
        \includegraphics[width=\textwidth]{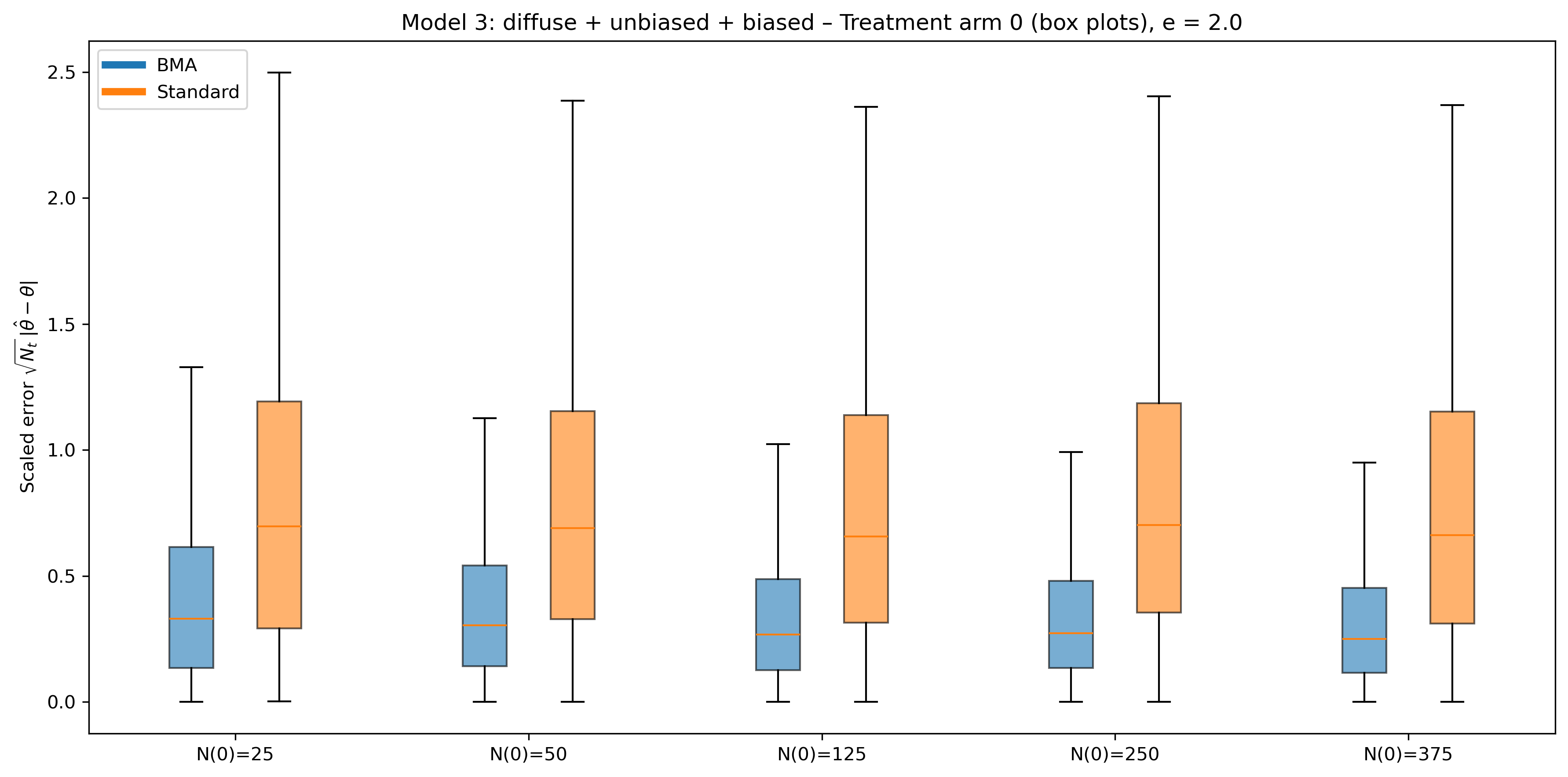}
        \caption{$E=2.0$}
    \end{subfigure}
    \caption{Scaled-Error Rates}
    \label{fig:boxplots_model3}
    \floatfoot{\footnotesize \textit{Notes:} Each panel reports box plots of the scaled absolute error in the control arm, $\sqrt{N(0)}\,|\hat\theta(0)-\theta(0)|$, across 1{,}000 Monte Carlo replications. For each $N(0)\in\{25,50,125,250,375\}$ (corresponding to $T\in\{50,100,250,500,750\}$), the figure shows the distribution of scaled errors for the BMA posterior mean estimator and the standard sample mean. The three panels vary the external-evidence parameter $e\in\{0.5,1,2\}$, which scales source precisions proportionally to $N(0)$. Model~3 includes a diffuse baseline source (fixed precision $\nu=1$), an informative unbiased source centered at the true mean, and an informative biased source whose initial mean is shifted by $+1$ relative to the truth. The data-generating process is $Y(0)\sim \mathcal{N}(1,1)$ and $Y(1)\sim \mathcal{N}(1.3,1)$. Outliers are suppressed in the box plots.}

\end{figure}

For each model, Figures~\ref{fig:model1}--\ref{fig:model3} report mean scaled absolute errors in arm $0$ for ours and the standard estimator across $T\in\{50,100,250,500,750\}$ (i.e., $N_{T}(0)\in\{25,50,125,250,375\}$). Each figure contains three panels corresponding to $e\in\{0.5,1,2\}$. Appendix Figures~\ref{fig:boxplots_model1}--\ref{fig:boxplots_model3} report the full distribution of scaled errors using box plots, to verify that mean effects are not driven by a small set of outliers.

The scaled absolute error captures the speed of learning. According to Theorem \ref{thm:learning.EV}, for models 1 and 3, our estimator will present a faster learning rate than the standard one --- faster by a factor of $1/(1+e)$. Whereas for model 2, our estimator will present a learning rate comparable to the standard one, despite not having unbiased sources.

\subsection*{Results}

\begin{figure}[!h]
\centering
    \begin{subfigure}{0.7\textwidth}
        \centering
        \includegraphics[width=\textwidth]{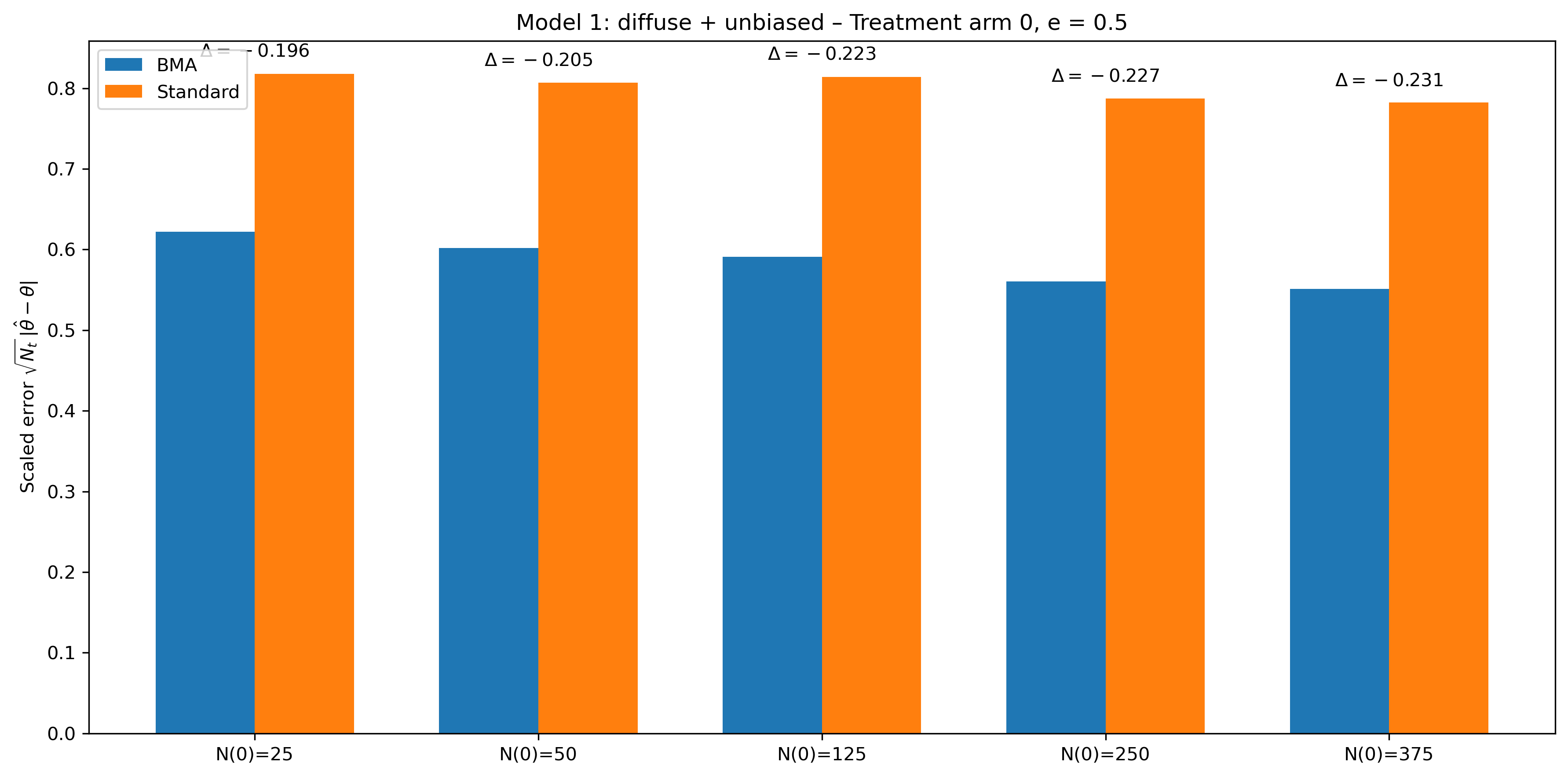}
        \caption{$E=0.5$}
        \label{fig:sub1}
    \end{subfigure}
    \hfill 
    \begin{subfigure}{0.7\textwidth}
        \centering
        \includegraphics[width=\textwidth]{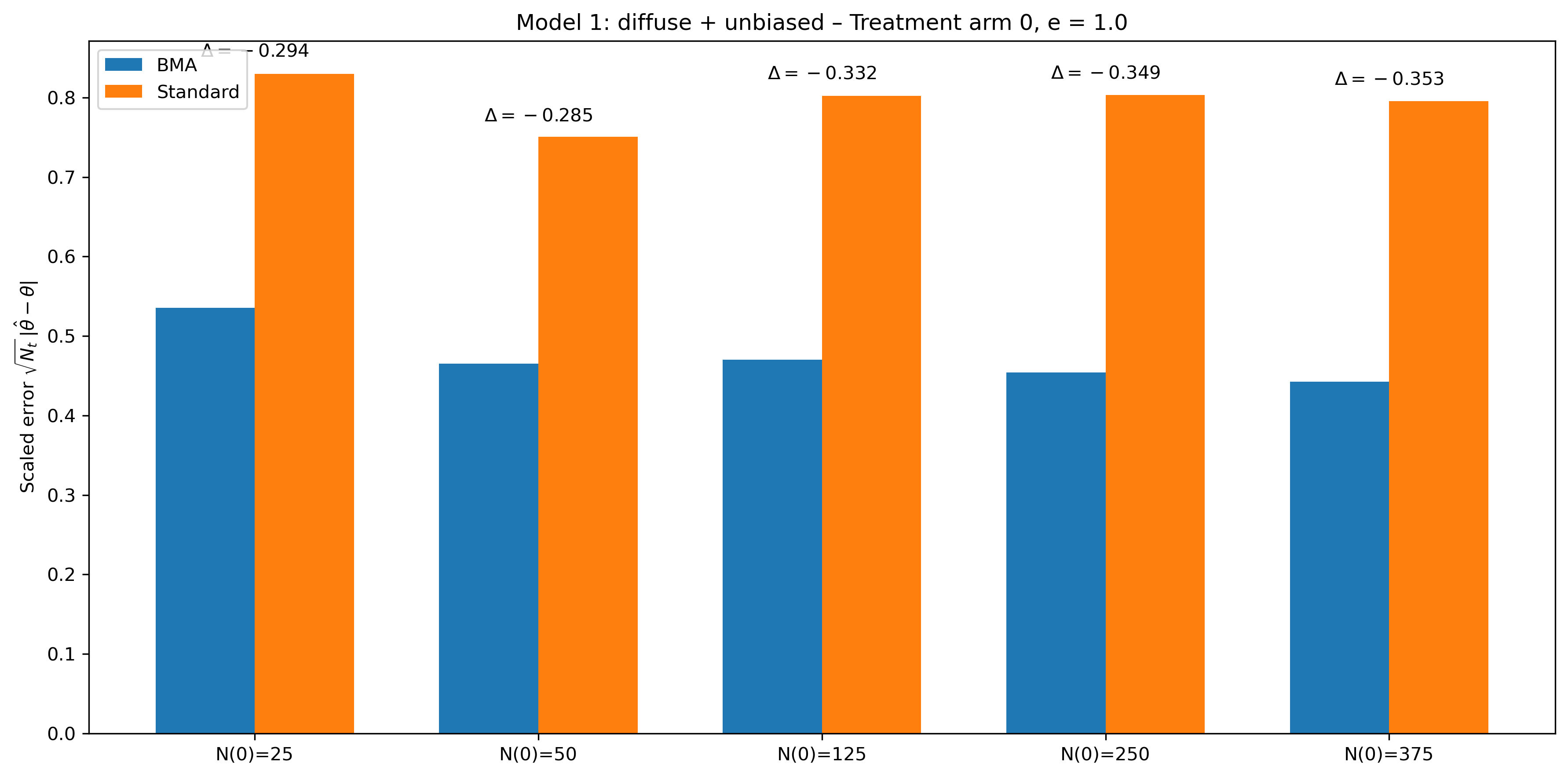}
        \caption{$E=1.0$}
        \label{fig:sub2}
    \end{subfigure}
    \hfill 
    \begin{subfigure}{0.7\textwidth}
        \centering
        \includegraphics[width=\textwidth]{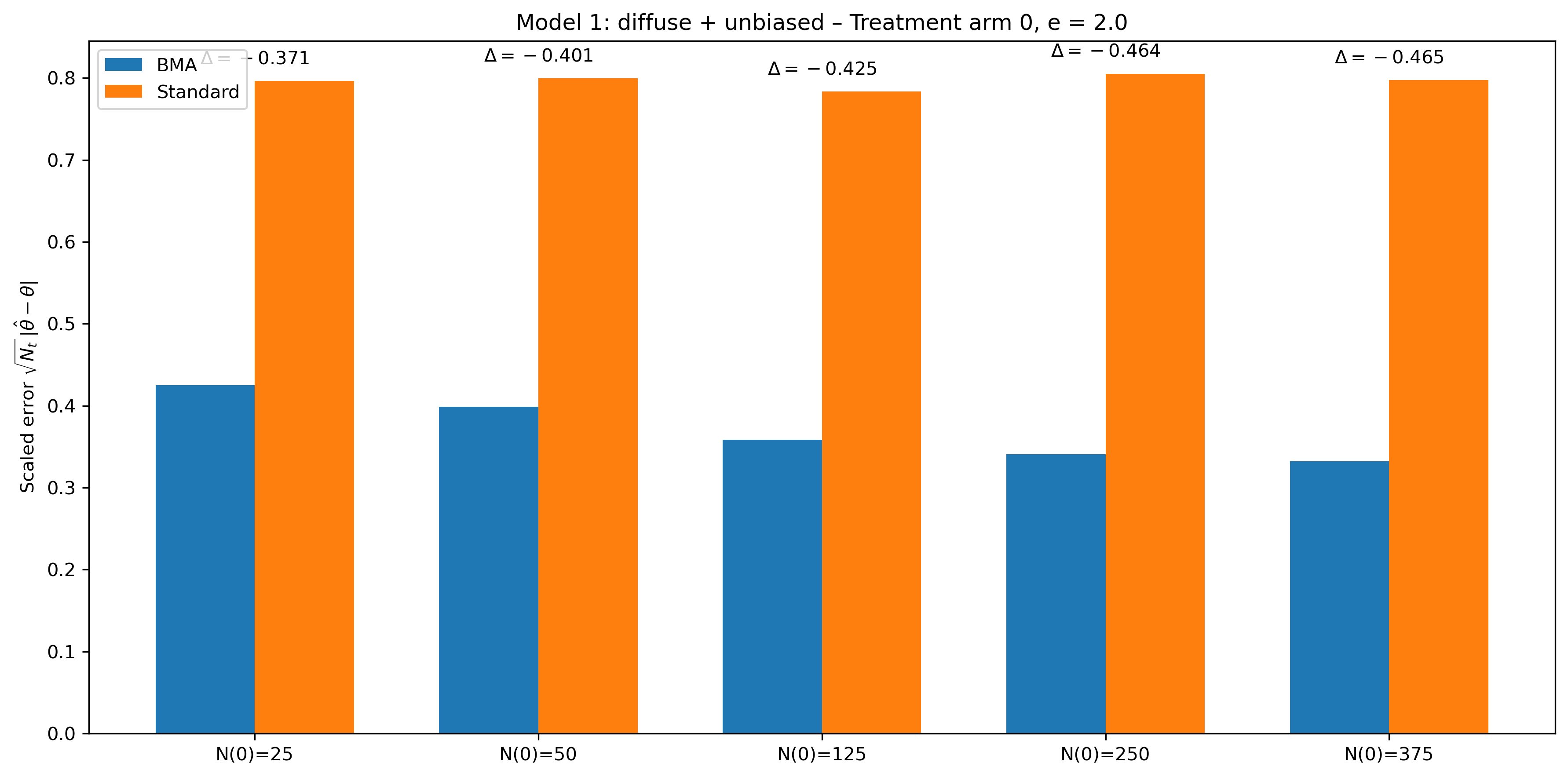}
        \caption{$E=2.0$}
        \label{fig:sub2}
    \end{subfigure}
    \caption{Scaled-Error Rates}
    \label{fig:model1}
    \floatfoot{\footnotesize \textit{Notes:} Each panel reports the mean scaled absolute error in the control arm, $\sqrt{N(0)}\,|\hat\theta(0)-\theta(0)|$, from 1{,}000 Monte Carlo replications. Bars compare the BMA posterior mean estimator (blue) to the standard sample mean (orange) across $N(0)\in\{25,50,125,250,375\}$ (corresponding to $T\in\{50,100,250,500,750\}$). The three panels vary the relative strength of the informative source, $e\in\{0.5,1,2\}$, which scales the source precision proportionally to $N(0)$. Model~1 includes a diffuse baseline source and an informative \emph{unbiased} source, both centered at the true mean. The data-generating process is $Y(0)\sim \mathcal{N}(1,1)$ and $Y(1)\sim \mathcal{N}(1.3,1)$. The $\Delta$ labels denote the difference in mean scaled error (BMA minus standard) at each sample size.}
\end{figure}

\paragraph{A model with an unbiased source.}
Figure~\ref{fig:model1} shows that our estimator delivers sizable gains when an informative source is correctly centered. Across sample sizes, the mean scaled error of the standard estimator is approximately stable (around $0.8$), consistent with Gaussian sampling and $\sqrt{N_T}$ scaling. In contrast, our estimator's errors are uniformly lower and decline modestly with $T$ within each $e$-panel. For example, mean scaled error for our estimator falls from $0.622$ at $N_{T}(0)=25$ to $0.551$ at $N_{T}(0)=375$ when $e=0.5$, from $0.536$ to $0.442$ when $e=1$, and from $0.425$ to $0.332$ when $e=2$.

The alpha-weight table reinforces this interpretation. Table~\ref{tab:alpha} shows that the posterior weight on the unbiased source rises with the experiment sample size: for instance, in Model~1 the average $\alpha$-weight on the unbiased source increases from about $0.71$--$0.75$ at $N_{T}(0)=25$ to about $0.90$--$0.91$ by $N_{T}(0)=375$ (with slightly higher weights for larger $e$). Thus, the improvements in Figure~\ref{fig:model1} reflect systematic reweighting toward the externally valid informative source as data accumulate.

\paragraph{A model with a biased source.}
Figure~\ref{fig:model2} shows that when the only informative alternative is biased, our estimator can perform worse in small samples---particularly when the biased source is not very informative in effective sample size (low $e$). At $N_{T}(0)=25$, mean scaled error of our estimator exceeds that of the sample mean for all $e$ (e.g., $0.879$ vs.\ $0.818$ for $e=0.5$; $0.864$ vs.\ $0.830$ for $e=1$; $0.804$ vs.\ $0.796$ for $e=2$). Intuitively, with limited experimental data, the posterior model weights can still place nontrivial mass on the biased source, and the resulting posterior mean inherits some of that bias, worsening finite-sample accuracy.

As $N_{T}(0)$ increases, our estimator approaches the performance of the standard estimator. By $N_{T}(0)=50$, our estimator and the sample mean are already close, and for larger sample sizes the two estimators are nearly indistinguishable in mean scaled error. This convergence is mirrored in the alpha weights: Table \ref{tab:alpha} shows that the mean $\alpha$-weight on the biased source in Model~2 is small even at $N_{T}(0)=25$ (about $0.09$ for $e=0.5$, $0.04$ for $e=1$, and $0.018$ for $e=2$) and becomes essentially zero by $N_{T}(0)=50$ and beyond. Thus, the small-sample underperformance arises precisely in the range where the biased source still receives some posterior weight and is most pronounced when the source is relatively weak (low $e$).

\paragraph{A model with an unbiased source and a biased source.}
Model~3 allows our estimator to choose among an unbiased informative source and a biased informative source, in addition to the diffuse baseline. Figure~\ref{fig:model3} shows that adding an unbiased competitor restores robustness: Our estimator performance closely tracks Model~1 for moderate and large samples, with only modest degradation in the smallest sample. For instance, at $N_{T}(0)=25$ mean scaled error is $0.667$ for $e=0.5$, $0.570$ for $e=1$, and $0.446$ for $e=2$, compared to $0.622$, $0.536$, and $0.425$ in Model~1; by $N_{T}(0)\ge 50$, mean scaled errors in Model~3 are essentially identical to those in Model~1.

The alpha table clarifies why. Table \ref{tab:alpha} shows that the biased source receives only a small initial weight in Model~3 (e.g., $0.034$ at $N_{T}(0)=25$ for $e=0.5$, and smaller for larger $e$) and this weight collapses quickly with $N_{T}(0)$. Meanwhile, the $\alpha$-weight on the unbiased source rises with $N_{T}(0)$ and converges to the same levels as in Model~1. As a result, Model~3 inherits the precision gains from the valid informative source while rapidly discarding the biased alternative.

\paragraph{Distributional evidence.}
Appendix Figures~\ref{fig:boxplots_model1}--\ref{fig:boxplots_model3} show that these patterns hold across the distribution of simulation outcomes, not only in means. In Model~1, our estimator’s error distribution is uniformly shifted downward relative to the standard estimator, with tighter interquartile ranges as $e$ increases. In Model~2, the main discrepancy occurs in the smallest sample size, where our estimator's distribution exhibits a modest rightward shift relative to the sample mean (especially for low $e$); for larger samples the distributions largely coincide. In Model~3, the small-sample penalty is small and disappears quickly, with the BMA distribution converging to the Model~1 benchmark as the biased source weight vanishes.

\paragraph{Summary.}
Across models, the figures and alpha weights jointly show that our theoretical results accurately captures the behavior of our estimator, even in small samples, and that our estimator’s finite-sample performance is governed by the interaction between (i) the informativeness of external sources (controlled by $e$), (ii) their external validity (bias vs.\ unbiasedness), and (iii) posterior weight concentration. When an unbiased informative source is available (Models~1 and~3), posterior weights tilt toward it and our estimator yields substantial efficiency gains. When only a biased informative source is available (Model~2), our estimator can underperform in small samples---particularly when the source effective sample size is small---but it rapidly learns to downweight the biased source, leading to performance that becomes nearly indistinguishable from the sample mean as $N(0)$ increases.

\section{Extensions}
\label{sec:extensions}

In this section we consider two extensions of the baseline model. The first replaces Gaussian outcomes with binary outcomes and leads to a Bernoulli specification. The second keeps the Gaussian environment but allows each source to use its own set of control variables.


The main message is that Proposition \ref{pro:alpha.asymp} and Theorem \ref{thm:learning.EV} continue to hold in both settings. As in the baseline model, the BMA weights $\alpha_t$ are governed by external validity, understood as the cost of reconciling the source prior value with the target parameter. The form of the cost, however, changes. This cost is naturally expressed in terms of Kullback-Leibler divergence, so its functional form changes outside the homoskedastic Gaussian benchmark: in the Bernoulli model it is no longer quadratic in the mean, and with source-specific controls it becomes matrix-valued because learning is no longer treatment-by-treatment. Even so, the core logic is unchanged: sources that fit the target environment better receive more weight.

\subsection{Binary outcomes: Bernoulli model}
\label{subsec:bernoulli}


\paragraph{Data and parameter.}
Outcomes satisfy $Y_t(d,x) \in \{0,1\}$, and the object of interest is the mean success probability
\[
\theta(d,x) := \mathbb E[Y(d,x)] \in (0,1).
\]
Let $N_t(d,x) := \sum_{i=1}^t \mathbf 1\{D_i(x)=d\}$ and $K_t(d,x) := \sum_{i=1}^t Y_i(d,x)\mathbf 1\{D_i(x)=d\}$, 
and denote the sample mean by $\bar Y_t(d,x) := \frac{K_t(d,x)}{N_t(d,x)}$. 

\paragraph{Prior sources and Updating.} Each source $s \in \mathcal{S} = \{0,\dots,L\}$ is represented by  a Beta prior over $\theta(d,x)$, $\mathrm{Beta}\big(a^s_{0}(d,x),\, b^s_{0}(d,x)\big)$.  As in the Gaussian case, we reparametrize priors by their mean and precision: $a^s_{0} = \nu^s \zeta^s_0$ and $b^s_{0} = \nu^s (1-\zeta^s_0)$. That is, these parameters are the successes and failures associated to the prior source $s$, so $\zeta^{s}_{0}(d,x)$ represents the prior value of the probability of outcome being equal to one, while $\nu^{s}(d,x)$ reflects the precision.

For each $\theta$, the outcome distribution is given by $p_\theta(y) = \theta^y (1-\theta)^{1-y},~ y \in \{0,1\}$.  Hence, conjugacy implies the posterior is $\mathrm{Beta}\big(a^s_{0}(d,x)+K_t(d,x),\;
b^s_{0}(d,x)+N_t(d,x)-K_t(d,x)\big)$ with posterior mean
\begin{align*}
	\zeta^s_t(d,x) 	&= \frac{a^s_{0}(d,x)+K_t(d,x)}{\nu^s(d,x)+N_t(d,x)} = \frac{N_t(d,x)}{N_t(d,x)+\nu^s(d,x)}\,\bar Y_t(d,x)
	\;+\;
	\frac{\nu^s(d,x)}{N_t(d,x)+\nu^s(d,x)}\,\zeta^s_0(d,x).
\end{align*}
This equation is analogous to expression \ref{eqn:zeta.o} for the Gaussian model. 

\paragraph{The BMA weights.} As above, the $\alpha$ weights depend on the integrated likelihood. For each source $s \in \mathcal{S}$, 
\[
\alpha^s_t(d,x) = \frac{\mathcal M^s_t(d,x)}{\sum_{s' \in \mathcal{S}} \mathcal M^{s'}_t(d,x)},
\]
where 
\[
\mathcal M^s_t(d,x)
=
\frac{
	B\!\left(a^s_{0}(d,x)+K_t(d,x),\;
	b^s_{0}(d,x)+N_t(d,x)-K_t(d,x)\right)
}{
	B\!\left(a^s_{0}(d,x),\; b^s_{0}(d,x)\right)
},
\]
where $B(\cdot,\cdot)$ denotes the Beta function. The BMA estimator is defined as before,
\[
\widehat\theta_t(d,x) := \sum_{s \in \mathcal{S}} \alpha^s_t(d,x)\,\zeta^s_t(d,x).
\]

Finally, we impose the same nonstandard asymptotics as in the Gaussian case: Allowing $\nu^{s}(d,x)$ to be indexed by $t$ and $\frac{\nu^s_t(d,x)}{N_t(d,x)} \to c_s(d,x) \in \mathbb R_+,~\text{a.s.}$.  Sources with $c_s(d,x)=0$ are asymptotically diffuse, while $c_s(d,x)>0$ corresponds to sources whose information content is comparable to the experiment.

\paragraph{External Validity Measure.}  
The external validity of source $s$ for $(d,x)$ at instance $t$ is given by
	\begin{equation}
	\label{eq:EV-bern}
	\mathbb{EV}^s_{\mathrm{Bern},t}(d,x)
	:= -N_{t}(d,x)  	\Psi^{s}\!\big(\theta(d,x), \zeta_{0}^{s}(d,x)\big),
\end{equation}
where
\begin{align}\label{eqn:Bern.KL2}
\Psi^{s}\!\big(\theta(d,x), \zeta_{0}^{s}(d,x)\big)
:= \inf_{u\in(0,1)}\left\{	\mathrm{KL}\!\big(\theta(d,x)\,\|\,u\big)
+ c^{s}(d,x)\,\mathrm{KL}\!\big(\zeta^{s}_0(d,x)\,\|\,u\big)
\right\}.
\end{align}

Expression \ref{eq:EV-bern} is a generalization of expression \ref{eqn:EV} to non-Gaussian frameworks wherein the KL divergence is not quadratic. The quantity $\Psi^{s}(\theta(d,x),\zeta_0^{s}(d,x))$ therefore represents the (asymptotic) penalty incurred when the source posterior is required to reconcile its prior belief $\zeta_0^{s}(d,x)$ with the target truth $\theta(d,x)$. Larger values of $c^{s}(d,x)$ (more dogmatic sources) magnify the cost of discrepancy. 

However as expression \ref{eqn:EV} in the Gaussian model, expression \ref{eq:EV-bern} also acts as a notion of distance between $\theta(d,x)$ and $\zeta^{s}_{0}(d,x)$. Indeed, it is not hard to show that for $\Psi^{s}(\theta(d,x),\zeta_0^{s}(d,x)) =0$ if $\theta(d,x)=\zeta_0^{s}(d,x)$ and positive otherwise.

\paragraph{Theoretical Guarantees for the Bernoulli Model.} We conclude by extending all our results extend to the binary outcomes case.  The next result is analogous to Proposition \ref{pro:Bweights.EV}. 

\begin{proposition}[Posterior weights and external validity: Bernoulli case]
	\label{prop:EV.Bern}
For any $(d,x)\in \mathbb D\times \mathbb X$,
	\begin{equation}
		\label{eq:prop31-bern}
	\alpha^{s}_{t}(d,x) = \frac{  e^{-N_{t}(d,x) \Psi^{s}\!\big(\theta(d,x), \zeta_{0}^{s}(d,x)\big) + o_{as}(1)  )  }   }{\sum_{s' \in \mathcal M}  e^{-N_{t}(d,x) \Psi^{s}\!\big(\theta(d,x), \zeta_{0}^{s}(d,x)\big) + o_{as}(1)  )  } } .  
	\end{equation}
\end{proposition}

\begin{proof}
	See Appendix \ref{app:Bernoulli}.
\end{proof}

The concept of unbiased source remains unchanged in this new setup, because, as pointed out above, $\Psi^{s}\!\big(\theta(d,x), \zeta_{0}^{s}(d,x)\big)$ is naught only if the source is unbiased (or diffuse). Therefore, an analogous result to Proposition \ref{pro:alpha.asymp} holds: Bias sources will be discarded exponentially fast --- at rate given by $N_t(d,x) \Psi^{s}(\theta(d,x),  \zeta^{s}_0(d,x))$ --- in favor of unbiased ones (or diffused should all sources be biased). Consequently, Theorem \ref{thm:learning.EV} also holds for the Bernoulli model.

In addition to extending Theorem~\ref{thm:learning.EV} --- and our theory --- to binary outcomes, the main takeaway of this section is that the posterior weights continue to be governed by the same external validity logic as in the Gaussian case, even though the squared-Euclidean metric is replaced by the information-theoretic divergence $\Psi^{s}(\theta(d,x),\zeta_0^{s}(d,x))$, which emerges from the joint KL projection of the target and source beliefs. Thus, the geometry of comparison changes  but the economic content of the weights remains identical: sources are rewarded or penalized according to how costly it is, in information terms, to reconcile their prior belief with the target environment, with the penalty scaled by their effective precision $c^{s}(d,x)$.

\subsection{A Gaussian Linear Regression model}
\label{subsec:regression}

\paragraph{Model.} Let $\boldsymbol{\theta}:=(\theta(0),\ldots,\theta(M))^{\top}$ denote the vector of treatment-specific mean parameters, let $Z_{t} = (1\{ D_{t} = 0 \},\ldots , 1\{D_{t} = M\})^{\top}$ be the $(M+1)\times 1$ treatment-indicator vector, and let $W_t^{s}\in\mathbb R^{p_s}$ be the predetermined vector of controls used by source $s$, with associated nuisance coefficient $\gamma^{s}\in\mathbb R^{p_s}$. We work with the linear Gaussian regression
\begin{align}
	Y_{t} = \boldsymbol{\theta}^{\top}  Z_{t}  + ( \gamma^{s}  )^{\top} W^{s}_{t}	+	\varepsilon^{s}_t,
	\qquad d\in\mathbb D,
	\label{eq:controls.model-stacked} 
\end{align}
where $\varepsilon_t^s \sim \mathcal N(0,1)$. The parameter of interest is $\boldsymbol{\theta}$, so the average treatment effect of arm $d$ relative to the baseline arm is $\theta(d)-\theta(0)$. Allowing $W_t^s$ and $\gamma^s$ to vary across sources captures the fact that different prior studies may use different controls, while keeping the treatment means comparable across sources. As in the benchmark Gaussian model, normality is imposed for analytical convenience; for identification of the conditional mean, the essential requirement is a mean-zero error with finite variance.

\paragraph{Likelihoods, Sources, and Posteriors.}  The posterior in  this model is analogous to the one for no-controls but with an important caveat. Due to the fact the coefficients $\gamma^{s}$ can be common across treatment, learning does not longer take place ``treatment-by-treatment" but jointly.  We now formalize this. 

For each source $s \in \mathcal{S}$, let 
\[
X_{t}^{s}:=\big[\,Z_{t}\;\; W_{t}^{s}\,\big]\in\mathbb R^{1\times((M+1)+p_s)},
\qquad 
\beta^{s}:=\begin{pmatrix}\boldsymbol{\theta} \\ \gamma^{s}\end{pmatrix}\in\mathbb R^{((M+1)+p_s)  \times 1}.
\]
Then expression \eqref{eq:controls.model-stacked} implies the Gaussian working likelihood
\begin{align}
\prod_{i=1}^{t} 	\phi \!\Big(Y_{t};\; X_{t}^{s}\beta^{s},\; 1 \Big) = (2\pi)^{-t/2}
\exp\!\left(
-\frac{1}{2}
\sum_{i=1}^{t}
\big(Y_i - X_i^{s}\beta^{s}\big)^2
\right).
	\label{eq:likelihood}
\end{align}

Each source $s$ is represented by Gaussian prior over its model-specific parameter $\beta^{s}$:
\begin{align}
\mathcal N\!\big(\beta_{0}^{s},\,\Sigma_{0,t}^{s}\big),
	\qquad 
	\beta_{0}^{s}:=\begin{pmatrix}\boldsymbol{\zeta}_0^{s}\\ \eta_0^{s}\end{pmatrix},
	\label{eq:prior-beta}
\end{align}
where $\boldsymbol{\zeta}_0^{s}\in\mathbb R^{M+1}$ is the source mean for $\theta$ and $\eta_0^{s}\in\mathbb R^{p_s}$ is the source mean for $\gamma^{s}$. We allow the prior covariance $\Sigma_{0,t}^{s}$ to depend on $t$ so as to reproduce the same ``growing information in the sources'' regime as in the benchmark model. 

Specifically,\footnote{Here and throughout, for any vector $X$, $\operatorname{Diag}[X]$ denotes a diagonal matrix with diagonal components given by the vector $X$.}
\begin{align}
	\Sigma_{0,t}^{s}
	=
	\begin{pmatrix}
	\Sigma^{s}_{\theta\theta,t} & 0 \\
		0 &  \Sigma^{s}_{\gamma\gamma,t}
	\end{pmatrix}
= 	\begin{pmatrix}
\operatorname{Diag}[1/\nu^{s}_{t}] & 0 \\
	0 &  \operatorname{Diag}[1/\lambda^{s}_{t}]
\end{pmatrix}
\end{align}
where $\lambda^{s}_{t}$ is a $p_{s} \times 1$ vector uniformly bounded away from zero.

The term $\operatorname{Diag}[1/\nu^{s}_{t}]$ is completely analogous to $\nu^{s}_{t}(d,x)$ in the no-controls case. The vector $\lambda^{s}_{t}$ represents the prior precision associated to the control coefficients. We assume that $\lambda_{t}^{s}/t = o(1/\sqrt{t})$, so the effect of this prior precision vanishes asymptotically. This assumption is only used to simplify the asymptotic expressions.

By conjugacy, for each source $s\in\mathcal S$, the posterior over $\beta^{s}$ is Gaussian,
\[
\beta^s \mid Y_{1:t},D_{1:t},W_{1:t}^s \sim \mathcal N(\beta_t^s,\Sigma_t^s),
\]
where $\Sigma_t^s$ is the posterior covariance matrix and the posterior mean is $\beta_t^s=((\boldsymbol\zeta_t^s)^\top,(\eta_t^s)^\top)^\top$.\footnote{See Appendix \ref{app:Gaussian.controls} for a proof of this result and the expressions for the posterior mean and covariance matrix.}

To isolate the treatment-effect component, partition the posterior precision matrix as
\begin{align}
	(\Sigma_t^s)^{-1}
	=
	\begin{pmatrix}
		A_t^s & B_t^s \\
		(B_t^s)^\top & C_t^s
	\end{pmatrix},
\end{align}
where $A_t^s$ is the treatment-treatment block, $B_t^s$ is the treatment-control block, and $C_t^s$ is the control-control block. Then Appendix \ref{app:Gaussian.controls} provides expressions for these terms, in the main text we only need the induced expression for the treatment posterior mean.

\paragraph{Estimator.} Our goal is to estimate $\boldsymbol{\theta}$. Without prior sources, the standard estimator is the OLS coefficient on the treatment indicators in a regression of $Y_t$ on $(Z_t,W_t)$, or equivalently the Frisch-Waugh-Lovell residualized estimator that partials out the controls before averaging within treatment arms. Based on this intuition, a natural source-specific estimator uses the same logic while incorporating the prior information. This is formalized in the next lemma. Define $N_t := (N_t(0),\ldots,N_t(M))^\top$ and $\boldsymbol m_t := (m_t(0),\ldots,m_t(M))^\top$ with $m_t(d) := \frac{1}{N_t(d)}\sum_{i=1}^t 1\{D_i=d\}Y_i$.
\begin{lemma}\label{lem:controls.zeta.posterior}
	For any instance $t$ and any source $s$, 
	\begin{align}
		\boldsymbol{\zeta}^{s}_{t} = 
		(T^{s}_{t})^{-1} \left( 	\operatorname{Diag}[\nu^{s}_{t}]  \boldsymbol{\zeta}_0^s +   \operatorname{Diag}[N_{t}]  \boldsymbol{m}_{t}  -  B_t^s (C_t^s)^{-1} \left( \operatorname{Diag}[\lambda^{s}_{t}]   \eta_0^s +    \sum_{i=1}^{t} (W^{s}_{i})^{\top} Y_{i} \right)      \right)
	\end{align}
	where $T_t^s := A_t^s - B_t^s (C_t^s)^{-1} (B_t^s)^\top $.
\end{lemma}

\begin{proof}
	See Appendix \ref{app:Gaussian.controls}. 
\end{proof}


Lemma \ref{lem:controls.zeta.posterior} is the special case of the more general block posterior system that is relevant for our treatment-effect analysis; the full statement is reported in Appendix \ref{app:Gaussian.controls}. The expression combines a vector-valued version of \eqref{eqn:zeta.o}, where each component averages the prior mean $\zeta_0^s$ and the empirical treatment mean $m_t$, with the Frisch-Waugh-Lovell adjustment that partials out the controls. If the precision matrix $(\Sigma_t^s)^{-1}$ were diagonal, the treatment effects would update independently across arms. With controls, however, the off-diagonal block $B_t^s$ captures the empirical link between treatment assignments and controls, so updating must be done jointly.


In particular, the term
\[
B_t^s (C_t^s)^{-1}
\left(
\operatorname{Diag}[\lambda_t^s]\eta_0^s
+
\sum_{i=1}^{t} (W_i^s)^\top Y_i
\right)
\]
subtracts the part of $	\operatorname{Diag}[\nu^{s}_{t}]  \boldsymbol{\zeta}_0^s +   \operatorname{Diag}[N_{t}]  \boldsymbol{m}_{t}  $ that is explained by the controls. The matrix $T_t^s$ is the Schur complement of the control block and represents the effective information about the treatment effects after accounting for the controls. In this sense, the posterior treatment effects correspond to a precision-weighted update based on treatment information that has been residualized with respect to the controls, mirroring the role of the Frisch--Waugh--Lovell theorem in classical regression.

The estimator of $\boldsymbol{\theta}$ takes the ensemble of estimators $(\boldsymbol{\zeta}^{s}_{t})_{s \in \mathcal{S}}$ and combines them using BMA weights,
\begin{align}
	\widehat{\boldsymbol{\theta}}_t := \sum_{s\in\mathcal S}\alpha_t^{s} \boldsymbol{\zeta}_t^{s}.
	\label{eq:controls.bma-theta}
\end{align}
where $\alpha_t^{s}
	\;:=\;
	\frac{\pi_s\, m_t^{s}(Y_{1:t}\mid D_{1:t},w_{1:t}^{s})}
	{\sum_{r\in\mathcal S}\pi_r\, m_t^{r}(Y_{1:t}\mid D_{1:t},w_{1:t}^{r})}$, with $m_t^{s}(Y_{1:t}\mid D_{1:t},w_{1:t}^{s})
:=\int  \prod_{i=1}^{t} \phi(Y_{i} ;  X^{s}_{i} \beta , 1 )
	\phi (\beta;\beta_{0}^{s},\Sigma_{0,t}^{s})\,d\beta$ 
being the source marginal likelihood. Unlike the no-controls case, these source weights are not treatment-specific because the controls make posterior updating joint across treatment arms.

\paragraph{Theoretical guarantees.} In this section we present the theoretical guarantees for this model. The next proposition links the weights of the source to the External Validity measure. 

\begin{proposition}\label{pro:control.weights.EV}
	For any instance $t$ and any source $s \in \mathcal{S}$,
	\begin{align*}
		\alpha^{s}_{t} \propto \exp\!\left\{
			-\frac t2
			\Big[
			(\boldsymbol{\theta}^{s}-\boldsymbol{\zeta}_{0}^{s})^\top \bar{H}^{s}
			(	\boldsymbol{\theta}^{s}-\boldsymbol{\zeta}_{0}^{s} )
		+ \frac{1}{t} \left( 2d_{s} \log t  +  \log | \bar{Q}^{s} \Sigma^{s}_{0,t} | \right)
			+ o_{as}(1)
			\Big]
			\right\},
	\end{align*}
where $ H^{s} : = (\operatorname{Diag}[c^{s}\cdot\delta])^{1/2}\Big(I - (\operatorname{Diag}[c^{s}\cdot\delta])^{1/2}K^{-1}(\operatorname{Diag}[c^{s}\cdot\delta])^{1/2}\Big)(\operatorname{Diag}[c^{s}\cdot\delta])^{1/2}$ with
$ K
	:=
	\operatorname{Diag}[1+c^s] E[(Z)(Z)^{\top}]
	-
	E[Z \boldsymbol W^{s}]
	\big(E[W^{s}(W^{s})^\top]\big)^{-1}
	E[Z \boldsymbol W^{s}]^{\top}$. 
\end{proposition}

\begin{proof}
	See Appendix \ref{app:contro.weights.EV}.
\end{proof}

The matrix $\bar H^{s}$ represents the asymptotic curvature of the marginal likelihood with respect to deviations of the treatment parameters from the benchmark value $\beta_0^{s}$. It shows that this curvature arises from the interaction between the prior precision contributed by source $s$ and the information about the treatment parameters available in the experimental data after accounting for the nuisance parameters. The matrix $\operatorname{Diag}[c^{s}\cdot\delta]$ captures the prior precision of the source, while $K$ represents the information matrix for the treatment parameters after partialling out the controls via the Schur complement --- $K$ measures how informative the experiment is about treatment after accounting for the controls. Consequently, $\bar H^{s}$ can be interpreted as the effective prior precision of source $s$ once it is filtered through the experimental information about the treatment parameters.

The term $ \frac{1}{t} \left( 2d_{s} \log t  +  \log | \bar{Q}^{s} (\Sigma^{s}_{0,t}/t) | \right) $ is the standard ``log $t$" penalization term obtained in BMA regression models (e.g. BIC criteria, \cite{Fernandez2001}) but extended to incorporate the fact that prior precision may not be vanishing. 

Thus, the leading term of the source weights for the Gaussian regression with controls is given by
\begin{align*}
	- \left(     \left( \boldsymbol{\theta} - \boldsymbol{\zeta}^{s}_{0} \right)^{\top}  (t\bar{H}^{s})  \left( 	\boldsymbol{\theta} - \boldsymbol{\zeta}^{s}_{0} 	 \right)  + \left( 2d_{s} \log t  +  \log | \bar{Q}^{s} (\Sigma^{s}_{0,t}/t) | \right)\right) .
\end{align*}

To shed more light on this expression, consider the absence of controls. In this case the matrix $K$ simplifies to $K = \operatorname{Diag}[(1+c^s)\cdot\delta]$.  Substituting this expression into the definition of $\bar H^s$ yields (up to negligible $o_{as}(1)$ terms)
\[
(t\bar H^s)
=
\operatorname{Diag}\left[ t \delta \cdot \frac{c^s}{1+c^s} \right] = \operatorname{Diag}\left[ N_{t} \cdot \frac{c^s}{1+c^s} \right] = \operatorname{Diag}\left[ \frac{\nu^{s}_{t} }{1+c^s} \right] 
\]
Thereby yielding a leading term in the EV measure of $  \left( \boldsymbol{\theta} - \boldsymbol{\zeta}^{s}_{0} \right)^{\top} \operatorname{Diag}\left[ \frac{\nu^{s}_{t} }{1+c^s} \right]   \left( 	\boldsymbol{\theta} - \boldsymbol{\zeta}^{s}_{0} 	 \right) $ which is a vector-valued version of the EV measure for the no-controls case given in expression \ref{eqn:EV}.

\medskip

We now show that an analogous result to Theorem \ref{thm:learning.EV} holds in this framework. To see this, we provide a characterization of $	\boldsymbol{\zeta}_t^s$ based on Lemma \ref{lem:controls.zeta.posterior}.

\begin{lemma}
	\label{lem:zeta-fwl-asymp}
	Fix a source $s\in\mathcal S$ and an instance $t$. Define
	\begin{align*}
	V_t^s := \operatorname{Diag}[\nu_t^s/t],~\Pi_t := \operatorname{Diag}[N_t/t],~	M_t^s := \Pi_t - (B_t^s/t) (C_t^s/t)^{-1} (B_t^s/t)^\top.
	\end{align*}
	Then
	\begin{align}
		\boldsymbol{\zeta}_t^s
		&=
		(V_t^s+M_t^s)^{-1}
		\left(
		V_t^s\boldsymbol{\zeta}_0^s
		+
		M_t^s\boldsymbol{\theta}
		+
		r_t^s
		\right),
		\label{eq:zeta-fwl-asymp}
	\end{align}
where $r_t^s :=
\Pi_t \bar{\boldsymbol{\varepsilon}}_t^s
-
R_t^s (Q_t^s)^{-1}
\left(
\operatorname{Diag}[\lambda_t^s/t]\eta_0^s
+
t^{-1}\sum_{i=1}^t (W_i^s)^\top \varepsilon_i^s
\right)$ and \\ $\bar{\boldsymbol{\varepsilon}}_t^s : = ((N_{t}(0))^{-1} \sum_{i=1}^{t} 1\{D_{i} = 0\} \varepsilon_{i}^{s}(0), \ldots, (N_{t}(M))^{-1} \sum_{i=1}^{t} 1\{D_{i} = M \}  \varepsilon_{i}^{s}(M) )^{\top}$.
\end{lemma}

\begin{proof}
	See Appendix \ref{app:Gaussian.controls}.
\end{proof}

Suppose $\mathcal{U}$ is non-empty. Then, by proposition \ref{pro:control.weights.EV}, biased sources will be given a weight that converge to zero at least at rate $1/t$. Hence,
\begin{align*}
	\left \Vert   \hat{\boldsymbol{\theta}}_{t} - \boldsymbol{\theta} \right \Vert  \leq    \left \Vert   \sum_{s \in \mathcal{S}} \alpha^{s}_{t}  \{  \boldsymbol{\zeta}^{s}_{t}  -   \boldsymbol{\theta} \} \right \Vert  =  \sum_{s \in \mathcal{U}} \alpha^{s}_{t} \left \Vert  \boldsymbol{\zeta}^{s}_{t}  -   \boldsymbol{\theta} \right \Vert  + O_{as}(1/t),
\end{align*}

Any source in $\mathcal{U}$ satisfies, by definition, $ \boldsymbol{\zeta}^{s}_{0}  =   \boldsymbol{\theta} $. This and Lemma \ref{lem:zeta-fwl-asymp} imply that 
\begin{align*}
	\left \Vert   \hat{\boldsymbol{\theta}}_{t} - \boldsymbol{\theta} \right \Vert  = \left \Vert  \sum_{s \in \mathcal{U}} \alpha^{s}_{t} 	(V_t^s+M_t^s)^{-1} r_{t}^{s}   \right \Vert   + O_{as}(1/t).
\end{align*}

By definition of $(V_t^s+M_t^s)$ and strong LLN it is easy to see that 
\begin{align*}
V_t^s+M_t^s = &	(\operatorname{Diag}[c^{s}] + I)E[(Z)(Z)^{\top}]
-
E[Z \boldsymbol W^{s}]
\big(E[W^{s}(W^{s})^\top]\big)^{-1}
E[Z \boldsymbol W^{s} ]^{\top}+ o_{as}(1) \\
= & \operatorname{Diag}[c^{s}] E[(Z)(Z)^{\top}] + Var(Z - P_{\boldsymbol{W}^{s}} Z) + o_{as}(1)
\end{align*}
where $P_{\boldsymbol{W}^{s}}$ is the projection matrix onto $\boldsymbol{W}^{s}$. 

The fact that $\lambda^{s}_{t}/t = o(t^{-1/2})$ and the strong LLN imply that $r_{t} = o_{as}(\ell(t)/\sqrt{t})$ where $\ell$ is as in Theorem \ref{thm:learning.EV}. Moreover, $E[(Z)(Z)^{\top}]  = t \operatorname{Diag}[\delta]$. Putting all this results together, we obtain that
\begin{align}
	\left \Vert   \hat{\boldsymbol{\theta}}_{t} - \boldsymbol{\theta} \right \Vert  = o_{as} \left(   \frac{\ell(t)}{\sqrt{t} }   \sum_{s\in \mathcal{U}} \alpha^{s}_{t}   \left \Vert  (\operatorname{Diag}[c^{s} \cdot \delta] + Var(Z - P_{\boldsymbol{W}^{s}} Z)  )^{-1}  \right \Vert \right)  + O_{as}(1/t).
\end{align}

This expression is completely analogous to the conclusion obtained in Theorem \ref{thm:learning.EV} for the no controls case. The standard rate of convergence is given $ \frac{\ell(t)}{\sqrt{t} }    \left \Vert  (Var(Z - P_{\boldsymbol{W}^{s}} Z)  )^{-1}  \right \Vert$, and this is precisely the rate we obtain if $\mathcal{U}$ were empty but we have at least one diffuse source. However, if there are unbiased non-diffuse source, our rate is faster than the standard one due to the $\operatorname{Diag}[c^{s}] E[(Z)(Z)^{\top}] $ factor.

\section{Concluding Remarks}
\label{sec:conclusion}

This paper studies learning about treatment effects in experiments (both RCTs and sequential ones) when each  study starts multiple prior sources --- past pilots, related studies, or expert assessments --- whose external validity is uncertain. We formalize this environment by treating each source as a distinct model within Bayesian model averaging (BMA), and we introduce a nonstandard asymptotic framework in which the precision of each source can grow with the sample size of the ongoing experiment. This scaling captures empirically relevant settings where prior studies may be as informative as the current trial, and where standard asymptotics provide little guidance about how posterior weights should evolve.

Within this framework, we characterize posterior weights and the estimator's convergence rate through an external-validity index that depends jointly on bias and effective precision. The resulting rate exhibits an oracle property: as formalized in Theorem \ref{thm:learning.EV}, if at least one source is unbiased, posterior weight concentrates asymptotically on the unbiased set and the estimator converges strictly faster than the benchmark rate based only on the new experiment, with the speedup determined by the effective precisions of the valid sources. At the same time, the same theorem delivers robustness at the level of rates: when all informative sources are biased, the presence of a deliberately conservative (diffuse) prior forces posterior weight onto that source, and the estimator reverts to the standard convergence rate, avoiding contamination by precise but misspecified information. Simulations illustrate how these oracle and robustness properties manifest in finite samples and quantify the magnitude of the resulting efficiency gains and protections.

\bibliography{learning}

\newpage

\section*{Figures \& Tables}

\begin{table}[!h]
\centering
\begin{threeparttable}
\caption{Posterior Model Weights}
 \label{tab:alpha}
    \centering \begin{tabular}{lccccc}
\toprule
 & T=50 & T=100 & T=250 & T=500 & T=750 \\
\midrule
\multicolumn{6}{c}{Model 1: $\alpha$-weight on Unbiased Source} \\
$e=0.5$ & 0.714 & 0.777 & 0.843 & 0.884 & 0.903 \\
$e=1.0$ & 0.734 & 0.797 & 0.857 & 0.894 & 0.911 \\
$e=2.0$ & 0.746 & 0.801 & 0.861 & 0.897 & 0.911 \\
\midrule
\multicolumn{6}{c}{Model 2: $\alpha$-weight on \emph{Biased} Source} \\
$e=0.5$ & 0.090 & 0.006 & 0.000 & 0.000 & 0.000 \\
$e=1.0$ & 0.040 & 0.001 & 0.000 & 0.000 & 0.000 \\
$e=2.0$ & 0.018 & 0.000 & 0.000 & 0.000 & 0.000 \\
\midrule
\multicolumn{6}{c}{Model 3: $\alpha$-weight on Unbiased} \\
$e=0.5$ & 0.692 & 0.775 & 0.843 & 0.884 & 0.903 \\
$e=1.0$ & 0.725 & 0.797 & 0.857 & 0.894 & 0.911 \\
$e=2.0$ & 0.742 & 0.801 & 0.861 & 0.897 & 0.911 \\
\multicolumn{6}{c}{Model 3: $\alpha$-weight on Biased} \\
$e=0.5$ & 0.034 & 0.002 & 0.000 & 0.000 & 0.000 \\
$e=1.0$ & 0.017 & 0.001 & 0.000 & 0.000 & 0.000 \\
$e=2.0$ & 0.009 & 0.000 & 0.000 & 0.000 & 0.000 \\
\bottomrule
\end{tabular}
    \begin{tablenotes}[flushleft]
    \footnotesize
    \item \textit{Notes:} This table reports average posterior model weights ($\alpha$-weights) assigned to selected sources in the control arm ($d=0$) across 1{,}000 MC replications. Columns vary the experiment sample size $T\in\{50,100,250,500,750\}$ (so $N_{T}(0)=T/2$). Rows vary the external-evidence parameter $e\in\{0.5,1,2\}$, which scales the precision (effective sample size) of informative sources proportionally to $N_{T}(0)$. Model~1 reports the $\alpha$-weight on the informative unbiased source; Model~2 reports the $\alpha$-weight on the informative biased source (initial mean shifted by $+1$); and Model~3 reports $\alpha$-weights on the informative unbiased and biased sources. The diffuse baseline source (fixed precision $\nu=1$) receives the residual weight. The data-generating process is $Y(0)\sim \mathcal{N}(1,1)$ and $Y(1)\sim \mathcal{N}(1.3,1)$.
    \end{tablenotes}
\end{threeparttable}
\end{table}

\pagebreak
\newpage

\pagebreak
\clearpage

\pagebreak
\newpage
\clearpage

\renewcommand\thesection{A.\arabic{section}}
\renewcommand\thesubsection{\thesection.\arabic{subsection}} \setcounter{section}{0}

\appendix
\setcounter{page}{1}

\begin{center}
	\Huge{Supplemental Material}
\end{center}

\section{Almost Sure Concentration Rates}
\label{app:AlmostSure}

\begin{lemma}\label{lem:N.diverge}
	Suppose for each $(d,x)\in\mathbb{D}\times\mathbb{X}$,  $\sum_{i=1}^{t} \delta_{i}(d|x)$ diverges a.s. as $t \to \infty$. Then, 
	\[
	N_t(d,x) := \sum_{i=1}^t 1\{D_i(x)=d\} \to \infty \quad \text{a.s. as } t\to\infty.
	\]
\end{lemma}

\begin{proof}
	Fix $(d,x)\in\mathbb{D}\times\mathbb{X}$ and define the events
	\[
	A_i(d,x) := \{D_i(x)=d\}, \qquad i\ge 1.
	\]
	Let $\mathcal{F}_i := \sigma\big((D_s(x),Y_s(x))_{s\le i}\big)$ denote the history up to time $i$. By definition of the policy rule,
	\[
	\mathbb{P}\big(A_i(d,x)\mid \mathcal{F}_{i-1}\big) = \delta_i(d\mid x), \qquad i\ge 1.
	\]
	By assumption $\sum_{i=1}^\infty \delta_i(d\mid x) = \infty$ a.s.. 	Hence,
	\[
	\sum_{i=1}^\infty \mathbb{P}\big(A_i(d,x)\mid \mathcal{F}_{i-1}\big)
	= \sum_{i=1}^\infty \delta_i(d\mid x) = \infty \quad \text{a.s.}
	\]
	
	By the conditional (Lévy) Borel--Cantelli lemma (\cite{williams1991probability} Theorem 12.15), it follows that
	\[
	\mathbb{P}\big(A_i(d,x) \ \text{i.o.}\big) = 1,
	\]
	that is, with probability one the event $\{D_i(x)=d\}$ occurs infinitely often. Therefore,
	\[
	N_t(d,x) = \sum_{i=1}^t  1\{A_i(d,x)\} \to \infty \quad \text{a.s.},
	\]
	which proves the claim.
\end{proof}

Let  $\lambda  \colon [1,\infty) \rightarrow \mathbb{R}_{+}$ be any increasing function such that $\int_{1}^{\infty} \lambda(x)^{-2} dx < \infty  $. The function $x \mapsto \lambda(x)$ must diverge faster than $x \mapsto \sqrt{x} \sqrt{\log(x)}$ --- it can be chosen to be $x \mapsto \sqrt{x} \sqrt{\log(x) } (1 + \log (x)) $ or $x \mapsto \sqrt{x} \log(x) $. To capture this we set $\lambda(x) = \sqrt{x} \ell(x)$ where $\ell : [1,\infty) \to \mathbb{R}_{+}$ is increasing and such that $\lim_{x \to \infty} \ell(x)/\sqrt{\log x} = \infty$.

\begin{lemma}\label{lem:Y.ASrate}
	For any $(d,x) \in \mathbb{D} \times \mathbb{X}$, 
	\begin{align*}
		 \left | \frac{\sum_{i=1}^{t} 1\{ D_{i}(x) = d \} (Y_{i}(d,x)-\theta(d,x))}{\lambda(N_{t}(d,x) )} \right | = o_{as}(1).
	\end{align*}
\end{lemma}

\begin{proof}[Proof of Lemma \ref{lem:Y.ASrate}]
	We show that $S_{t} : = \sum_{i=1}^{t} \frac{ 1\{ D_{i}(x) = d \} (Y_{i}(d,x)-\theta(d,x))   }{\lambda(N_{i}(d,x))}$ converges a.s.. To do this, observe that $E_{i}[S_{i+1}] = S_{i}$ since $E_{i} \left[ 1\{ D_{i}(x) = d \}  \frac{  (Y_{i+1}(d,x)-\theta(d,x))  }{\lambda(N_{i+1}(d,x))} \right] = 0$ as $N_{i+1}(d,x)$ and $D_{i+1}(x)$ are measurable with respect to the $\sigma$-algebra generated by $(Y_{j}(d,x))_{j\leq i}$.  Therefore, $(S_{t})_{t}$ is a Martingale with respect to such $\sigma$-algebra. 
	
	Thus, by Doob's Martingale Convergence Theorem to prove the claim it suffices to show that $\sup_{t} E[|S_{t}|^{2}] < \infty$. To do this,
	\begin{align*}
		E\left[ |S_{t}|^{2} \right] = &	E\left[  \sum_{i=1}^{t} 1\{ D_{i}(x) = d \}  \left |  \frac{ (Y_{i}(d,x)-\theta(d,x)) }{\lambda(N_{i}(d,x))} \right |^{2} \right] \\
		& + E \left[   2 \sum_{i>j} 1\{ D_{i}(x) = d \}  1\{ D_{j}(x) = d \}  \left\langle  \frac{ (Y_{i}(d,x)-\theta(d,x)) }{\lambda(N_{i}(d,x))} ,    \frac{ (Y_{j}(d,x)-\theta(d,x)) }{\lambda(N_{j}(d,x))}  \right \rangle    \right] \\
		= & 	E\left[  \sum_{i=1}^{t} 1\{ D_{i}(x) = d \}  \frac{E_{i-1} [ \left | (Y_{i}(d,x)-\theta(d,x))   \right |^{2}  ]  }{|\lambda(N_{i}(d,x))|^{2}}  \right]\\
		\leq & \mathbb{L} E\left[ \sum_{i=1}^{t}  \frac{1\{ D_{i}(x) = d \} }{|\lambda(N_{i}(d,x))|^{2}}  \right]
	\end{align*}
	where the second line follows from the fact that $E\left[ 1\{ D_{i}(x) = d \}  1\{ D_{j}(x) = d \}  \left\langle  \frac{ (Y_{i}(d,x)-\theta(d,x)) }{\lambda(N_{i}(d,x))} ,    \frac{(Y_{j}(d,x)-\theta(d,x))  }{\lambda(N_{j}(d,x))}  \right \rangle   \right]  = E\left[    1\{ D_{j}(x) = d \}  \left\langle  E_{i-1} \left[ 1\{ D_{i}(x) = d \} \frac{ (Y_{i}(d,x)-\theta(d,x))  }{\lambda(N_{i}(d,x))}  \right]  ,    \frac{ (Y_{j}(d,x)-\theta(d,x))  }{\lambda(N_{j}(d,x))}  \right \rangle  \right] = 0$; the third line follows because $E_{i-1} [ \left | Y_{i}(d,x)-\theta(d,x)  \right |^{2}  ] \leq \mathbb{L} < \infty$ under Assumption \ref{ass:IID}.
	
	We now show that  $\lim_{t \rightarrow \infty} E\left[ \sum_{i=1}^{t}  \frac{1\{ D_{i}(x) = d \} }{|\lambda(N_{i}(d,x))|^{2}}  \right] < \infty$. To do this, note that $N_{i}(d,x)$ counts the number of times the even $D_{i}(x)=d$ occurred up to instance $i$. For any $k \{ 1, ..., N_{t}(d,x)   \}$, let $i(k) : = \min\{ i \leq t \colon  N_{i}(d,x) = k \} $ --- so $i(k)$ is the instance of the $k$-th case of treatment $d$. It follows that 
    \begin{align*}
    \sum_{i=1}^{t}  \frac{1\{ D_{i} = d \} }{|\lambda(N_{i}(d,x))|^{2}} & = \frac{1}{|\lambda(N_{i(1)}(d,x))|^{2}} + \frac{1}{|\lambda(N_{i(2)}(d,x))|^{2}} + ... = \sum_{k=1}^{N_{t}(d,x)} \frac{1}{\lambda(N_{i(k)}(d,x))^{2}} \\
    & = \sum_{k=1}^{N_{t}(d,x)} \frac{1}{\lambda(k)^{2}} \leq \sum_{k=1}^{\infty} \frac{1}{\lambda(k)^{2}} .
    \end{align*}
    Since $k \mapsto \lambda(k)$ is increasing in the relevant domain, the last expression implies that $ \sum_{i=1}^{t}  \frac{1\{ D_{i} = d \} }{|\lambda(N_{i}(d,x))|^{2}} \leq \int_{1}^{\infty} \lambda(x)^{-2} dx$ which is finite under our assumptions.
	
	Therefore, by Doob's Martingale Convergence Theorem, $S_{t}$ converges to $S_{\infty}$ almost surely (a.s.). Since $(N_{i}(d,x))_{i}$ is a divergent sequence a.s. (see Lemma \ref{lem:N.diverge}) and $\lambda$ is increasing, the statement of the lemma follows from Kronecker Lemma. 
\end{proof}

\begin{remark}\label{rem:concentration.Y}
Thus, this lemma implies that
 \begin{align*}
\frac{	\sum_{i=1}^{t} 1\{D_{i}(x) =d \} Y_{i}(d,x) }{N_{t}(d,x)} = \theta(d,x) + o_{as}\left( \frac{ \ell( N_{t}(d,x)) }{\sqrt{N_{t}(d,x)}} \right)
\end{align*}    
\end{remark}

	\section{Proofs of Lemmas}
    \label{app:supp.lemmas}

\begin{proof}[Proof of Lemma \ref{lem:alpha.properties}]
	
	Let $p_{\theta}$ denote a Gaussian PDF with mean $\theta$ and variance $1$. Note that
\begin{align*}
	&	\int \prod_{i=1}^{t} ( p_{\theta}(Y_{i}) )^{1\{ D_{i}(x) =d  \}} \phi(\theta; \zeta^{s}_{0}(d,x) , 1/\nu^{s}_{t}(d,x) ) d\theta \\
	= & \int (2\pi)^{-0.5 \sum_{i=1}^{t} 1\{ D_{i}(x) =d  \}   }\exp \left\{ - \frac{1}{2} \sum_{i=1}^{t} 1\{ D_{i}(x) =d  \} \left( Y_{i}(d,x) - m_{t}(d,x)    \right)^{2}   \right\}\\
	\times & \exp \left\{ - \frac{1}{2} \sum_{i=1}^{t} 1\{ D_{i}(x) =d  \} \left( m_{t}(d,x) - \theta    \right)^{2}   \right\} \\
	\times & \exp \left\{ - \sum_{i=1}^{t} 1\{ D_{i}(x) =d  \} \left( Y_{i}(d,x) - m_{t}(d,x)    \right) \left( m_{t}(d,x) - \theta    \right)   \right\}   \phi(\theta; \zeta^{s}_{0}(d,x) , 1/\nu_{t}^{s}(d,x) ) d\theta.
\end{align*}
Observe that $\sum_{i=1}^{t} 1\{ D_{i}(x) =d  \} \left( Y_{i}(d,x) - m_{t}(d,x)    \right) =  0$,  so, letting $N_{t}(d,x) : = \sum_{i=1}^{t} 1\{ D_{i}(x) =d  \}$ it follows that 
\begin{align*}
	\int 	\prod_{i=1}^{t} ( p_{\theta}(Y_{i}) )^{1\{ D_{i}(x) =d  \}}  \phi(\theta; \zeta^{s}_{0}(d,x) , 1/\nu^{s}_{t}(d,x) ) d\theta = & \frac{ \exp \left\{ - \frac{1}{2} \sum_{i=1}^{t} 1\{ D_{i}(x) =d  \} \left( Y_{i}(d,x) - m_{t}(d,x)    \right)^{2}   \right\}}{(2\pi)^{0.5 \sum_{i=1}^{t} 1\{ D_{i}(x) =d  \} + 0.5}  N_{t}(d,x)^{0.5} }  \\
	\times & \int (2\pi/N_{t}(d,x))^{-1/2} \exp \left\{ - \frac{1}{2}  \left( m_{t}(d,x) - \theta    \right)^{2} N_{t}(d,x)   \right\} \\	
	&  \phi(\theta; \zeta^{s}_{0}(d,x) , 1/\nu_{t}^{s}(d,x) ) d\theta.
\end{align*}

The expression of the integral can be viewed as a convolution between to Gaussian PDFs one indexed by $(  0,   1/N_{t}(d,x) )$ and $(  \zeta^{s}_{0}(d,x) , 1/\nu_{t}^{s}(d,x)  )$ resp, which in turn is equivalent to PDF of the sum of the corresponding random variables evaluated at $m_{t}(d,x)$. Therefore, 
\begin{align*}
	\int \prod_{i=1}^{t} ( p_{\theta}(Y_{i}) )^{1\{ D_{i}(x) =d  \}}  \phi(\theta; \zeta^{s}_{0}(d,x) , 1/\nu^{s}_{t}(d,x) ) d\theta  = C \phi( m_{t}(d,x) ; \zeta^{s}_{0}(d,x) , ( N_{t}(d,x)  + \nu^{s}_{t}(d,x)  )/( N_{t}(d,x)\nu^{s}_{t}(d,x) ) )
\end{align*}
where $C : = (2\pi)^{-0.5 \sum_{i=1}^{t} 1\{ D_{i}(x) =d  \} + 0.5}   N_{t}(d,x)^{-1/2} \exp \left\{ - \frac{1}{2} \sum_{i=1}^{t} 1\{ D_{i}(x) =d  \} \left( Y_{i}(d,x) - m_{t}(d,x)    \right)^{2}   \right\}$ which, importantly, doesn't depend on the source $s$. 

Hence
\begin{align*}
	\alpha^{s}_{t}(d,x) = \frac{       \phi( m_{t}(d,x) ; \zeta^{s}_{0}(d,x) , ( N_{t}(d,x)  + \nu^{s}_{t}(d,x)  )/( N_{t}(d,x)\nu^{s}_{t}(d,x) ) )       }  {  \sum_{s=0}^{L}   \phi( m_{t}(d,x) ; \zeta^{s}_{0}(d,x) , ( N_{t}(d,x)  + \nu^{s}_{t}(d,x)  )/( N_{t}(d,x)\nu^{s}_{t}(d,x) ) )    }  
\end{align*} 
and  the desired result follows.
\end{proof}

\begin{proof}[Proof of Lemma \ref{lem:log.approx.v}]
	Write $1+c+o(1) = 1+c+\delta_n$ with $\delta_n=o(1)$. Then
	\[
	\log\!\left(\frac{v}{1+c+\delta_n}\right)
	= \log v - \log(1+c+\delta_n).
	\]
	By the mean value theorem, there exists $\xi_n$ between $1+c$ and 
	$1+c+\delta_n$ such that
	\[
	\log(1+c+\delta_n)-\log(1+c)
	= \frac{\delta_n}{\xi_n}.
	\]
	For $n$ large enough, $|\delta_n|\le (1+c)/2$, implying
	\[
	\xi_n \in \left[\frac{1+c}{2},\,\frac{3(1+c)}{2}\right]
	\qquad\text{and}\qquad
	\left|\frac{1}{\xi_n}\right|\le \frac{2}{1+c}.
	\]
	Hence
	\[
	\big|\log(1+c+\delta_n)-\log(1+c)\big|
	\le \frac{2}{1+c}\,|\delta_n|
	= o(1).
	\]
	Substituting this into the earlier identity gives
	\[
	\log\!\left(\frac{v}{1+c+\delta_n}\right)
	=
	\log v - \log(1+c) + o(1)
	=
	\log\!\left(\frac{v}{1+c}\right)
	+ o(1).
	\]
	
	For the second statement, assume $v=v_n$ satisfies $v_n \to v_\ast>0$ 
	or $v_n\to\infty$. Then any $o(1)$ term is automatically $o(v_n)$, and thus
	\[
	\log\!\left(\frac{v}{1+c+o(1)}\right)
	= \log\!\left(\frac{v}{1+c}\right)
	+ o(v),
	\]
	as claimed.
\end{proof}

\section{Proofs}
\label{sec:proofs}

In this section we present the proofs of Propositions \ref{pro:Bweights.EV} and \ref{pro:alpha.asymp}, and Theorem \ref{thm:learning.EV}. Throughout this section, let $\mathbb{EV}^{s}_{t}(d,x) : = E(bias^{s}(d,x),\nu^{s}_{t}(d,x)/(1+c^{s}(d,x)))$ for any source $s$, any treatment-covariate pair $(d,x)$, and any instance $t$.

\subsection{Proof of Proposition \ref{pro:Bweights.EV}}

The following lemma provides a representation for the weights.
\begin{lemma}\label{lem:alpha.properties}
	For any $(d,x) \in \mathbb{D}\times \mathbb{X}$, any $t \geq 1$, and any $s \in \mathcal{S}$,
	\begin{align*}
		\alpha^{s}_{t}(d,x) = \frac{       \phi( m_{t}(d,x) ; \zeta^{s}_{0}(d,x) , ( N_{t}(d,x)  + \nu^{s}_{t}(d,x)  )/( N_{t}(d,x)\nu^{s}_{t}(d,x) ) )       }  {  \sum_{s=0}^{L}   \phi( m_{t}(d,x) ; \zeta^{s}_{0}(d,x) , ( N_{t}(d,x)  + \nu^{s}_{t}(d,x)  )/( N_{t}(d,x)\nu^{s}_{t}(d,x) ) )    } ,
	\end{align*} 
	where $m_{t}(d,x) : = \sum_{i=1}^{t} 1\{ D_{i}(x) =d  \}  Y_{i}(d,x) /N_{t}(d,x)$. 
\end{lemma}

\begin{proof}
    See Appendix \ref{app:supp.lemmas}.
\end{proof}

The following lemma is also used in the proof.
\begin{lemma}\label{lem:log.approx.v}
	For any $c>0$ and $v>0$,  $\log\!\left(\frac{v}{1+c+o(1)}\right)
	=
	\log\!\left(\frac{v}{1+c}\right) + o(1)$. 
	Moreover, if $v=v_n$ satisfies either $v_n \to v_\ast>0$ or $v_n \to \infty$, then	$\log\!\left(\frac{v}{1+c+o(1)}\right)
	=
	\log\!\left(\frac{v}{1+c}\right) + o(v)$. 
\end{lemma} 

\begin{proof}
    See Appendix \ref{app:supp.lemmas}.
\end{proof}

\begin{proof}[Proof of Proposition \ref{pro:Bweights.EV}]
	
	Given Lemma \ref{lem:alpha.properties}, $	\alpha^{s}_{t}(d,x) = \frac{   e^{ l^{s}_{t}(d,x) }  }  {  \sum_{s=0}^{L}  e ^{ l_{t}^{s}(d,x) }    } $ where 
	\begin{align*}
		l^{s}_{t}(d,x) = &   \log  \phi( m_{t}(d,x) ; \zeta^{s}_{0}(d,x) , ( N_{t}(d,x)  + \nu^{s}_{t}(d,x)  )/( N_{t}(d,x)\nu^{s}_{t}(d,x) ) )   \\
		= &  - \frac{1}{2} \frac{\left(  m_{t}(d,x) - \zeta^{s}_{0}(d,x)   \right)^{2}}{( 1/\nu^{s}_{t}(d,x) + 1/N_{t}(d,x))} - \frac{1}{2} \log  \frac{ N_{t}(d,x)+ \nu^{s}_{t}(d,x)   }{ N_{t}(d,x) \nu^{s}_{t}(d,x) }.
	\end{align*}
	
Let $  \frac{\nu^{s}_{t}(d,x)}{N_{t}(d,x)} = :c^{s}_{t}(d,x)$ a.s. by expression \ref{eqn:c.limit}, $1/N_{t}(d,x) + 1/\nu^{s}_{t}(d,x) = \frac{1}{\nu^{s}_{t}(d,x)} (1+c^{s}_{t}(d,x) )$. Thus,
\begin{align*}
	\log  \frac{ N_{t}(d,x)+ \nu^{s}_{t}(d,x)   }{ N_{t}(d,x) \nu^{s}_{t}(d,x) } = -  \log \left( \frac{\nu^{s}_{t}(d,x)} {1+c^{s}_{t}(d,x) }   \right) = -  \log \left( \frac{\nu^{s}_{t}(d,x)}{1+c^{s}_{t}(d,x)} \right).  
\end{align*}

Also, by Lemma \ref{lem:Y.ASrate} in Appendix \ref{app:AlmostSure}, $m_{t}(d,x) = \theta(d,x) + o_{as}(1)$. Thus, 
\begin{align*}
	- \frac{1}{2} \frac{\left(  m_{t}(d,x) - \zeta^{s}_{0}(d,x)   \right)^{2}}{( 1/\nu^{s}_{t}(d,x) + 1/N_{t}(d,x))} 	= &  - \frac{\nu^{s}_{t}(d,x)}{2} \frac{\left(  \theta(d,x) - \zeta^{s}_{0}(d,x)   \right)^{2} + o_{as}(1) }{  1+ c^{s}_{t}(d,x)  } \\
	= & - \frac{\nu^{s}_{t}(d,x)}{2} \frac{\left(  \theta(d,x) - \zeta^{s}_{0}(d,x)   \right)^{2} }{  1+ c^{s}_{t}(d,x)  } (1+o_{as}(1)).
\end{align*}
Therefore,
	\begin{align*}
		l^{s}_{t}(d,x) = & \frac{1}{2} \left(  - \frac{\nu^{s}_{t}(d,x)}{1+c^{s}_{t}(d,x)}  \left(  \theta(d,x) - \zeta^{s}_{0}(d,x)   \right)^{2}(1+o_{as}(1)) + \log \frac{\nu^{s}_{t}(d,x) }{1+c^{s}_{t}(d,x)}   \right) \\
		= & \frac{1}{2} E \left( bias^{s}(d,x),\frac{\nu^{s}_{t}(d,x)}{1+c^{s}_{t}(d,x)} \right)  (1+o_{as}(1)) \\
		= & \frac{1}{2} E \left( bias^{s}(d,x),\frac{N_{t}(d,x)c^{s}_{t}(d,x)}{1+c^{s}_{t}(d,x)} \right)  (1+o_{as}(1)) \\
	\end{align*}	
\end{proof}

\subsection{Proof of Proposition \ref{pro:alpha.asymp} }

\begin{proof}[Proof of Proposition \ref{pro:alpha.asymp}]
	(1) Take any source $b \notin \mathcal{U}(d,x)$. By Proposition \ref{pro:Bweights.EV},
	\begin{align*}
		\frac{	\alpha_{t}^{b}(d,x)}{ 	\alpha_{t}^{s}(d,x) } = \frac{e^{0.5 \mathbb{EV}^{b}_{t}(d,x)(1 + o_{as}(1))  + o_{as}(1) }   }{e^{0.5 \mathbb{EV}^{s}_{t}(d,x)(1 + o_{as}(1))  + o_{as}(1)   }} = e^{ - 0.5 (\mathbb{EV}^{s}_{t}(d,x) - \mathbb{EV}^{b}_{t}(d,x))(1 + o_{as}(1))  + o_{as}(1) }. 
	\end{align*}
	
	Since $s \in \mathcal{U}(d,x)$, $\mathbb{EV}^{s}_{t}(d,x) = \log \nu^{s}_{t}(d,x) $ and it follows that 
	\begin{align*}
		\frac{	\alpha_{t}^{b}(d,x)}{ 	\alpha_{t}^{s}(d,x) } =  e^{  \log \frac{1}{\nu^{s}_{t}(d,x)}  + 0.5  \mathbb{EV}^{b}_{t}(d,x)(1 + o_{as}(1))  + o_{as}(1) }= \frac{1}{\nu^{s}_{t}(d,x)} e^{  0.5  \mathbb{EV}^{b}_{t}(d,x)(1 + o_{as}(1))  + o_{as}(1) }.
	\end{align*}
	
	On the other hand, since $b \notin \mathcal{U}(d,x)$, $ \mathbb{EV}^{b}_{t}(d,x) = \nu^{b}_{t}(d,x) \left( - bias^{s}(d,x)^{2}  + o( 1)  \right)$. If $b$ is a diffuse source ($\sup_{t}\nu^{b}_{t}(d,x) < \infty$), then $ \sup_{t}  \mathbb{EV}^{b}_{t}(d,x) < \infty$ and the desired result follows since $\frac{	\alpha_{t}^{b}(d,x)}{ 	\alpha_{t}^{s}(d,x) } = O_{as} \left(  \frac{1} {\nu^{s}_{t}(d,x)} \right)$. If $b$ is not diffuse, then  $ \mathbb{EV}^{b}_{t}(d,x) = - \nu^{b}_{t}(d,x)  \left( (bias^{b}(d,x))^{2}  + o_{as}( 1)  \right)$. Therefore, $\frac{	\alpha_{t}^{b}(d,x)}{ 	\alpha_{t}^{0}(d,x) } =  \frac{1}{\nu^{s}_{t}(d,x)} e^{  - 0.5 \nu^{b}_{t}(d,x) (bias^{b}(d,x))^{2}   (1 + o_{as}(1))  + o_{as}(1) }$.  	Since $s$ is arbitrary within $\mathcal{U}(d,x)$, we can choose the maximal one. 
	
	\medskip
	
	(2) The expression $\frac{	\alpha_{t}^{s}(d,x)}{ 	\alpha_{t}^{diff}(d,x) } = \frac{e^{  0.5  \mathbb{EV}^{s}_{t}(d,x)(1 + o_{as}(1))  + o_{as}(1) }}{\nu^{diff}_{t}(d,x)} $	is still valid. In this case, the LHS decays exponentially fast with $\nu^{s}_{t}(d,x)$ because $s$ is non-diffuse and biased.
	\end{proof}

\subsection{Proof of Theorem \ref{thm:learning.EV}}

The proof of Theorem \ref{thm:learning.EV} relies on two insights. First, observe that 
\begin{align*}
	| \widehat{\theta}_{t}(d,x) - \theta(d,x) | \leq & \sum_{s=0}^{L}  \alpha^{s}_{t}(d,x) \frac{ \nu^{s}_{t}(d,x)}{\nu^{s}_{t}(d,x) + N_{t}(d,x)}   |bias^{s}(d,x)|  \\
	& + \sum_{s=0}^{L}  \alpha^{s}_{t}(d,x) \frac{N_{t}(d,x) }{\nu^{s}_{t}(d,x) + N_{t}(d,x)} \left| \frac{\sum_{i=1}^{t} 1\{D_{i}(x) =d \} (Y_{i}(d,x) - \theta(d,x))  }{N_{t}(d,x)} \right|.
\end{align*}
And, by Lemma \ref{lem:Y.ASrate} and remark \ref{rem:concentration.Y} in Appendix \ref{app:AlmostSure}, 
\begin{align}\label{eqn:proof.main.1}
	| \widehat{\theta}_{t}(d,x) - \theta(d,x) | \leq & \sum_{s=0}^{L}  \alpha^{s}_{t}(d,x) \frac{ \nu^{s}_{t}(d,x)}{\nu^{s}_{t}(d,x) + N_{t}(d,x)}   |bias^{s}(d,x)|  \\
	& + \sum_{s=0}^{L}  \alpha^{s}_{t}(d,x) \frac{N_{t}(d,x) }{\nu^{s}_{t}(d,x) + N_{t}(d,x)}  o_{as} \left( \frac{ \ell(N_{t}(d,x)) }{\sqrt{N_{t}(d,x)}}  \right).
\end{align}

The second insight, relies on Proposition \ref{pro:alpha.asymp}, which established an (asymptotic) representation for the weights in terms of the concept of external validity. Part (1) of this proposition implies that for any real-valued sequence $(a^{s}_{t})_{s =0}^{L}$,\footnote{In order to ease the notational burden, in this section we omit the dependence on $(d,x)$ of $\mathcal{U}$.} 
\begin{align*}
	\sum_{s=0}^{L} \alpha^{s}_{t}(d,x) a^{s}_{t} =	& \sum_{s \in \mathcal{U} } \alpha^{s}_{t}(d,x) a^{s}_{t}  + \sum_{s \notin \mathcal{U} } \alpha^{s}_{t}(d,x) a^{s}_{t} = \sum_{s \in \mathcal{U} } \alpha^{s}_{t}(d,x) a^{s}_{t}  + O_{as} \left( \frac{\sum_{s \notin \mathcal{U} } e^{  - 0.25 \nu^{s}_{t}(d,x) (bias^{s}(d,x))^{2}  } a^{s}_{t} }{\max_{u \in \mathcal{U}} \nu^{u}_{t}(d,x)} \right).
\end{align*}

By taking $a^{s}_{t} = \frac{ \nu^{s}_{t}(d,x)}{\nu^{s}_{t}(d,x) + N_{t}(d,x)}   |bias^{s}(d,x)|  $ it follows that 
\begin{align*}
	\sum_{s=0}^{L} \alpha^{s}_{t}(d,x) \frac{ \nu^{s}_{t}(d,x)}{\nu^{s}_{t}(d,x) + N_{t}(d,x)}   |bias^{s}(d,x)|  = & O_{as} \left( \frac{\sum_{s \notin \mathcal{U} }  \frac{ \nu^{s}_{t}(d,x)}{\nu^{s}_{t}(d,x) + N_{t}(d,x)}   |bias^{s}(d,x)|  e^{  - 0.25 \nu^{s}_{t}(d,x) (bias^{s}(d,x))^{2}  }  }{\max_{u \in \mathcal{U}} \nu^{u}_{t}(d,x)} \right).
\end{align*}

For each biased source, if $(\nu^{s}_{t(k)}(d,x))_{k}$ diverges along some subsequence $(t(k))_{k}$, then
\begin{align*}
	\frac{ \nu^{s}_{t(k)}(d,x)}{\nu^{s}_{t(k)}(d,x)/ N_{t(k)}(d,x) + 1}   |bias^{s}(d,x)|  e^{  - 0.25 \nu^{s}_{t(k)}(d,x) (bias^{s}(d,x))^{2}  }=o_{as}(1).
\end{align*}
If $(\nu^{s}_{t(k)}(d,x))_{k}$ remains uniformly bounded along some subsequence $(t(k))_{k}$, then  the same quanitity is of order $O_{as}(1)$, and it is also uniformly bounded as a function of the bias. Thus,
\begin{align*}
	\frac{ \nu^{s}_{t}(d,x)}{\nu^{s}_{t}(d,x) + N_{t}(d,x)}   |bias^{s}(d,x)|  e^{  - 0.25 \nu^{s}_{t}(d,x) (bias^{s}(d,x))^{2}  } = &  N_{t}(d,x)^{-1} \frac{ \nu^{s}_{t}(d,x)}{\frac{\nu^{s}_{t}(d,x)}{ N_{t}(d,x) } + 1}   |bias^{s}(d,x)|  e^{  - 0.25 \nu^{s}_{t}(d,x) (bias^{s}(d,x))^{2}  } \\
	= & O_{as} \left(  N_{t}(d,x)^{-1}    \right). 
\end{align*}
Since the number of sources is finite, $\sum_{s=0}^{L} \alpha^{s}_{t}(d,x) \frac{ \nu^{s}_{t}(d,x)}{\nu^{s}_{t}(d,x) + N_{t}(d,x)}   |bias^{s}(d,x)|  = O_{as} \left(  N_{t}(d,x)^{-1}  \right)$. 

Therefore, expression \ref{eqn:proof.main.1} implies,
\begin{align*}
	| \widehat{\theta}_{t}(d,x) - \theta(d,x) |= &  \sum_{s=0}^{L}  \alpha^{s}_{t}(d,x) \frac{N_{t}(d,x) }{\nu^{s}_{t}(d,x) + N_{t}(d,x)}  o_{as} \left( \frac{ \ell(N_{t}(d,x)) }{\sqrt{N_{t}(d,x)}}  \right) + O_{as} \left(  N_{t}(d,x)^{-1}  \right)\\
	= &   o_{as} \left( \frac{ \ell(N_{t}(d,x)) }{\sqrt{N_{t}(d,x)}}  \left(  \sum_{s \in \mathcal{U} }  \alpha^{s}_{t}(d,x) \frac{N_{t}(d,x) }{\nu^{s}_{t}(d,x) + N_{t}(d,x)} + \sum_{s \notin \mathcal{U} }  \alpha^{s}_{t}(d,x) \frac{N_{t}(d,x) }{\nu^{s}_{t}(d,x) + N_{t}(d,x)}    \right) \right) \\
	&  + O_{as} \left(  N_{t}(d,x)^{-1}  \right) \\
	= &   o_{as} \left( \frac{ \ell(N_{t}(d,x)) }{\sqrt{N_{t}(d,x)}}  \left(  \sum_{s \in \mathcal{U} } \frac{\alpha^{s}_{t}(d,x)}{ \sum_{s' \in \mathcal{U}}  \alpha^{s'}_{t}(d,x) } \left(  1+c^{s}(d,x)  \right)^{-1}   \right) \right)  + O_{as} \left(  N_{t}(d,x)^{-1}  \right)
\end{align*}
where the third equality follows because $\sum_{s \notin \mathcal{U} } \alpha^{s}_{t}(d,x) = o_{as}(1)$ by Proposition \ref{pro:alpha.asymp} and because $\frac{N_{t}(d,x) }{\nu^{s}_{t}(d,x) + N_{t}(d,x)} = \frac{1}{1+c^{s}(d,x)}(1+o_{as}(1))$ by our asymptotic framework.

If $\mathcal{U}$ is empty, but there exists a diffuse source, we employ Proposition \ref{pro:alpha.asymp}(2) and obtain:
\begin{align*}
	& \sum_{s=0}^{L} \alpha^{s}_{t}(d,x) \frac{ \nu^{s}_{t}(d,x)}{\nu^{s}_{t}(d,x) + N_{t}(d,x)}   |bias^{s}(d,x)|  \\
	= & O_{as} \left(\sum_{s \notin Diff }  \frac{ \nu^{s}_{t}(d,x)    |bias^{s}(d,x)|  }{\nu^{s}_{t}(d,x) + N_{t}(d,x)}  e^{  - 0.25 \nu^{s}_{t}(d,x) (bias^{s}(d,x))^{2}  } + \sum_{s \in Diff } \alpha^{s}_{t}(d,x)  \frac{ \nu^{s}_{t}(d,x)   |bias^{s}(d,x)|  }{\nu^{s}_{t}(d,x) + N_{t}(d,x)}  \right)\\
	= & O_{as} \left(\max_{s \in Diff} |bias^{s}(d,x)|/N_{t}(d,x) \right),
\end{align*}
where the second equality follows from analogous arguments to those above and the fact that $\sup_{t} \nu^{s}_{t}(d,x) < \infty$ for any $s \in Diff$.

Therefore, putting these results together it follows that if $\mathcal{U}$ is non empty,
\begin{align*}
	| \widehat{\theta}_{t}(d,x) - \theta(d,x) | =  o_{as}\left( \frac{ \ell(N_{t}(d,x)) }{\sqrt{N_{t}(d,x)}}  \mathcal{A}_{t}(\mathcal{U}) \right)  + O_{as} \left(  N_{t}(d,x)^{-1}    \right)
\end{align*}
where $ \mathcal{A}_{t}(\mathcal{U})   : =  \sum_{s \in \mathcal{U}}  \frac{1}{(1+c^{s}(d,x)} \frac{\alpha^{s}_{t}(d,x)}{\sum_{s \in \mathcal{U}} \alpha^{s}_{t}(d,x)} $. And if $\mathcal{U} = \emptyset$, but there exists a diffuse source,
\begin{align*}
	| \widehat{\theta}_{t}(d,x) - \theta(d,x) | =  o_{as}\left( \frac{ \ell(N_{t}(d,x)) }{\sqrt{N_{t}(d,x)}}  \right)  + O_{as} \left( \frac{\max_{s \in Diff} |bias^{s}(d,x)|}{ N_{t}(d,x)}   \right).
\end{align*}

	\subsection{Proofs for Section \ref{subsec:bernoulli}} 
	\label{app:Bernoulli}

	\paragraph{Laplace Principle for the KL Integral.}	Let
	\begin{align}
		I_t
		:=
		\int_{0}^{1}
		\exp\Big\{
		- N_t\, \Phi_t(u)
		\Big\}\,du,
		\qquad
		\Phi_t(u)
		:=
		\mathrm{KL}(m_t\|u)
		+
		c_t^s\,\mathrm{KL}(\zeta_0^s\|u),
	\end{align}
	where
	\[
	c_t^s := \frac{\nu_t^s}{N_t},
	\qquad
	m_t = \theta + o_{as}(1),
	\qquad
	c_t^s = c^s + o_{as}(1),
	\]
	and $\mathrm{KL}(p\|u)$ denotes the Bernoulli Kullback--Leibler divergence
	\begin{align}
		\mathrm{KL}(p\|u)
		=
		p \log\frac{p}{u}
		+
		(1-p)\log\frac{1-p}{1-u},
		\qquad p,u\in(0,1).
	\end{align}
	
	Define the limiting objective
	\begin{align}
		\Phi(u)
		:=
		\mathrm{KL}(\theta\|u)
		+
		c^s\,\mathrm{KL}(\zeta_0^s\|u),
		\qquad u\in(0,1).
	\end{align}
	
	\begin{lemma}[Laplace principle] \label{lem:Laplace}
		Assume $\theta\in(0,1)$, $\zeta_0^s\in(0,1)$, $c^s\ge0$, and $N_t\to\infty$. Then, almost surely,
		\begin{align}
			\lim_{t\to\infty}
			\frac{1}{N_t}
			\log I_t
			=
			-
			\inf_{u\in(0,1)}
			\Phi(u).
		\end{align}
		Moreover, $\Phi$ is strictly convex on $(0,1)$ and admits a unique minimizer
		\begin{align}
			u^\star
			=
			\frac{\theta + c^s \zeta_0^s}{1+c^s},
		\end{align}
		so that
		\begin{align}
			\inf_{u\in(0,1)} \Phi(u)
			=
			\Phi(u^\star).
		\end{align}
	\end{lemma}
	
	\begin{proof}
		Work on the almost sure event on which $m_t\to\theta$ and $c_t^s\to c^s$.
		
		\medskip
		\noindent
		\textbf{Step 1: Uniform convergence on interior compacts.} Fix $\varepsilon\in(0,1/2)$ and restrict attention to $u\in[\varepsilon,1-\varepsilon]$.
		The map $(p,u)\mapsto \mathrm{KL}(p\|u)$ is continuous on
		$[0,1]\times[\varepsilon,1-\varepsilon]$ and Lipschitz in $p$ on that set.
		Since $m_t\to\theta\in(0,1)$, eventually $m_t$ stays in a compact subset of $(0,1)$.
		Hence
		\begin{align}
			\sup_{u\in[\varepsilon,1-\varepsilon]}
			|\Phi_t(u)-\Phi(u)|
			\longrightarrow 0
			\qquad \text{a.s.}
		\end{align}
		
		\medskip
		\noindent
		\textbf{Step 2: Upper bound for the limit superior.}	Observe that 
		\begin{align}
			I_t
			\le
			\int_0^1
			e^{-N_t\Phi_t(u)}\,du
			\le
			e^{-N_t \inf_{u\in [0,1]} \Phi_t(u)}.
		\end{align}
		Therefore $\frac{1}{N_t}\log I_t
			\le
			-\inf_{u\in [0,1]} \Phi_t(u)$. 
		
		We now show that the infimum is achieved at $u^{\ast}_{t}$ and that $(u^{\ast}_{t})_{t}$ is uniformly bounded away from 0 and 1, at least for sufficiently large $t$. To do this, by the arguments in Step 4, $u^{\ast}_{t} = \frac{m_{t} + c^{s}_{t} \zeta_{0}^{s}}{1+c^{s}_{t}}$. Since $m_{t} \to \theta \in (0,1)$ and $\zeta_{0}^{s} \in (0,1)$, it follows that that  $(u^{\ast}_{t})_{t}$ is uniformly bounded away from 0 and 1, at least for sufficiently large $t$.
		
		Thus, there exists an $\varepsilon>0$ under which we can apply step 1 and obtain that $\Phi_t(u^{\ast}_{t})\to\Phi(u^{\ast}_{t})$ and
		\begin{align}
			\limsup_{t\to\infty}
			\frac{1}{N_t}\log I_t
			\le -\inf_{u\in [0,1]} 
			\Phi(u).
		\end{align}

		\medskip
		\noindent
		\textbf{Step 3: Lower bound for the limit inferior.} Fix $\delta>0$ and choose $u_\delta\in(0,1)$ such that
		\[
		\Phi(u_\delta)
		\le
		\inf_{u\in (0,1)}\Phi(u)
		+
		\delta.
		\]
		By continuity of $\Phi$, there exists a neighborhood $B_\delta\subset(0,1)$ of
		$u_\delta$ such that
		\[
		\Phi(u)
		\le
		\inf_{u\in (0,1)}\Phi(u)
		+
		2\delta
		\quad
		\text{for all } u\in B_\delta.
		\]
		By uniform convergence on compact sets,
		for $t$ large enough (a.s.),
		\[
		\sup_{u\in B_\delta}
		|\Phi_t(u)-\Phi(u)|
		\le
		\delta,
		\]
		so that
		\[
		\Phi_t(u)
		\le
		\inf_{u\in (0,1)}\Phi(u)
		+
		3\delta
		\quad
		\text{for all } u\in B_\delta.
		\]
		Hence
		\begin{align}
			I_t
			\ge
			\int_{B_\delta}
			e^{-N_t\Phi_t(u)}\,du
			\ge
			|B_\delta|\,
			e^{-N_t(\inf\Phi+3\delta)}.
		\end{align}
		It follows that
		\begin{align}
			\liminf_{t\to\infty}
			\frac{1}{N_t}\log I_t
			\ge
			-
			\inf_{u\in (0,1)}\Phi(u)
			-
			3\delta.
		\end{align}
		Letting $\delta\downarrow 0$ gives
		\begin{align}
			\liminf_{t\to\infty}
			\frac{1}{N_t}\log I_t
			\ge
			-
			\inf_{u\in(0,1)}\Phi(u).
		\end{align}
		
		\medskip
		Combining the upper and lower bounds and noting that $\lim_{u \rightarrow \{0,1\}}\Phi(u) = \infty$ establishes the desired limit.
		
		\medskip
		\noindent
		\textbf{Step 4: Identification of the minimizer.}	For Bernoulli KL,
		\begin{align}
			\frac{d}{du}\mathrm{KL}(p\|u)
			=
			-
			\frac{p}{u}
			+
			\frac{1-p}{1-u}.
		\end{align}
		Therefore
		\begin{align}
			\Phi'(u)
			=
			-
			\frac{\theta + c^s\zeta_0^s}{u}
			+
			\frac{(1-\theta)+c^s(1-\zeta_0^s)}{1-u}.
		\end{align}
		Solving $\Phi'(u)=0$ yields
		\begin{align}
			u^\star
			=
			\frac{\theta + c^s\zeta_0^s}{1+c^s}.
		\end{align}
		Moreover,
		\begin{align}
			\Phi''(u)
			=
			\frac{\theta + c^s\zeta_0^s}{u^2}
			+
			\frac{(1-\theta)+c^s(1-\zeta_0^s)}{(1-u)^2}
			>
			0
			\quad \text{for } u\in(0,1),
		\end{align}
		so $\Phi$ is strictly convex and the minimizer is unique.
	\end{proof}

	\begin{proof}[Proof of Proposition \ref{prop:EV.Bern}]
		Fix $(d,x)$ and suppress $(d,x)$ to lighten notation.
		Let $N:=N_t$, $K:=K_t$, and write the empirical mean $m_t:=K/N$.
		By the strong law (conditional on the treatment--covariate cell), $m_t\to \theta$ a.s.
		
		For source $s$, the (integrated) marginal likelihood is Beta--binomial:
		\[
		\mathcal M_t^s
		=
		\frac{B(a_s+K,b_s+N-K)}{B(a_s,b_s)},
		\qquad
		a_s:=\nu_s\zeta_s,\quad b_s:=\nu_s(1-\zeta_s),
		\]
		where $\nu_s:=\nu_t^s$ and $\zeta_s:=\zeta_0^s$.
		Since $\alpha_t^s=\mathcal M_t^s/\sum_{r}\mathcal M_t^r$, it suffices to obtain a sharp
		(asymptotically almost sure) expansion for $\log \mathcal M_t^s$ up to terms that do not depend on $s$.

		So \begin{align*}
			\mathcal M^{s}_{t} = & \frac{\int_{0}^{1} \exp \left\{ (a_{s} + K) \log u  + (b_s + N - K) \log (1-u)    \right\} du }{\int_{0}^{1} \exp \left\{ a_{s}  \log u  + b_s \log (1-u)     \right\} du} \\
			= & \frac{\int_{0}^{1} \exp \left\{ K \log u  + (N - K) \log (1-u)    \right\} \exp \left\{ a_{s} \log u  + b_s \log (1-u)    \right\}  du }{\int_{0}^{1} \exp \left\{ a_{s}  \log u  + b_s \log (1-u)     \right\} du} 
		\end{align*}
		
		Let \(\mathrm{KL}(p\|u)\) be the Bernoulli KL:
		\[
		\mathrm{KL}(p\|u)
		=
		p\log\frac{p}{u}
		+
		(1-p)\log\frac{1-p}{1-u}.
		\]
		Then for any \(u\in(0,1)\),
		\[
		K\log u+(N-K)\log(1-u)
		=
		N_t \Big( m_t \log m_t+(1-m_t)\log(1-m_t)\Big) 
		-
		N_t \mathrm{KL}(m_t \|u).
		\]
		Therefore, 
		\begin{align}
				\mathcal M^{s}_{t} 
			=
			\exp\!\Big\{
			N_t \big[m_t\log m_{t} +(1-m_t)\log(1-m_t)\big]
			\Big\}
			\cdot
			\mathbb E^{s} \left[
			\exp\!\big\{-N_t \mathrm{KL}(m_t\|U)\big\}
			\right], \label{eqn:M.charac.1}
		\end{align}
		where
		\[
		U \sim 	\mathrm{Beta}\big(\nu_t^s\zeta_0^s+1,\;
		\nu_t^s(1-\zeta_0^s)+1\big).
		\]

		Using the Beta density
		\begin{align}
			\mathbb E^{s}\!\left[e^{-N_t\mathrm{KL}(m_{t} \|U)}\right]
			&=
			\frac{1}{B(\nu^{s}_{t}\zeta^{s}_0 +1,\nu^{s}_{t} (1-\zeta^{s}_0)+1)}
			\int_0^1
			e^{-N_{t} \mathrm{KL}(m_{t} \|u)}\,
			u^{\nu^{s}_{t} \zeta^{s}_0 }(1-u)^{\nu^{s}_{t} (1-\zeta^{s}_0)}\,du.
			\label{eq:Etheta-int}
		\end{align}

		Define \(h(q):=q\log q+(1-q)\log(1-q)\). Then for all \(q,u\in(0,1)\), $q\log u+(1-q)\log(1-u)=h(q)-\mathrm{KL}(q\|u)$. 	Hence
		\[
		u^{\nu^{s}_{t}  \zeta^{s}_0 }(1-u)^{\nu^{s}_{t} (1- \zeta^{s}_0)}
		=
		\exp\Big\{\nu^{s}_{t} \big[ \zeta^{s}_0 \log u+(1- \zeta^{s}_0)\log(1-u)\big]\Big\}
		=
		\exp\Big\{\nu^{s}_{t}  h( \zeta^{s}_0 )- \nu^{s}_{t} \mathrm{KL}( \zeta^{s}_0 \|u)\Big\}.
		\]
		Plugging into \eqref{eq:Etheta-int} gives the exact identity
		\begin{align}
			\mathbb E^{s}\!\left[e^{-N_{t} \mathrm{KL}(m_{t} \|U)}\right]
			&=
			\frac{e^{\nu^{s}_{t} h( \zeta^{s}_0 )}}{B( \nu^{s}_{t}   \zeta^{s}_0 +1,  \nu^{s}_{t} (1- \zeta^{s}_0)+1)}
			\int_0^1
			\exp\Big\{- N_t \left(    \mathrm{KL}(m_{t} \|u) + \frac{\nu^{s}_{t}}{N_{t}}  \mathrm{KL}( \zeta^{s}_0 \|u) \right)  \Big\}\,du
			\label{eq:Etheta-singleexp}
		\end{align}
		
		By Lemma \ref{lem:Laplace}, the fact that $N_{t}$ diverges a.s. (Lemma \ref{lem:N.diverge}), $\nu^{s}_{t}/N_{t} = c^{s} + o_{as}(1)$, and $m_{t} = \theta + o_{as}(1)$ (Lemma \ref{lem:Y.ASrate} and its remark), and $\theta, \zeta^{s}_{0} \in (0,1)$, 
	\begin{align*}
		N_{t}^{-1} \log 	\int_0^1
		\exp\Big\{- N_t \left(    \mathrm{KL}(m_{t} \|u) + \frac{\nu^{s}_{t}}{N_{t}}  \mathrm{KL}( \zeta^{s}_0 \|u) \right)  \Big\}\,du = - \inf_{u\in(0,1)}  \left(    \mathrm{KL}(\theta \|u) + c^{s} \mathrm{KL}( \zeta^{s}_0 \|u) \right) + o_{as}(1). 
	\end{align*}
	
	Thus,
			\begin{align}
		\mathcal M^{s}_{t} 
		=&
		\exp\!\Big\{
		N_t \big[m_t\log m_{t} +(1-m_t)\log(1-m_t)\big]
		\Big\}\\
		& \times
	\frac{e^{\nu^{s}_{t} h( \zeta^{s}_0 )}}{B( \nu^{s}_{t}   \zeta^{s}_0 +1,  \nu^{s}_{t} (1- \zeta^{s}_0)+1)}
	e^{- N_{t} (\inf_{u\in(0,1)}  \left(    \mathrm{KL}(\theta \|u) + c^{s} \mathrm{KL}( \zeta^{s}_0 \|u) \right) + o_{as}(1)) }
	\end{align}
	
	Observe that
	\begin{align}
		B\!\Big(\nu_t^s \zeta_0^s+1,\;\nu_t^s(1-\zeta_0^s)+1\Big)
		=
		\int_0^1
		u^{\nu_t^s \zeta_0^s}(1-u)^{\nu_t^s(1-\zeta_0^s)}\,du  =
		\int_0^1
		\exp\Big\{
		\nu_t^s\, f^s(u)
		\Big\}\,du,
		\label{eqn:beta.laplace.def}
	\end{align}
	
 So, by Lemma \ref{lem:Laplace}, but applied to $I_{t} = \int_0^1
 \exp\Big\{
 \nu_t^s\, f^s(u)
 \Big\}\,du$
	\begin{align}\label{eqn:B.approx.final}
		\lim_{\nu^{s}_{t} \to \infty} 	\frac{1}{\nu^{s}_{t}} \log 	B\!\Big(\nu_t^s \zeta_0^s+1,\;\nu_t^s(1-\zeta_0^s)+1\Big) = \max_{u \in (0,1)} f^{s}(u) = h(\zeta^{s}_{0}).
	\end{align}

		Therefore, by plugging in this result in expression \ref{eqn:M.charac.1} it follows that
	\begin{align}
		\mathcal{M}^{s}_{t} = \exp\{ N_{t} h(m_{t})  \} e^{-N_{t} (- \inf_{u\in(0,1)}  \left(    \mathrm{KL}(\theta \|u) + c^{s} \mathrm{KL}( \zeta^{s}_0 \|u) \right) + o_{as}(1)  )  },
	\end{align}
		which implies 
		\begin{align*}
			\alpha^{s}_{t}(d,x) = \frac{  e^{-N_{t} (- \inf_{u\in(0,1)}  \left(    \mathrm{KL}(\theta \|u) + c^{s} \mathrm{KL}( \zeta^{s}_0 \|u) \right) + o_{as}(1)  )  }   }{\sum_{s' \in \mathcal M}  e^{-N_{t} (- \inf_{u\in(0,1)}  \left(    \mathrm{KL}(\theta \|u) + c^{s'} \mathrm{KL}( \zeta^{s'}_0 \|u) \right) + o_{as}(1)  )  } }   
		\end{align*}
		as desired.

	\end{proof}
	
\subsection{Proofs of Section \ref{subsec:regression}}  
\label{app:Gaussian.controls}

Throughout this section, for any vector $X$ of dimension $p$, let 	 $\operatorname{Diag}[X] : = \operatorname{Diag}\{  X_{1} , \ldots , X_{d}  \}$ be a diagonal $p \times p$ matrix with $X$ as elements of the diagonal.

\begin{proof}[Posterior block system for the Gaussian regression model]
	Let $Y_{1:t}:=(Y_1,\ldots,Y_t)^\top$ and $X_{1:t}^s$ be the $t\times(M+1+p_s)$ matrix with $i$th row
	$(X_i^s)$. Under the Gaussian working model (with unit noise variance), the likelihood is
	\[
	p(Y_{1:t}\mid \beta^s , s )  \propto \exp\!\left(-\frac12\|Y_{1:t}-X_{1:t}^s\beta^s\|_2^2\right).
	\]
	The prior density is
	\[
	\pi(\beta^s)  \propto  \exp\!\left(-\frac12(\beta^s-\beta_0^s)^\top(\Sigma_{0,t}^s)^{-1}(\beta^s-\beta_0^s)\right).
	\]
	Hence the posterior density is proportional to the product:
	\[
	\pi(\beta^s\mid Y_{1:t} , s)  \propto 
	\exp\!\left(
	-\frac12\|Y_{1:t}-X_{1:t}^s\beta^s\|_2^2
	-\frac12(\beta^s-\beta_0^s)^\top(\Sigma_{0,t}^s)^{-1}(\beta^s-\beta_0^s)
	\right).
	\]
	Expand the exponent and collect the terms that depend on $\beta^s$:
	\begin{align*}
		&\|Y_{1:t}-X_{1:t}^s\beta^s\|_2^2 + (\beta^s-\beta_0^s)^\top(\Sigma_{0,t}^s)^{-1}(\beta^s-\beta_0^s)\\
		&=
		(Y_{1:t}^\top Y_{1:t}) -2 (X_{1:t}^s)^\top Y_{1:t}\cdot \beta^s + (\beta^s)^\top (X_{1:t}^s)^\top X_{1:t}^s\,\beta^s\\
		&\quad + (\beta^s)^\top(\Sigma_{0,t}^s)^{-1}\beta^s -2(\Sigma_{0,t}^s)^{-1}\beta_0^s\cdot \beta^s + (\beta_0^s)^\top(\Sigma_{0,t}^s)^{-1}\beta_0^s.
	\end{align*}
	Dropping the constants that do not depend on $\beta^s$, the posterior kernel is
	\[
	\exp\!\left(
	-\frac12 (\beta^s)^\top\Big((\Sigma_{0,t}^s)^{-1}+(X_{1:t}^s)^\top X_{1:t}^s\Big)\beta^s
	+ \Big((\Sigma_{0,t}^s)^{-1}\beta_0^s+(X_{1:t}^s)^\top Y_{1:t}\Big)^\top \beta^s
	\right),
	\]
	which is the kernel of a multivariate normal distribution with precision
	\[
	(\Sigma_t^s)^{-1}=(\Sigma_{0,t}^s)^{-1}+(X_{1:t}^s)^\top X_{1:t}^s,
	\]
	and mean satisfying
	\[
	\beta_t^s=\Sigma_t^s (\Sigma_{0,t}^s)^{-1}\beta_0^s+\Sigma_t^s (X_{1:t}^s)^\top Y_{1:t}.
	\]
	Finally, observe that
	\[
	(X_{1:t}^s)^\top X_{1:t}^s = \sum_{i=1}^t (X_i^s)^\top X_i^s = t\,\widehat Q_t^s,
	\qquad
	(X_{1:t}^s)^\top Y_{1:t} = \sum_{i=1}^t (X_i^s)^\top Y_i = t\,\widehat r_t^s.
	\]
	
	So, 
	\begin{align*}
		\beta_t^s=\Sigma_t^s (\Sigma_{0,t}^s)^{-1}\beta_0^s+ ( \Sigma_t^s) t \widehat{r}^{s}_{t}
	\end{align*}
	and $(\Sigma^{s}_{t} )^{-1}  = (\Sigma^{s}_{0,t} )^{-1} + t \hat{Q}^{s}_{t} $.

	Observe that 
	\begin{align*}
			t \widehat r_t^s	=
		\begin{pmatrix}
	  \sum_{i=1}^t (Z_i^s)^\top Y_i   \\
	 \sum_{i=1}^t (W_i^s)^\top Y_i
	\end{pmatrix}.
	\end{align*}
Moreover, \begin{align*}
		 \sum_{i=1}^t (Z_i^s)^\top Y_i  = 
		 \begin{pmatrix}
		 	 \sum_{i=1}^t 1\{ D_{i} = 0 \} Y_i  \\
		 	 \vdots \\
		 	 \sum_{i=1}^t 1\{ D_{i} = M \} Y_i  	 	 
		 \end{pmatrix}
	 = 	 \operatorname{Diag}[N_{t}]	 \begin{pmatrix}
	 	\frac{1}{N_{t}(0)} \sum_{i=1}^t 1\{ D_{i} = 0 \} Y_i  \\
	 	\vdots \\
	 	\frac{1}{N_{t}(M)}  	\sum_{i=1}^t 1\{ D_{i} = M \} Y_i  	. 	 
	 \end{pmatrix}
\end{align*}
	
\end{proof}

\begin{lemma}\label{lem:charac.Sigma}
	For any instance $t$ and any source $s$, 
	\begin{align}
		(\Sigma_t^s)^{-1}= 	
		\begin{pmatrix}
			A_t^s & 	B^{s}_t \\
			(B^{s}_{t})^\top & C_t^s
		\end{pmatrix},
	\end{align}
	where 
	\begin{align}
		A_t^s &=  \operatorname{Diag}\{ \nu^{s}_{t}(0) + N_{t}(0) , \ldots ,  \nu^{s}_{t}(M) + N_{t}(M)  \},
		\\[6pt]
		B^{s}_{t} & =  			\begin{pmatrix}
			\sum_{i=1}^t 1\{D_i=0\}(W_i^{s})^\top\\
			\vdots\\
			\sum_{i=1}^t 1\{D_i=M\}(W_i^{s})^\top
		\end{pmatrix}\\
		C_t^s &= \sum_{i=1}^{t}  \{  \Sigma^{s}_{\gamma\gamma,t}/t 	+  	 (W_i^s)^\top W_i^s \}. 
	\end{align}

Moreover,
\begin{align*}
	(t \Sigma_t^s)^{-1} 
	& = 	\begin{pmatrix}
		\operatorname{Diag}[ \delta \cdot (c^{s}  + 1) ]    &\operatorname{Diag}[ \delta  ]  E[\boldsymbol{W}^{s} | D ]  \\
		(   E[\boldsymbol{W}^{s} |D]  )^{\top} \operatorname{Diag}[ \delta  ]   &   E[ ( W^{s} ) (W^{s})^{\top} ] 
	\end{pmatrix}  
	+ o_{as}(1) \\
	& = : (\bar{\Sigma}^{s})^{-1} + o_{as}(1),
\end{align*}
where $E[\boldsymbol{W}^{s}|D] $ is a $M \times p_{S}$ matrix given by,
\begin{align*}
	E[\boldsymbol{W}^{s}|D] 	= 	\begin{pmatrix}
		E[ (W^{s})^{\top} \mid D=0  ] \\
		\vdots\\
		E[ (W^{s})^{\top} \mid D=M  ]
	\end{pmatrix}
\end{align*}
\end{lemma}

\begin{proof}
	Observe that
	\begin{align*}
		\hat{Q}^{ZZ}_{t} &= t^{-1} \sum_{i=1}^{t} Z^{\top}_{t} Z_{t} =  \operatorname{Diag}\{ N_{t}(0)/t , \ldots, N_{t}(M) /t \}.
	\end{align*}
	Also, since  $Z_i = \big(1\{D_i=0\},\ldots,1\{D_i=M\}\big)$,  we have
	\begin{align}
		\hat Q^{ZW,s}_t
		&=
		\begin{pmatrix}
			\frac{1}{t}\sum_{i=1}^t 1\{D_i=0\}(W_i^{s})^\top\\
			\vdots\\
			\frac{1}{t}\sum_{i=1}^t 1\{D_i=M\}(W_i^{s})^\top
		\end{pmatrix}.
	\end{align}
	So
	\begin{align}
		\hat Q^{ZW,s}_t
		&=
		\operatorname{Diag}\!\left(
		\frac{N_t(0)}{t},\ldots,\frac{N_t(M)}{t}
		\right)
		\begin{pmatrix}
			\bar W^{s}_t(0)^\top\\
			\vdots\\
			\bar W^{s}_t(M)^\top
		\end{pmatrix}.
	\end{align}
	where
	\begin{align}
		\bar W^{s}_t(d) := \frac{1}{N_t(d)}\sum_{i=1}^t 1\{D_i=d\} W_i^{s}
		\quad \text{(for } N_t(d)>0\text{)}.
	\end{align}
	
	Then
	\begin{align*}
		(\Sigma_t^s)^{-1}
		&=
		(\Sigma_{0,t}^s)^{-1}
		+
		t\,\widehat Q_t^s
		\\[6pt]
		&=
		\begin{pmatrix}
			\Sigma_{\theta\theta,t}^s & 0 \\
			0 & \Sigma_{\gamma\gamma,t}^s
		\end{pmatrix}
		+
		\begin{pmatrix}
			\operatorname{Diag}[N_t]
			&
			t \hat Q^{ZW,s}_t 
			\\
			t  ( \hat {Q}^{ZW,s}_t)^{\top} 
			&
			\sum_{i=1}^t (W_i^s)^\top W_i^s
		\end{pmatrix}
		\\[10pt]
		&=
		\begin{pmatrix}
			A_t^s & 	\hat Q^{ZW,s}_t \\
		( 	\hat Q^{ZW,s}_t)^\top & C_t^s
		\end{pmatrix},
	\end{align*}
	where 
	\begin{align}
		A_t^s &=  \operatorname{Diag}  [ \nu^{s}_{t} + N_{t}(0)], 
		\\[6pt]
		C_t^s &=  \Sigma^{s}_{\gamma\gamma,t}	+  \sum_{i=1}^{t} 	 (W_i^s)^\top (W_i^s) .
	\end{align}

Moreover,	\begin{align}
	t \hat Q^{ZW,s}_t
	&=
	\operatorname{Diag}[N_t]
	\begin{pmatrix}
		\bar W^{s}_t(0)^\top\\
		\vdots\\
		\bar W^{s}_t(M)^\top
	\end{pmatrix} = : 	\operatorname{Diag}[N_t] \bar{\boldsymbol{W}}^{s}_{t}
\end{align}
So,
	\begin{align*}
	(t \Sigma_t^s)^{-1}
	&=
	\begin{pmatrix}
		A_t^s/t & 	\hat Q^{ZW,s}_t/t \\
		( 	\hat Q^{ZW,s}_t)^\top/t & C_t^s/t
	\end{pmatrix} \\
& = 	\begin{pmatrix}
	\operatorname{Diag}[\nu^{s}/t + N_t/t ] & \operatorname{Diag}[N_t/t] \bar{\boldsymbol{W}}^{s}_{t} \\
( \bar{\boldsymbol{W}}^{s}_{t})^{\top}  \operatorname{Diag}[N_t/t]  & t^{-1} C^{s}_{t}.
\end{pmatrix}  
\end{align*}

By the fact that $N_{t}$ diverges (see Lemma \ref{lem:N.diverge} in Appendix \ref{app:AlmostSure}) and $\lambda_{t}/t = o(1)$, it follows that 
\begin{align*}
	\operatorname{Diag}[\nu^{s}_{t}/t + N_t/t ] =  &	\operatorname{Diag}[ \delta \cdot c^{s} ] + 	\operatorname{Diag}[ \delta]  + o_{as}(1) \\
	  \operatorname{Diag}[N_t/t] \bar{\boldsymbol{W}}^{s}_{t} = & 	\operatorname{Diag}[ \delta  ]  E[\boldsymbol{W}^{s}|D] + o_{as}(1)  \\
	   t^{-1} C^{s}_{t} = &  E[ (W^{s})  (W^{s})^{\top} ] + o_{as}(1), 
\end{align*}
where $E[\boldsymbol{W}^{s}|D] $ is a $M \times p_{S}$ matrix given by,
\begin{align*}
E[\boldsymbol{W}^{s}|D] 	= 	\begin{pmatrix}
		 E[ (W^{s})^{\top} \mid D=0  ] \\
		\vdots\\
	E[ (W^{s})^{\top} \mid D=M  ]
	\end{pmatrix}
\end{align*}

Hence,
	\begin{align*}
	(t \Sigma_t^s)^{-1}
	= 	\begin{pmatrix}
		\operatorname{Diag}[ \delta \cdot (c^{s}  + 1) ]   &\operatorname{Diag}[ \delta  ]  E[\boldsymbol{W}^{s}|D]  \\
		(   E[\boldsymbol{W}^{s}|D]  )^{\top} \operatorname{Diag}[ \delta  ]   &  E[ W^{s} (W^{s})^{\top} ] 
	\end{pmatrix}  
+ o_{as}(1). 
\end{align*}

\end{proof}

\begin{proof}[Proof of Lemma \ref{lem:controls.zeta.posterior}]
	
	For any vector $X$, let 	 $\operatorname{Diag}[X] : = \operatorname{Diag}\{  X_{1} , \ldots , X_{d}  \}$. Under this notation   $\operatorname{Diag}[N_{t}] : = \operatorname{Diag}\{  N_{t}(0) , \ldots , N_{t}(M)  \}$ and $\operatorname{Diag}[\nu^{s}_{t}] = \Omega_{\theta\theta,t}^{s}$. Also, let 	$\boldsymbol{m}_{t} : = (m_{t}(0),\ldots,m_{t}(M))^{\top}$ where $m_{t}(d) : = (N_{t}(d))^{-1} \sum_{i=1}^{n} 1\{D_{i} = d\} Y_{i}(d)$ for any $d \in \mathbb{D}$.  
	
	\begin{align}
		\beta^{s}_{t} = \Sigma^{s}_{t}
		\begin{pmatrix}
			\Omega_{\theta\theta,t}^s \zeta_0^s +  \sum_{i=1}^{t} Z^{\top}_{i} Y_{i}  \\
			\Omega_{\gamma\gamma,t}^s \eta_0^s +  t   \left(  t^{-1} \sum_{i=1}^{t} (W^{s}_{i})^{\top} Y_{i} \right)  
		\end{pmatrix}
	\end{align}
	
	Observe that $\sum_{i=1}^{t} Z^{\top}_{i} Y_{i}  = \operatorname{Diag}[N_{t}] \boldsymbol{m}_{t}$. So,
	\begin{align}
		\beta^{s}_{t} = \Sigma^{s}_{t}
		\begin{pmatrix}
			\operatorname{Diag}[\nu^{s}_{t}]  \boldsymbol{\zeta}_0^s +   \operatorname{Diag}[N_{t}]  \boldsymbol{m}_{t} \\
			t \left( \operatorname{Diag}[\lambda^{s}_{t}/t]   \eta_0^s +     t^{-1} \sum_{i=1}^{t} (W^{s}_{i})^{\top} Y_{i} \right)  
		\end{pmatrix}
	\end{align}

	Let 
	\[
	T_t^s
	:=
	A_t^s - B_t^s (C_t^s)^{-1} (B_t^s)^\top .
	\]
	
	\[
	\Sigma^{s}_{t} 
	=
	\begin{pmatrix}
		(T_t^s)^{-1}
		&
		-(T_t^s)^{-1} B_t^s (C_t^s)^{-1}
		\\[8pt]
		-(C_t^s)^{-1}(B_t^s)^\top (T_t^s)^{-1}
		&
		(C_t^s)^{-1}
		+
		(C_t^s)^{-1}(B_t^s)^\top (T_t^s)^{-1} B_t^s (C_t^s)^{-1}
	\end{pmatrix}.
	\]
	
	In particular, 
	\begin{align*}
		\boldsymbol{\zeta}^{s}_{t} = 
		(T^{s}_{t})^{-1} \left( 	\operatorname{Diag}[\nu^{s}_{t}]  \boldsymbol{\zeta}_0^s +   \operatorname{Diag}[N_{t}]  \boldsymbol{m}_{t}  -  B_t^s (C_t^s)^{-1} \left( \operatorname{Diag}[\lambda^{s}_{t}]   \eta_0^s +    \sum_{i=1}^{t} (W^{s}_{i})^{\top} Y_{i} \right)      \right)
	\end{align*}
	
\end{proof}

\subsubsection{Proof of Proposition \ref{pro:control.weights.EV}}

\label{app:contro.weights.EV}

We establish a series of lemmas used in the proof of Proposition \ref{pro:control.weights.EV}; their proofs are relegated to the end of the section. 

\begin{lemma}\label{lem:ols-center-decomp}
	For any source $s$, let $Y:=(Y_{1},\ldots,Y_{t})^\top$ and let $X^{s}$ denote the associated $t\times k$ design matrix whose $i$-th row is $(X_i^{s})^\top$. Suppose $(X^{s})^\top X^{s}$ is invertible, and define the OLS estimator
	\[
	\hat{\beta}_{t}^{s}:=\big((X^{s})^\top X^{s}\big)^{-1}(X^{s})^\top Y.
	\]
	Then, for every $\beta\in\mathbb R^{k}$,
	\begin{align}
		\prod_{i=1}^{t}\phi(Y_i;X_i^{s\top}\beta,1)
		\propto
		\exp\!\left\{
		-\frac12
		(Y-X^{s}\hat{\beta}_{t}^{s})^\top (Y-X^{s}\hat{\beta}_{t}^{s})
		\right\}
		\exp\!\left\{
		-\frac12
		(\beta-\hat{\beta}_{t}^{s})^\top (X^{s})^\top X^{s}(\beta-\hat{\beta}_{t}^{s})
		\right\}.
		\label{eq:ols-center-likelihood}
	\end{align}
\end{lemma}

\begin{lemma}\label{lem:marginal-likelihood-ols-center}
	For any source $s$, let $Y:=(Y_{1},\ldots,Y_{t})^\top$ and let $X^{s}$ denote the associated $t\times k$ design matrix whose $i$-th row is $(X_i^{s})^\top$. Suppose $(X^{s})^\top X^{s}$ is invertible and define the OLS estimator $\hat{\beta}_{t}^{s}:=\big((X^{s})^\top X^{s}\big)^{-1}(X^{s})^\top Y$. 	Then
	\begin{align}
		m_t^{s}(Y_{1:t}\mid D_{1:t},w_{1:t}^{s})
		&:=\int  \prod_{i=1}^{t} \phi(Y_{i} ;  X^{s}_{i} \beta , 1 )
		\phi (\beta;\beta_{0}^{s},\Sigma_{0,t}^{s})\,d\beta \nonumber \\
		&=
		(2\pi)^{-t/2}
		|(X^{s})^\top X^{s}|^{-1/2}
		\exp\!\left\{
		-\frac12
		(Y-X^{s}\hat{\beta}_{t}^{s})^\top(Y-X^{s}\hat{\beta}_{t}^{s})
		\right\} \nonumber\\
		&\quad\times
		\left|
		\Sigma_{0,t}^{s}+\big((X^{s})^\top X^{s}\big)^{-1}
		\right|^{-1/2}
		\nonumber\\
		&\quad\times
		\exp\!\left\{
		-\frac12
		(\hat{\beta}_{t}^{s}-\beta_{0}^{s})^\top
		\left(
		\Sigma_{0,t}^{s}+\big((X^{s})^\top X^{s}\big)^{-1}
		\right)^{-1}
		(\hat{\beta}_{t}^{s}-\beta_{0}^{s})
		\right\}.
		\label{eq:marginal-likelihood-ols}
	\end{align}
\end{lemma}

\begin{lemma}\label{lem:characterization-ms-ols-center}
	For any source $s$, let $Y:=(Y_{1},\ldots,Y_{t})^\top$ and let $X^{s}$ denote the associated $t\times k$ design matrix whose $i$-th row is $(X_i^{s})^\top$. Suppose $(X^{s})^\top X^{s}$ is invertible and that
	\[
	Y=X^{s}\beta^{s}+\varepsilon^{s}.
	\]
	Define
	\[
	\hat{\beta}_{t}^{s}:=\big((X^{s})^\top X^{s}\big)^{-1}(X^{s})^\top Y,
	\qquad
	\Delta_t^s:=\hat{\beta}_{t}^{s}-\beta^{s},
	\]
	and
	\[
	V_t^s:=\Sigma_{0,t}^{s}+\big((X^{s})^\top X^{s}\big)^{-1},
	\qquad
	M_t^s:=(V_t^s)^{-1}.
	\]
	Then the marginal likelihood
	\begin{align}
		m_t^{s}(Y_{1:t}\mid D_{1:t},w_{1:t}^{s})
		:=
		\int  \prod_{i=1}^{t} \phi(Y_{i} ;  X^{s}_{i} \beta , 1 )
		\phi (\beta;\beta_{0}^{s},\Sigma_{0,t}^{s})\,d\beta
	\end{align}
	admits the characterization
	\begin{align}
		m_t^{s}(Y_{1:t}\mid D_{1:t},w_{1:t}^{s})
		&=
		(2\pi)^{-t/2}
		|(X^{s})^\top X^{s}|^{-1/2}
		|V_t^s|^{-1/2}\\
		& \times \exp\!\left\{
		-\frac12
		\Big[
		(\beta^{s}-\beta_{0}^{s})^\top
		M_t^s
		(\beta^{s}-\beta_{0}^{s})
		+
			(\varepsilon^{s})^\top\varepsilon^{s} + o_{as}(t)
		\Big]
		\right\},
		\label{eq:ms-characterization}
	\end{align}
%
\end{lemma}

\begin{lemma}\label{lem:Mt-expansion}
	Let $M_t^s 	:= 	\left( 	\Sigma_{0,t}^{s}+\big((X^{s})^\top X^{s}\big)^{-1}	\right)^{-1}$, 
	and suppose
	\[
	(X^{s})^\top X^{s}=t\hat Q_t^s,
	\qquad
	\hat Q_t^s=\bar Q^s+o_{as}(1).
	\]
	Assume moreover that
	\[
	\left(\frac{\Sigma_{0,t}^{s}}{t}\right)^{-1}
	=
	\Lambda^s+o_{as}(1),
	\qquad
	\Lambda^s
	=
	\begin{pmatrix}
		D^s & 0\\
		0 & 0
	\end{pmatrix},
	\qquad
	D^s:=\operatorname{Diag}(c^s\cdot \delta),
	\]
	and partition
	\[
	\bar Q^s=
	\begin{pmatrix}
		\bar Q^s[1,1] & \bar Q^s[1,2]\\
		\bar Q^s[2,1] & \bar Q^s[2,2]
	\end{pmatrix}.
	\]
	Then
	\begin{align}
		\frac{M_t^s}{t}
		&=
		\begin{pmatrix}
			D^s-
			D^s
			\Big(
			\bar Q^s[1,1]+D^s-\bar Q^s[1,2](\bar Q^s[2,2])^{-1}\bar Q^s[2,1]
			\Big)^{-1}
			D^s
			&
			0\\[1.2ex]
			0&0
		\end{pmatrix}
		+o_{as}(1).
		\label{eq:Mt-expansion}
	\end{align}
\end{lemma}

\begin{lemma}\label{lem:determinant-factor-expansion}
	Let
	\[
	V_t^s:=\Sigma_{0,t}^s+\big((X^s)^\top X^s\big)^{-1},~
	\Lambda_t^s:=\left(\frac{\Sigma_{0,t}^s}{t}\right)^{-1},~and~
	(X^s)^\top X^s= : t\hat Q_t^s.
	\]
	Then
	\begin{align}
		\big|(X^s)^\top X^s\big|^{-1/2}\,|V_t^s|^{-1/2}
		&=
		\left|I+\Sigma_{0,t}^s (X^s)^\top X^s\right|^{-1/2}
		\label{eq:det-factor-identity-1}
		\\
		&=
		\left|I+t^2(\Lambda_t^s)^{-1}\hat Q_t^s\right|^{-1/2}
		\label{eq:det-factor-identity-2}
		\\
		&=
		|\Lambda_t^s|^{1/2}\,
		\big|\Lambda_t^s+t^2\hat Q_t^s\big|^{-1/2}.
		\label{eq:det-factor-identity-3}
	\end{align}
	If, in addition, $\hat Q_t^s=\bar Q^s+o_{as}(1)$ and $\Lambda_t^s=\Lambda^s+o_{as}(1)$, then
	\begin{align} 	\label{eq:det-factor-asymptotic}
		\big|(X^s)^\top X^s\big|^{-1/2}\,|V_t^s|^{-1/2}
		&=
		t^{-d_s}\,
		|\Lambda_t^s|^{1/2}
		\left|
		\bar Q^s+t^{-2}\Lambda_t^s+o_{as}(1)
		\right|^{-1/2}\\ 
		&=	t^{-d_s}	|\Lambda_t^s|^{1/2} 	|\bar Q^s|^{-1/2} 	\big(1+o_{as}(1)\big).
		\label{eq:det-factor-asymptotic-simple}
	\end{align}
\end{lemma}

\begin{proof}[Proof of Proposition \ref{pro:control.weights.EV}]
	
	By Lemma \ref{lem:characterization-ms-ols-center},
		\begin{align}
		m_t^{s}(Y_{1:t}\mid D_{1:t},w_{1:t}^{s})
		&=
		(2\pi)^{-t/2}
		|(X^{s})^\top X^{s}|^{-1/2}
		|V_t^s|^{-1/2}\\
		& \times \exp\!\left\{
		-\frac12
		\Big[
		(\beta^{s}-\beta_{0}^{s})^\top
		M_t^s
		(\beta^{s}-\beta_{0}^{s})
		+
		(\varepsilon^{s})^\top\varepsilon^{s} + o_{as}(t)
		\Big]
		\right\}.
	\end{align}
	
	By Lemma \ref{lem:Mt-expansion}, 
			\begin{align}
		m_t^{s}(Y_{1:t}\mid D_{1:t},w_{1:t}^{s})
		&=
		(2\pi)^{-t/2}
		|(X^{s})^\top X^{s}|^{-1/2}
		|V_t^s|^{-1/2}\\
		& \times \exp\!\left\{
		-\frac t2
		\Big[
		(\boldsymbol{\theta}^{s}-\boldsymbol{\zeta}_{0}^{s})^\top \bar{H}^{s}
		(	\boldsymbol{\theta}^{s}-\boldsymbol{\zeta}_{0}^{s} )
		+
		t^{-1}(\varepsilon^{s})^\top\varepsilon^{s} + o_{as}(1)
		\Big]
		\right\},
	\end{align}
where 
\begin{align*}
	\bar{H}^{s} : 	= & D_c
	-
	D_c
	\Big(
	\operatorname{Diag}[(1+c^s)\cdot\delta]
	-
	\bar Q^s[1,2](\bar Q^s[2,2])^{-1}\bar Q^s[2,1]
	\Big)^{-1}  D_c.
\end{align*}

	Under Lemma \ref{lem:determinant-factor-expansion}, 
	\begin{align*}
	\log 	\big|(X^s)^\top X^s\big|^{-1/2}\,|V_t^s|^{-1/2} = &	- d_{s} \log t  - \frac{1}{2} \log | \bar{Q}^{s} (\Sigma^{s}_{0,t}/t) | +	o_{as}(1) 
	\end{align*}
	
	Also, under our assumptions, $	t^{-1}(\varepsilon^{s})^\top\varepsilon^{s}  = (\sigma_{\varepsilon}^{s})^{2} + o_{as}(1)$, but this term is asymptotically dominated by $d_{s} \log t$. 	Therefore, 
		\begin{align}
		m_t^{s}(Y_{1:t}\mid D_{1:t},w_{1:t}^{s})
		 \propto & \exp \left\{  - \left[  d_{s} \log t  + \frac{1}{2} \log | \bar{Q}^{s} (\Sigma^{s}_{0,t}/t) | +	o_{as}(1)   \right]   \right\}  \\
		& \times \exp\!\left\{
		-\frac t2
		\Big[
		(\boldsymbol{\theta}^{s}-\boldsymbol{\zeta}_{0}^{s})^\top \bar{H}^{s}
		(	\boldsymbol{\theta}^{s}-\boldsymbol{\zeta}_{0}^{s} )
	  + o_{as}(1)
		\Big]
		\right\}.
	\end{align}
	
Finally, to show that
\begin{align*}
	K
	:=
	\operatorname{Diag}[(1+c^s)] E[(Z)(Z)^{\top}]
	-
	E[Z \boldsymbol W^{s}]
	\big(E[W^{s}(W^{s})^\top]\big)^{-1}
	E[Z \boldsymbol W^{s}]^{\top}
\end{align*}	
observe that $\operatorname{Diag}[\delta] = E[(Z)(Z)^{\top}]$ and $\bar{Q}^{s}[1,2] = \operatorname{Diag}[\delta] E[\boldsymbol{W}^{s} \mid D] = E[\boldsymbol{W}^{s}  D]$. 
\end{proof}

\subsubsection{Proofs of Supplemental Lemmas}

\begin{proof}[Proof of Lemma \ref{lem:ols-center-decomp}]
	Write $r_t^s:=Y-X^{s}\hat{\beta}_{t}^{s}$	for the OLS residual vector. Then
	\[
	Y-X^{s}\beta
	=
	Y-X^{s}\hat{\beta}_{t}^{s}-X^{s}(\beta-\hat{\beta}_{t}^{s})
	=
	r_t^s-X^{s}(\beta-\hat{\beta}_{t}^{s}).
	\]
	Therefore,
	\begin{align*}
		(Y-X^{s}\beta)^\top(Y-X^{s}\beta)
		&=
		\big(r_t^s-X^{s}(\beta-\hat{\beta}_{t}^{s})\big)^\top
		\big(r_t^s-X^{s}(\beta-\hat{\beta}_{t}^{s})\big) \\
		&=
		(r_t^s)^\top r_t^s
		+
		(\beta-\hat{\beta}_{t}^{s})^\top (X^{s})^\top X^{s}(\beta-\hat{\beta}_{t}^{s})
		-
		2(r_t^s)^\top X^{s}(\beta-\hat{\beta}_{t}^{s}).
	\end{align*}
	Since $\hat{\beta}_{t}^{s}$ is the OLS estimator, it satisfies the normal equations
	\[
	(X^{s})^\top(Y-X^{s}\hat{\beta}_{t}^{s})=0 \Rightarrow 	(X^{s})^\top r_t^s=0.
	\]
	Hence the cross term vanishes: $(r_t^s)^\top X^{s}(\beta-\hat{\beta}_{t}^{s})
	= 	\big((X^{s})^\top r_t^s\big)^\top(\beta-\hat{\beta}_{t}^{s}) =0$. 	Substituting this into the previous display yields
	\[
	(Y-X^{s}\beta)^\top(Y-X^{s}\beta)
	=
	(r_t^s)^\top r_t^s
	+
	(\beta-\hat{\beta}_{t}^{s})^\top (X^{s})^\top X^{s}(\beta-\hat{\beta}_{t}^{s}).
	\]
	So, 
	\begin{align*} 
		\prod_{i=1}^{t}\phi(Y_i;X_i^{s\top}\beta,1)
		& \propto
		\exp\!\left\{
		-\frac12 (Y-X^{s}\beta)^\top(Y-X^{s}\beta)
		\right\} \\
		&\propto
		\exp\!\left\{
		-\frac12
		(Y-X^{s}\hat{\beta}_{t}^{s})^\top (Y-X^{s}\hat{\beta}_{t}^{s})
		\right\} \\
		&\qquad\times
		\exp\!\left\{
		-\frac12
		(\beta-\hat{\beta}_{t}^{s})^\top (X^{s})^\top X^{s}(\beta-\hat{\beta}_{t}^{s})
		\right\},
	\end{align*}
	which proves \eqref{eq:ols-center-likelihood}.
\end{proof}

\begin{proof}[Proof of Lemma \ref{lem:marginal-likelihood-ols-center}]
	The Gaussian likelihood satisfies
	\[
	\prod_{i=1}^{t} \phi(Y_{i} ;  X^{s}_{i} \beta , 1 )
	=
	(2\pi)^{-t/2}
	\exp\!\left\{
	-\frac12 (Y-X^{s}\beta)^\top(Y-X^{s}\beta)
	\right\}.
	\]
	By Lemma \ref{lem:ols-center-decomp},
	\begin{align*}
		\prod_{i=1}^{t} \phi(Y_{i} ;  X^{s}_{i} \beta , 1 )
		&=
		(2\pi)^{-t/2}
		\exp\!\left\{
		-\frac12
		(Y-X^{s}\hat{\beta}_{t}^{s})^\top(Y-X^{s}\hat{\beta}_{t}^{s})
		\right\} \\
		&\quad\times
		\exp\!\left\{
		-\frac12
		(\beta-\hat{\beta}_{t}^{s})^\top (X^{s})^\top X^{s}(\beta-\hat{\beta}_{t}^{s})
		\right\}.
	\end{align*}
	
	Since $(X^{s})^\top X^{s}$ is positive definite,
	\[
	\exp\!\left\{
	-\frac12
	(\beta-\hat{\beta}_{t}^{s})^\top (X^{s})^\top X^{s}(\beta-\hat{\beta}_{t}^{s})
	\right\}
	=
	(2\pi)^{k/2}|(X^{s})^\top X^{s}|^{-1/2}
	\phi\!\left(
	\beta;
	\hat{\beta}_{t}^{s},
	\big((X^{s})^\top X^{s}\big)^{-1}
	\right).
	\]
	
	Substituting this identity into the definition of
	$m_t^{s}(Y_{1:t}\mid D_{1:t},w_{1:t}^{s})$ yields
	\begin{align*}
		m_t^{s}(Y_{1:t}\mid D_{1:t},w_{1:t}^{s})
		&=
		(2\pi)^{-t/2}(2\pi)^{k/2}|(X^{s})^\top X^{s}|^{-1/2}
		\exp\!\left\{
		-\frac12
		(Y-X^{s}\hat{\beta}_{t}^{s})^\top(Y-X^{s}\hat{\beta}_{t}^{s})
		\right\} \\
		&\quad\times
		\int
		\phi\!\left(
		\beta;
		\hat{\beta}_{t}^{s},
		\big((X^{s})^\top X^{s}\big)^{-1}
		\right)
		\phi(\beta;\beta_{0}^{s},\Sigma_{0,t}^{s})\,d\beta.
	\end{align*}
	
	Using the convolution identity for Gaussian densities,
	\[
	\int \phi(\beta;\mu_{1},\Sigma_{1})\phi(\beta;\mu_{2},\Sigma_{2})\,d\beta
	=
	\phi(\mu_{1};\mu_{2},\Sigma_{1}+\Sigma_{2}),
	\]
	with $\mu_{1}=\hat{\beta}_{t}^{s},~
	\Sigma_{1}=\big((X^{s})^\top X^{s}\big)^{-1},~
	\mu_{2}=\beta_{0}^{s},~and~
	\Sigma_{2}=\Sigma_{0,t}^{s}$, 	gives
	\[
	\int
	\phi\!\left(
	\beta;
	\hat{\beta}_{t}^{s},
	\big((X^{s})^\top X^{s}\big)^{-1}
	\right)
	\phi(\beta;\beta_{0}^{s},\Sigma_{0,t}^{s})\,d\beta
	=
	\phi\!\left(
	\hat{\beta}_{t}^{s};
	\beta_{0}^{s},
	\Sigma_{0,t}^{s}+\big((X^{s})^\top X^{s}\big)^{-1}
	\right).
	\]
	
	Substituting the expression of this Gaussian density and canceling the
	$(2\pi)^{k/2}$ factor yields \eqref{eq:marginal-likelihood-ols}.
\end{proof}

\begin{proof}[Proof of Lemma \ref{lem:characterization-ms-ols-center}]
	By Lemma \ref{lem:marginal-likelihood-ols-center},
	\begin{align}
		m_t^{s}(Y_{1:t}\mid D_{1:t},w_{1:t}^{s})
		&=
		(2\pi)^{-t/2}
		|(X^{s})^\top X^{s}|^{-1/2}
		|V_t^s|^{-1/2}
		\exp\!\left\{
		-\frac12
		(Y-X^{s}\hat{\beta}_{t}^{s})^\top(Y-X^{s}\hat{\beta}_{t}^{s})
		\right\}
		\nonumber\\
		&\quad\times
		\exp\!\left\{
		-\frac12
		(\hat{\beta}_{t}^{s}-\beta_{0}^{s})^\top
		M_t^s
		(\hat{\beta}_{t}^{s}-\beta_{0}^{s})
		\right\}.
		\label{eq:proof-start-ms-char}
	\end{align}
	Therefore,
	\begin{align}
		m_t^{s}(Y_{1:t}\mid D_{1:t},w_{1:t}^{s})
		&=
		(2\pi)^{-t/2}
		|(X^{s})^\top X^{s}|^{-1/2}
		|V_t^s|^{-1/2} \\
		& 	\times \exp\!\left\{
		-\frac12
		\Big[
		(Y-X^{s}\hat{\beta}_{t}^{s})^\top(Y-X^{s}\hat{\beta}_{t}^{s}) 	+
		(\hat{\beta}_{t}^{s}-\beta_{0}^{s})^\top
		M_t^s
		(\hat{\beta}_{t}^{s}-\beta_{0}^{s})
		\Big]
		\right\}.
		\label{eq:proof-start-ms-char-2}
	\end{align}
	
	Next, since $\hat{\beta}_{t}^{s}=\beta^{s}+\Delta_t^s$, we have $\hat{\beta}_{t}^{s}-\beta_{0}^{s} 	=	\beta^{s}-\beta_{0}^{s}+\Delta_t^s$, 	and thus
	\begin{align}
		(\hat{\beta}_{t}^{s}-\beta_{0}^{s})^\top M_t^s(\hat{\beta}_{t}^{s}-\beta_{0}^{s})
		&=
		(\beta^{s}-\beta_{0}^{s})^\top
		M_t^s
		(\beta^{s}-\beta_{0}^{s})
		+
		2(\beta^{s}-\beta_{0}^{s})^\top
		M_t^s
		\Delta_t^s
		+
		(\Delta_t^s)^\top M_t^s\Delta_t^s.
		\label{eq:second-qf-char}
	\end{align}
	Also, using $Y=X^{s}\beta^{s}+\varepsilon^{s}$, $Y-X^{s}\hat{\beta}_{t}^{s} 	= 	\varepsilon^{s}-X^{s}\Delta_t^s$.  Hence
	\begin{align}
		(Y-X^{s}\hat{\beta}_{t}^{s})^\top(Y-X^{s}\hat{\beta}_{t}^{s})
		&=
		(\varepsilon^{s})^\top\varepsilon^{s}
		-2(\varepsilon^{s})^\top X^{s}\Delta_t^s
		+
		(\Delta_t^s)^\top (X^{s})^\top X^{s}\Delta_t^s.
		\label{eq:first-qf-char-a}
	\end{align}
	Because $\hat{\beta}_{t}^{s}$ is the OLS estimator, it satisfies $(X^{s})^\top(Y-X^{s}\hat{\beta}_{t}^{s})=0$.  Substituting $Y=X^{s}\beta^{s}+\varepsilon^{s}$ and
	$\hat{\beta}_{t}^{s}=\beta^{s}+\Delta_t^s$ yields $(X^{s})^\top\varepsilon^{s}=(X^{s})^\top X^{s}\Delta_t^s$. Therefore, $(\varepsilon^{s})^\top X^{s}\Delta_t^s 	= 	(\Delta_t^s)^\top (X^{s})^\top\varepsilon^{s} 	=
	(\Delta_t^s)^\top (X^{s})^\top X^{s}\Delta_t^s$, and \eqref{eq:first-qf-char-a} becomes
	\begin{align}
		(Y-X^{s}\hat{\beta}_{t}^{s})^\top(Y-X^{s}\hat{\beta}_{t}^{s})
		&=
		(\varepsilon^{s})^\top\varepsilon^{s}
		-
		(\Delta_t^s)^\top (X^{s})^\top X^{s}\Delta_t^s.
		\label{eq:first-qf-char}
	\end{align}
	Combining \eqref{eq:second-qf-char} and \eqref{eq:first-qf-char}, we obtain
	\begin{align}
		&\quad
		(Y-X^{s}\hat{\beta}_{t}^{s})^\top(Y-X^{s}\hat{\beta}_{t}^{s})
		+
		(\hat{\beta}_{t}^{s}-\beta_{0}^{s})^\top
		M_t^s
		(\hat{\beta}_{t}^{s}-\beta_{0}^{s})
		\nonumber\\
		&=
		(\beta^{s}-\beta_{0}^{s})^\top
		M_t^s
		(\beta^{s}-\beta_{0}^{s})
		+
		(\varepsilon^{s})^\top\varepsilon^{s}
		+
		2(\beta^{s}-\beta_{0}^{s})^\top M_t^s\Delta_t^s
		+
		(\Delta_t^s)^\top
		\Big(
		M_t^s-(X^{s})^\top X^{s}
		\Big)\Delta_t^s
		\nonumber\\
		&=
		(\beta^{s}-\beta_{0}^{s})^\top
		M_t^s
		(\beta^{s}-\beta_{0}^{s})
		+
		R_t^s.
		\label{eq:sum-qf-char}
	\end{align}
	Substituting \eqref{eq:sum-qf-char} into \eqref{eq:proof-start-ms-char-2} yields
	\begin{align}
		m_t^{s}(Y_{1:t}\mid D_{1:t},w_{1:t}^{s})
		&=
		(2\pi)^{-t/2}
		|(X^{s})^\top X^{s}|^{-1/2}
		|V_t^s|^{-1/2}\\
		& \times \exp\!\left\{
		-\frac12
		\Big[
		(\beta^{s}-\beta_{0}^{s})^\top
		M_t^s
		(\beta^{s}-\beta_{0}^{s})
		+ R_{t}^{s}
		\Big]
		\right\},
		\label{eq:ms-characterization}
	\end{align}
	where
	\begin{align}
		R_t^s
		:=
		(\varepsilon^{s})^\top\varepsilon^{s}
		+
		2(\beta^{s}-\beta_{0}^{s})^\top M_t^s\Delta_t^s
		+
		(\Delta_t^s)^\top
		\Big(
		M_t^s-(X^{s})^\top X^{s}
		\Big)\Delta_t^s.
		\label{eq:Rt-characterization}
	\end{align}
	
	Observe that
	\[
	R_t^s-(\varepsilon^{s})^\top\varepsilon^{s}
	=
	2(\beta^{s}-\beta_{0}^{s})^\top M_t^s\Delta_t^s
	+
	(\Delta_t^s)^\top
	\Big(
	M_t^s-(X^{s})^\top X^{s}
	\Big)\Delta_t^s.
	\]
	
	By standard properties of OLS estimators,
	\begin{align}
		\Delta_t^s=O_{as}(t^{-1/2}),
		\qquad
		\|M_t^s\|=O(t),
		\qquad
		\|(X^{s})^\top X^{s}\|=O(t),
		\qquad
		\|\beta^{s}-\beta_{0}^{s}\|=O(1),
		\label{eq:rate-conditions-R}
	\end{align}

	For the linear term, by Cauchy--Schwarz and \eqref{eq:rate-conditions-R},
	\begin{align}
		\left|
		2(\beta^{s}-\beta_{0}^{s})^\top M_t^s\Delta_t^s
		\right|
		&\leq
		2\|\beta^{s}-\beta_{0}^{s}\|\,\|M_t^s\|\,\|\Delta_t^s\|
		=
		O(1)\,O(t)\,O_{as}(t^{-1/2})
		=
		O_{as}(\sqrt{t}).
		\label{eq:bound-linear-R}
	\end{align}
	For the quadratic term, again by norm inequalities and \eqref{eq:rate-conditions-R},
	\begin{align}
		\left|
		(\Delta_t^s)^\top
		\Big(
		M_t^s-(X^{s})^\top X^{s}
		\Big)\Delta_t^s
		\right|
		&\leq
		\|\Delta_t^s\|^2
		\,
		\|M_t^s-(X^{s})^\top X^{s}\| \leq
		\|\Delta_t^s\|^2
		\Big(
		\|M_t^s\|+\|(X^{s})^\top X^{s}\|
		\Big)
		\nonumber\\
		&=
		O_{as}(t^{-1})
		\cdot O(t)
		=
		O_{as}(1).
		\label{eq:bound-quadratic-R}
	\end{align}
	Combining \eqref{eq:bound-linear-R} and \eqref{eq:bound-quadratic-R}, we conclude that
	\[
	R_t^s-(\varepsilon^{s})^\top\varepsilon^{s}
	=
	O_{as}(\sqrt{t}).
	\]
	Since $t^{-1/2}\to 0$, $R_t^s-(\varepsilon^{s})^\top\varepsilon^{s} 	= 	o_{as}(t)$, 	and thus 
	\begin{align}
		m_t^{s}(Y_{1:t}\mid D_{1:t},w_{1:t}^{s})
		&=
		(2\pi)^{-t/2}
		|(X^{s})^\top X^{s}|^{-1/2}
		|V_t^s|^{-1/2}\\
		& \times \exp\!\left\{
		-\frac12
		\Big[
		(\beta^{s}-\beta_{0}^{s})^\top
		M_t^s
		(\beta^{s}-\beta_{0}^{s})
		+ (\varepsilon^{s})^\top\varepsilon^{s} 	+	o_{as}(t)
		\Big]
		\right\}.
		\label{eq:ms-characterization}
	\end{align}

\end{proof}

\begin{proof}[Proof of Lemma \ref{lem:Mt-expansion}]
	Using $(X^{s})^\top X^{s}=t\hat Q_t^s$ we may write $M_t^s
	=
	\left(
	\Sigma_{0,t}^{s}+\big(t\hat Q_t^s\big)^{-1}
	\right)^{-1}$. 	From the identity
	\[
	(A^{-1}+B^{-1})^{-1}=A-A(A+B)^{-1}A,
	\]
	applied with $A:=(\Sigma_{0,t}^{s})^{-1}$ and $B:=t\hat Q_t^s$, 	we obtain
	\[
	M_t^s
	=
	(\Sigma_{0,t}^{s})^{-1}
	-
	(\Sigma_{0,t}^{s})^{-1}
	\Big(
	(\Sigma_{0,t}^{s})^{-1}+t\hat Q_t^s
	\Big)^{-1}
	(\Sigma_{0,t}^{s})^{-1}.
	\]
	Dividing by \(t\) yields
	\begin{align}
		\frac{M_t^s}{t}
		&=
		\frac{(\Sigma_{0,t}^{s})^{-1}}{t}
		-
		\frac{(\Sigma_{0,t}^{s})^{-1}}{t}
		\Big(
		\frac{(\Sigma_{0,t}^{s})^{-1}}{t}+\hat Q_t^s
		\Big)^{-1}
		\frac{(\Sigma_{0,t}^{s})^{-1}}{t}.
		\label{eq:Mt-over-t-prelimit}
	\end{align}
	Now define
	\[
	\Lambda_t^s:=\frac{(\Sigma_{0,t}^{s})^{-1}}{t}
	=
	\left(\frac{\Sigma_{0,t}^{s}}{t}\right)^{-1}.
	\]
	Then \eqref{eq:Mt-over-t-prelimit} becomes
	\begin{align}
		\frac{M_t^s}{t}
		&=
		\Lambda_t^s-\Lambda_t^s(\Lambda_t^s+\hat Q_t^s)^{-1}\Lambda_t^s.
		\label{eq:Mt-over-t-Lambda}
	\end{align}
	By assumption, $\Lambda_t^s=\Lambda^s+o_{as}(1)$ and $\hat Q_t^s=\bar Q^s+o_{as}(1)$. Hence, by continuity of matrix multiplication and inversion,
	\begin{align}
		\frac{M_t^s}{t}
		&=
		\Lambda^s-\Lambda^s(\Lambda^s+\bar Q^s)^{-1}\Lambda^s+o_{as}(1).
		\label{eq:Mt-over-t-limit-form}
	\end{align}
	
	Next, since
	\[
	\Lambda^s=
	\begin{pmatrix}
		D^s & 0\\
		0 & 0
	\end{pmatrix},
	\qquad
	\bar Q^s=
	\begin{pmatrix}
		\bar Q^s[1,1] & \bar Q^s[1,2]\\
		\bar Q^s[2,1] & \bar Q^s[2,2]
	\end{pmatrix},
	\]
	we have
	\[
	\Lambda^s+\bar Q^s
	=
	\begin{pmatrix}
		\bar Q^s[1,1]+D^s & \bar Q^s[1,2]\\
		\bar Q^s[2,1] & \bar Q^s[2,2]
	\end{pmatrix}.
	\]
	Let
	\[
	(\Lambda^s+\bar Q^s)^{-1}
	=
	\begin{pmatrix}
		G_{11}^s & G_{12}^s\\
		G_{21}^s & G_{22}^s
	\end{pmatrix}.
	\]
	Since \(\Lambda^s\) has only a nonzero \((1,1)\) block, we get
	\[
	\Lambda^s(\Lambda^s+\bar Q^s)^{-1}\Lambda^s
	=
	\begin{pmatrix}
		D^sG_{11}^sD^s & 0\\
		0 & 0
	\end{pmatrix}.
	\]
	Therefore
	\begin{align}
		\Lambda^s-\Lambda^s(\Lambda^s+\bar Q^s)^{-1}\Lambda^s
		&=
		\begin{pmatrix}
			D^s-D^sG_{11}^sD^s & 0\\
			0 & 0
		\end{pmatrix}.
		\label{eq:block-reduction}
	\end{align}
	
	It remains to identify \(G_{11}^s\). By the block inverse formula,
	\[
	G_{11}^s
	=
	\Big(
	\bar Q^s[1,1]+D^s-\bar Q^s[1,2](\bar Q^s[2,2])^{-1}\bar Q^s[2,1]
	\Big)^{-1}.
	\]
	Substituting this expression into \eqref{eq:block-reduction}, and then combining with \eqref{eq:Mt-over-t-limit-form}, gives
	\[
	\frac{M_t^s}{t}
	=
	\begin{pmatrix}
		D^s-
		D^s
		\Big(
		\bar Q^s[1,1]+D^s-\bar Q^s[1,2](\bar Q^s[2,2])^{-1}\bar Q^s[2,1]
		\Big)^{-1}
		D^s
		&
		0\\[1.2ex]
		0&0
	\end{pmatrix}
	+o_{as}(1),
	\]
	which proves \eqref{eq:Mt-expansion}.
\end{proof}

\begin{proof}[Proof of Lemma \ref{lem:determinant-factor-expansion}]
	Using $V_t^s=\Sigma_{0,t}^s+\big((X^s)^\top X^s\big)^{-1}$, we have $|V_t^s|
	=
	\left|\big((X^s)^\top X^s\big)^{-1}\right|
	\left|I+(X^s)^\top X^s\Sigma_{0,t}^s\right|$. Therefore,
	\[
	\big|(X^s)^\top X^s\big|^{-1/2}\,|V_t^s|^{-1/2}
	=
	\left|I+(X^s)^\top X^s\Sigma_{0,t}^s\right|^{-1/2}.
	\]
	Using $|I+AB|=|I+BA|$, this yields \eqref{eq:det-factor-identity-1}.
	
	Next, since $\Sigma_{0,t}^s=t(\Lambda_t^s)^{-1}$  and $(X^s)^\top X^s=t\hat Q_t^s$,  we obtain	$\Sigma_{0,t}^s (X^s)^\top X^s 	=	t^2(\Lambda_t^s)^{-1}\hat Q_t^s$,  	which proves \eqref{eq:det-factor-identity-2}.
	
	Now use $\left|I+A^{-1}B\right| 	=	|A|^{-1}|A+B|$, with $A=\Lambda_t^s$ and $B=t^2\hat Q_t^s$. Then
	\[
	\left|I+t^2(\Lambda_t^s)^{-1}\hat Q_t^s\right|
	=
	|\Lambda_t^s|^{-1}\,
	\big|\Lambda_t^s+t^2\hat Q_t^s\big|,
	\]
	and so \eqref{eq:det-factor-identity-3} follows.
	
	Finally, if $t\hat Q_t^s=t\bar Q^s+o_{as}(t)$, 	then
	\[
	\Lambda_t^s+t^2\hat Q_t^s
	=
	\Lambda_t^s+t^2\bar Q^s+o_{as}(t^2)
	=
	t^2\Big(\bar Q^s+t^{-2}\Lambda_t^s+o_{as}(1)\Big).
	\]
	Taking determinants gives
	\[
	\big|\Lambda_t^s+t^2\hat Q_t^s\big|^{-1/2}
	=
	t^{-d_s}
	\left|
	\bar Q^s+t^{-2}\Lambda_t^s+o_{as}(1)
	\right|^{-1/2},
	\]
	which proves \eqref{eq:det-factor-asymptotic}.
\end{proof}

\subsection{Proof of Lemma \ref{lem:zeta-fwl-asymp}}

\begin{proof}[Proof of Lemma \ref{lem:zeta-fwl-asymp}]
	By Lemma \ref{lem:controls.zeta.posterior},
	\begin{align}
		\boldsymbol{\zeta}^{s}_{t}
		= &
		(T_t^s/t)^{-1}
		\Bigg(
		\operatorname{Diag}[\nu_t^s/t]\boldsymbol{\zeta}_0^s
		+
		\operatorname{Diag}[N_t/t]
		\big(
		\boldsymbol{\theta}
		+
		(\bar{\boldsymbol W}_t^s)^{\top}\gamma^s
		+
		\bar{\boldsymbol\varepsilon}_t^s
		\big)
		\Bigg) \\
		& -	(T_t^s/t)^{-1}
		\Bigg(
		(B_t^s/t)(C_t^s/t)^{-1}
		\left(
		\operatorname{Diag}[\lambda_t^s/t]\eta_0^s
		+
		t^{-1}\sum_{i=1}^t (W_i^s)^\top Y_i
		\right)
		\Bigg).
		\label{eq:start-zeta}
	\end{align}
	
	Define,
	\[
	V_t^s = \operatorname{Diag}[\nu_t^s/t], \qquad
	\Pi_t = \operatorname{Diag}[N_t/t], \qquad
	R_t^s = B_t^s/t, \qquad
	Q_t^s = C_t^s/t.
	\]
	Moreover,
	\[
	T_t^s/t
	=
	V_t^s + \Pi_t - R_t^s (Q_t^s)^{-1} (R_t^s)^\top
	=
	V_t^s + M_t^s .
	\]
	
	Next, under the model
	\[
	Y_i = \theta^\top Z_i + (W_i^s)^\top \gamma^s + \varepsilon_i^s,
	\]
	we have
	\begin{align}
		t^{-1}\sum_{i=1}^t (W_i^s)^\top Y_i
		&=
		(R_t^s)^\top \boldsymbol{\theta}
		+
		Q_t^s \gamma^s
		+
		t^{-1}\sum_{i=1}^t (W_i^s)^\top \varepsilon_i^s .
		\label{eq:Wy-expand}
	\end{align}
	Also,
	\[
	\Pi_t (\bar{\boldsymbol W}_t^s)^\top \gamma^s = R_t^s \gamma^s .
	\]
	
	Substituting these identities into \eqref{eq:start-zeta} yields
	\begin{align*}
		\boldsymbol{\zeta}_t^s
		&=
		(V_t^s+M_t^s)^{-1}
		\Bigg[
		V_t^s\boldsymbol{\zeta}_0^s
		+
		\Pi_t \boldsymbol{\theta}
		+
		R_t^s \gamma^s
		+
		\Pi_t \bar{\boldsymbol{\varepsilon}}_t^s
		\\
		&\hspace{2cm}
		-
		R_t^s (Q_t^s)^{-1}
		\left(
		\operatorname{Diag}[\lambda_t^s/t]\eta_0^s
		+
		(R_t^s)^\top \boldsymbol{\theta}
		+
		Q_t^s \gamma^s
		+
		t^{-1}\sum_{i=1}^t (W_i^s)^\top \varepsilon_i^s
		\right)
		\Bigg].
	\end{align*}
	
	Since
	\[
	R_t^s\gamma^s - R_t^s (Q_t^s)^{-1} Q_t^s \gamma^s = 0,
	\]
	the $\gamma^s$ terms cancel. Moreover,
	\[
	\Pi_t \boldsymbol{\theta}
	-
	R_t^s (Q_t^s)^{-1} (R_t^s)^\top \boldsymbol{\theta}
	=
	M_t^s \boldsymbol{\theta}.
	\]
	
	Therefore,
	\begin{align*}
		\boldsymbol{\zeta}_t^s
		&=
		(V_t^s+M_t^s)^{-1}
		\Bigg[
		V_t^s\boldsymbol{\zeta}_0^s
		+
		M_t^s\boldsymbol{\theta}
		+
		\Pi_t\bar{\boldsymbol{\varepsilon}}_t^s
		-
		R_t^s(Q_t^s)^{-1}
		\left(
		\operatorname{Diag}[\lambda_t^s/t]\eta_0^s
		+
		t^{-1}\sum_{i=1}^t (W_i^s)^\top \varepsilon_i^s
		\right)
		\Bigg].
	\end{align*}
	
	Recognizing the remainder $r_t^s$ defined above yields \eqref{eq:zeta-fwl-asymp}. The limit statement follows by continuity of matrix inversion when $V^s+M^s$ is nonsingular.
\end{proof}

\end{document}